\documentclass[aps,prd,showpacs,twocolumn,superscriptaddress,nofootinbib,preprintnumbers]{revtex4-1} 

\makeatletter
\def\p@subsection{}
\makeatother
\usepackage{color,graphicx,amsmath,amssymb,lipsum,bm}
\usepackage[colorlinks=true,citecolor=blue,linkcolor=blue,urlcolor=blue, backref=false,pdfborder={0 0 0}]{hyperref}
\usepackage[normalem]{ulem}
\usepackage[utf8]{inputenc} 
\usepackage{mathtools}
\usepackage{float}
\usepackage{ulem}
\usepackage{diagbox}
\usepackage[dvipsnames]{xcolor}
\usepackage{mathrsfs}
\usepackage{multirow}

\newcommand{\be}{\begin{equation}}
\newcommand{\ee}{\end{equation}}
\newcommand{\beqa}{\begin{eqnarray}}
\newcommand{\eeqa}{\end{eqnarray}}

\newcommand{\bseq}{\begin{subequations}}
\newcommand{\eseq}{\end{subequations}}

\renewcommand{\ln}{\mathop{\rm ln}\nolimits}

\newcommand{\di}{\mathrm d}

\def\gsim{\raise0.3ex\hbox{$\;>$\kern-0.75em\raise-1.1ex\hbox{$\sim\;$}}}
\def\lsim{\raise0.3ex\hbox{$\;<$\kern-0.75em\raise-1.1ex\hbox{$\sim\;$}}}
\def\beqn#1{\begin{equation}\label{#1}}
\def\eeqn{\end{equation}}
\def\beqa#1{\begin{eqnarray}\label{#1}}
\def\eeqa{\end{eqnarray}}

\def\Z2{$\mathcal{Z_2}$}

\newcommand {\ignore}[1]{}

\renewcommand{\arraystretch}{1.3}

\begin{document}

\title{Probing Cosmology through Higher-Order CMB Lensing Statistics}

\author{Shu-Fan Chen} 
\email{sc5848@columbia.edu
 } 
\affiliation{Astrophysics Laboratory, Columbia University, New York, NY 10027, USA}
\affiliation{Department of Physics, Columbia University New York, NY 10027, USA}
\affiliation{Department of Astronomy \& Astrophysics, Columbia University New York, NY 10027, USA}

\author{J.~Colin Hill}
\affiliation{Astrophysics Laboratory, Columbia University, New York, NY 10027, USA}
\affiliation{Department of Physics, Columbia University New York, NY 10027, USA}

\author{Zolt\'an Haiman}
\affiliation{Institute of Science and Technology Austria (ISTA), Am Campus 1, Klosterneuburg 3400, Austria}
\affiliation{Department of Astronomy \& Astrophysics, Columbia University New York, NY 10027, USA}
\affiliation{Department of Physics, Columbia University New York, NY 10027, USA}


\begin{abstract}
We investigate the cosmological information in higher-order statistics of the cosmic microwave background (CMB) lensing convergence field for a near-term experiment with noise properties similar to the Simons Observatory (SO). Using a fully field-level forward-modeling pipeline based on ray-traced simulations from the {\tt MassiveNuS} suite and realistic SO-like CMB lensing reconstruction, we naturally include nonlinear structure formation, post-Born effects, and higher-order reconstruction noise. We measure several non-Gaussian statistics, including Minkowski functionals, peak and minima counts, moments, and wavelet-scattering coefficients. We train Gaussian-process emulators to model each statistic's dependence on the matter density fraction $\Omega_m$, the scalar power spectrum amplitude $A_s$, and the neutrino mass sum $M_\nu$. We quantify the relative information gain these statistics provide beyond the lensing power spectrum and identify which are most robust to reconstruction noise. We find that morphology-based statistics, particularly Minkowski functionals and peak/minima counts, offer significant complementary constraining power: combining all non-Gaussian statistics with the power spectrum yields reductions of 40\% and 38\% in the marginalized uncertainties on $\Omega_m$ and $A_s$, respectively, and a 70\% reduction in the one-sided uncertainty on $M_\nu$. These gains remain non-negligible even when the power spectrum is extended to larger scales and combined with primary CMB and BAO data, with Minkowski functionals providing an additional 11\% improvement in $\sigma(M_\nu)$ and 35\% in $\sigma(\Omega_m)$ beyond the extended power spectrum. By contrast, moments and wavelet-scattering coefficients provide more limited gains at SO noise levels. Our results highlight the potential of non-Gaussian statistics to enhance cosmological constraints from SO and future CMB surveys.
\end{abstract}

\maketitle

\section{Introduction}
Gravitational lensing of the cosmic microwave background (CMB) provides a powerful probe of the projected matter distribution and the growth of structure over cosmic time. As CMB photons propagate from the last scattering surface, deflections induced by intervening large-scale structure imprint coherent mode couplings in the observed temperature and polarization fields (see~\cite{Lewis:2006fu} for a review). These signatures enable the reconstruction of the CMB lensing convergence field, a line-of-sight integral of the matter fluctuation field, yielding tight constraints on key cosmological parameters~\cite{Hu:2001kj}. With the upcoming wide-area, high-resolution, low-noise measurements from the Simons Observatory (SO)~\cite{SimonsObservatory:2018koc,SimonsObservatory:2025wwn}, the fidelity of CMB lensing reconstruction will improve significantly, opening new opportunities for precision cosmology.

Current CMB lensing analyses rely primarily on the two-point function or angular power spectrum of the reconstructed convergence field, $C_{\ell}^{\kappa\kappa}$, as demonstrated in recent measurements from ACT~\cite{ACT:2023dou,ACT:2023kun}, SPT~\cite{Wu:2019hek}, SPT-3G~\cite{SPT:2023jql}, \emph{Planck}~\cite{Planck:2018lbu,Carron:2022eyg}, and their joint analyses~\cite{ACT:2025qjh}. However, nonlinear structure formation generates significant non-Gaussian features, particularly on the small angular scales to which upcoming data will be sensitive, which are not captured by the lensing power spectrum alone. Refs.~\cite{Namikawa:2016jff,Chen:2021vba} study the additional information coming from the bispectrum and find at least $35\%$ improvements on the constraint on neutrino mass when they include it in addition to the power spectrum. A variety of higher-order and morphology-based statistics, including Minkowski functionals~\cite{Minkowski1903}, peak and minima counts~\cite{Bardeen:1985tr,1987MNRAS.226..655B}, low-order moments, and wavelet-scattering coefficients~\cite{2011arXiv1101.2286M}, provide complementary probes of these non-Gaussian structures and have already improved cosmological constraints in galaxy weak-lensing and large-scale-structure analyses (see Refs.~\cite{Cheng:2020qbx,Valogiannis:2022xwu,Sabyr:2024kcu,Liu:2014fzc,Petri:2013ffb,Liu:2018dsw,Coulton:2018ebd,Li:2018owg,Coulton:2019enn,Marques:2018ctl} for example). Ref.~\cite{Liu:2016nfs} has studied the one-point PDF and peak counts of CMB lensing maps, but their potential for CMB lensing has not yet been comprehensively assessed under realistic reconstruction noise for SO, SPT-3G~\cite{SPT-3G:2014dbx}, PICO~\cite{NASAPICO:2019thw}, or other future experiments.

In this work, we take a fully field-level approach to CMB lensing inference. Rather than relying solely on analytical modeling of the lensing power spectrum or perturbative expansions of estimator noise, we forward-model the entire reconstruction pipeline using ray-traced convergence fields from $N$-body simulations and simulated CMB temperature and polarization maps expected from SO-like observations. This procedure naturally incorporates all higher-order effects --- such as post-Born corrections~\cite{Pratten:2016dsm,Barthelemy:2020igw}, nonlinear mode coupling, and the full hierarchy of lensing reconstruction noise biases~\cite{Madhavacheril:2020ido} --- without requiring separate analytical treatments. Because all statistics are measured directly from the reconstructed fields themselves, the resulting constraints reflect the true information content of the lensing maps in the presence of realistic noise, filtering, and non-Gaussian structure.

Building on this framework, our analysis aims to quantify how a broad set of non-Gaussian statistics respond to variations in the matter density fraction $\Omega_m$, the primordial scalar power spectrum amplitude $A_s$, and the sum of neutrino masses $M_\nu$. Because our simulation maps can cover only a small area in the sky, our focus is on the \emph{relative} information content provided by different statistics rather than the absolute constraints achievable with an SO-like experiment.  Higher-order and morphology-based statistics are expected to probe complementary aspects of the reconstructed convergence field compared to those probed by the power spectrum, with much of the additional information coming from the nonlinear regime.  Thus, small-area but high-resolution maps are useful in assessing the information content in these statistics.  In addition, their cosmological sensitivity and robustness in the presence of realistic SO-like reconstruction noise remain insufficiently understood.

With these goals in mind, we assess which statistics are most informative for CMB lensing and under what conditions they provide meaningful gains beyond the standard power spectrum. In particular, we investigate how Minkowski functionals, peak and minima counts, low-order moments, and wavelet-scattering coefficients capture the non-Gaussian structure introduced by nonlinear growth and lensing reconstruction. We also explore their ability to constrain the neutrino mass when analyzed together with $C_\ell^{\kappa\kappa}$. By conducting this comparison within a consistent field-level forward-modeling pipeline, our aim is to clarify the extent to which non-Gaussian statistics can enhance cosmological analyses with SO and other upcoming experiments, and to identify the most promising avenues for incorporating them into future CMB lensing studies.

The remainder of this paper is organized as follows. In Section~\ref{sec:simulation}, we describe the simulations and the forward-modeling pipeline used to generate the reconstructed convergence maps for our forecasts. Section~\ref{sec:higherOrderStatistics} introduces the higher-order statistics employed in our analysis, and Section~\ref{sec:emulator} outlines the Gaussian-process emulators used to model their cosmological dependence. In Section~\ref{sec:likelihood}, we present the likelihood framework, followed by our main forecasting results in Section~\ref{sec:forecast}. We conclude in Section~\ref{sec:conclusion}.

\section{Simulations}\label{sec:simulation}
\subsection{N-body Simulations and Ray-Tracing}
\begin{figure*}
    \centering
    \includegraphics[width=0.32\linewidth]{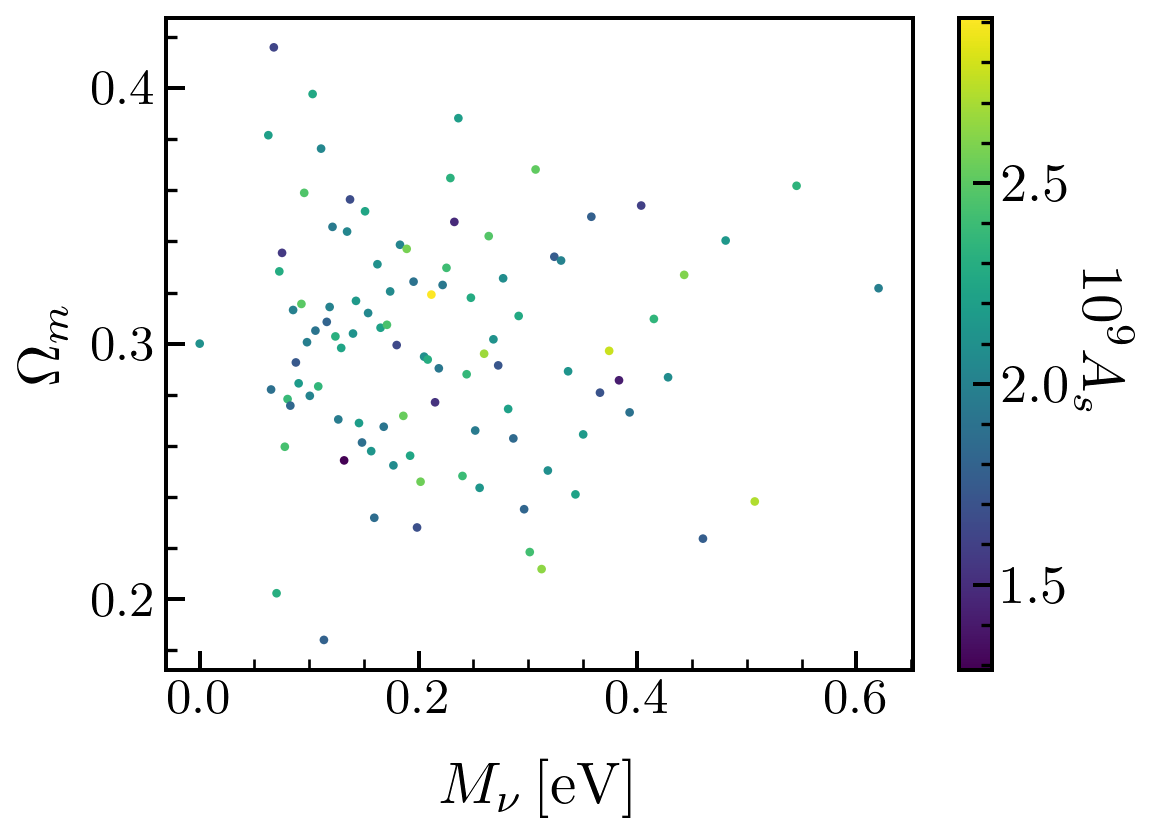}
    \includegraphics[width=0.32\linewidth]{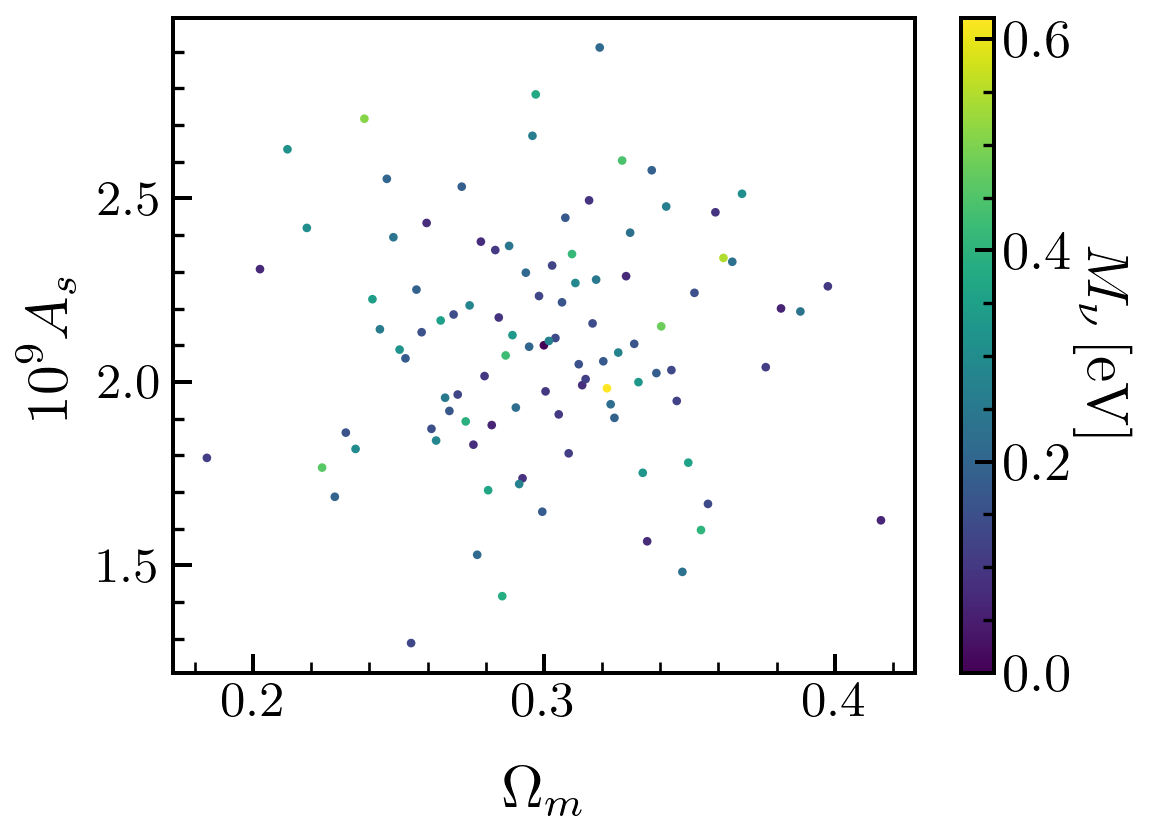}
    \includegraphics[width=0.32\linewidth]{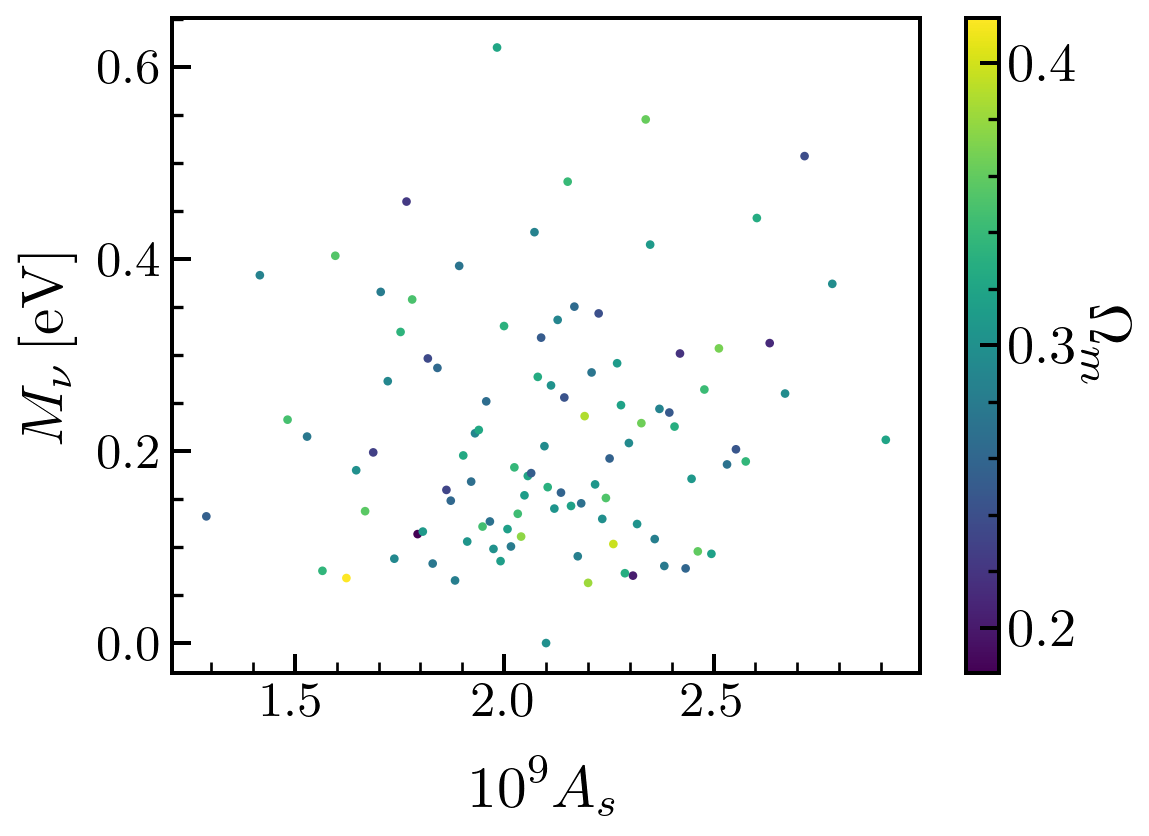}
    \caption{The 101 sampled cosmologies from the {\tt MassiveNuS} suite. Each panel shows a 2D projection of the sampled parameter space, illustrating the distribution of $(M_\nu, \Omega_m, 10^9 A_s)$ across the simulation set.}
    \label{fig:massivenus}
\end{figure*}

The simulations used in this work are drawn from the Cosmological Massive Neutrino Simulation suite ({\tt MassiveNuS})~\cite{Liu:2017now}\footnote{\url{https://columbialensing.github.io/\#massivenus}}, a fully ray-traced N-body simulation suite designed to model the effects of massive neutrinos on large-scale structure. The simulations are run using the public code {\tt Gadget-2}~\cite{Springel:2005mi} with $1024^3$ cold dark matter (CDM) particles in a $(512 \,\,\mathrm{Mpc}/h)^3$ box, evolving the matter density field from $z=99$ to $z=0$. Initial conditions are generated with a modified version of {\tt NGenIC}~\cite{Springel:2005mi} that enables the inclusion of massive neutrinos directly in Fourier space. Massive neutrinos themselves are modeled using the fast linear-response algorithm developed in Refs.~\cite{Ali-Haimoud:2012fzp,Bird:2018all}, in which the neutrino perturbations follow linear theory while their clustering responds to the nonlinear gravitational potential sourced by the CDM. This approximation is accurate down to $k = 10\,h/\mathrm{Mpc}$ over the range $z = 45$ to $z = 0$, and for neutrinos with degenerate mass eigenstates. The {\tt MassiveNuS} suite varies three cosmological parameters: $\Omega_m$, $A_s$, and $M_\nu$, while keeping the remaining $\Lambda$CDM parameters fixed at $h = 0.7$, $\Omega_b = 0.046$, and $n_s = 0.97$. A flat geometry is assumed, i.e., $\Omega_m + \Omega_\Lambda = 1$. Figure~\ref{fig:massivenus} shows the 101 cosmologies sampled in the suite, and Table~\ref{tab:sim_range} lists the parameter ranges, which we adopt as the natural prior volume for our emulator. We also set our fiducial cosmology to $\Omega_m=0.3$, $10^9A_s=2.1$, and $M_\nu=0.1$ eV. 

\begin{table}[!t]
    \centering
    \begin{tabular}{lcc}
        \hline\hline
        \textbf{Parameter} & \textbf{Lower Bound} & \textbf{Upper Bound} \\
        \hline
        $M_\nu\,[\mathrm{eV}]$   & 0      & 0.620 \\
        $\Omega_m$               & 0.184  & 0.416 \\
        $10^{9}A_s$              & 1.289  & 2.911 \\
        \hline\hline
    \end{tabular}
    \caption{Ranges of cosmological parameters spanned by the {\tt MassiveNuS} simulation suite, which varies the total neutrino mass $M_\nu$, matter density fraction $\Omega_m$, and primordial scalar amplitude $A_s$.}
    \label{tab:sim_range}
\end{table}

To generate the lensing convergence maps, simulation snapshots are stored every $126\,\mathrm{Mpc}/h$ from $z=45$ down to $z=0$. For each snapshot, we apply a random shift and rotation to the simulation box and then extract a slice of thickness $126\,\mathrm{Mpc}/h$ --- matching the snapshot spacing --- to construct a lens plane. We then ray-trace $4096^2$ regularly spaced light rays from the observer at $z=0$ back to the source plane at the last scattering surface ($z \approx 1100$). For a simulation box of side length $512\,\mathrm{Mpc}/h$, the resulting maps span approximately $3.5^\circ$ on a side. This full pipeline is implemented using the public {\tt LensTools} package~\cite{Petri:2016zrq}\footnote{\url{https://lenstools.readthedocs.io/en/latest/}}. As shown in Ref.~\cite{Petri:2016wlu}, the use of random shifts and rotations enables the generation of up to 10,000 pseudo-independent lensing convergence maps from a single N-body initial condition realization.

\subsection{Lensed CMB maps}
To forward-model the noise associated with CMB lensing reconstruction, we begin by generating Gaussian realizations of unlensed CMB temperature ($\bar{T}$) and polarization ($\bar{E}$/$\bar{B}$) fields, each of which is then to be lensed by one of the lensing convergence maps produced from the N-body simulations. These realizations are drawn from (unlensed) primary CMB power spectra computed with {\tt CAMB}~\cite{Lewis:1999bs}\footnote{\url{http://camb.info}} at the fiducial cosmology. Throughout this work, we fix the cosmology of the unlensed CMB to the fiducial model in order to minimize any cosmology-dependent variations in the lensing reconstruction noise.

\begin{figure}
    \centering
    \includegraphics[width=0.9\linewidth]{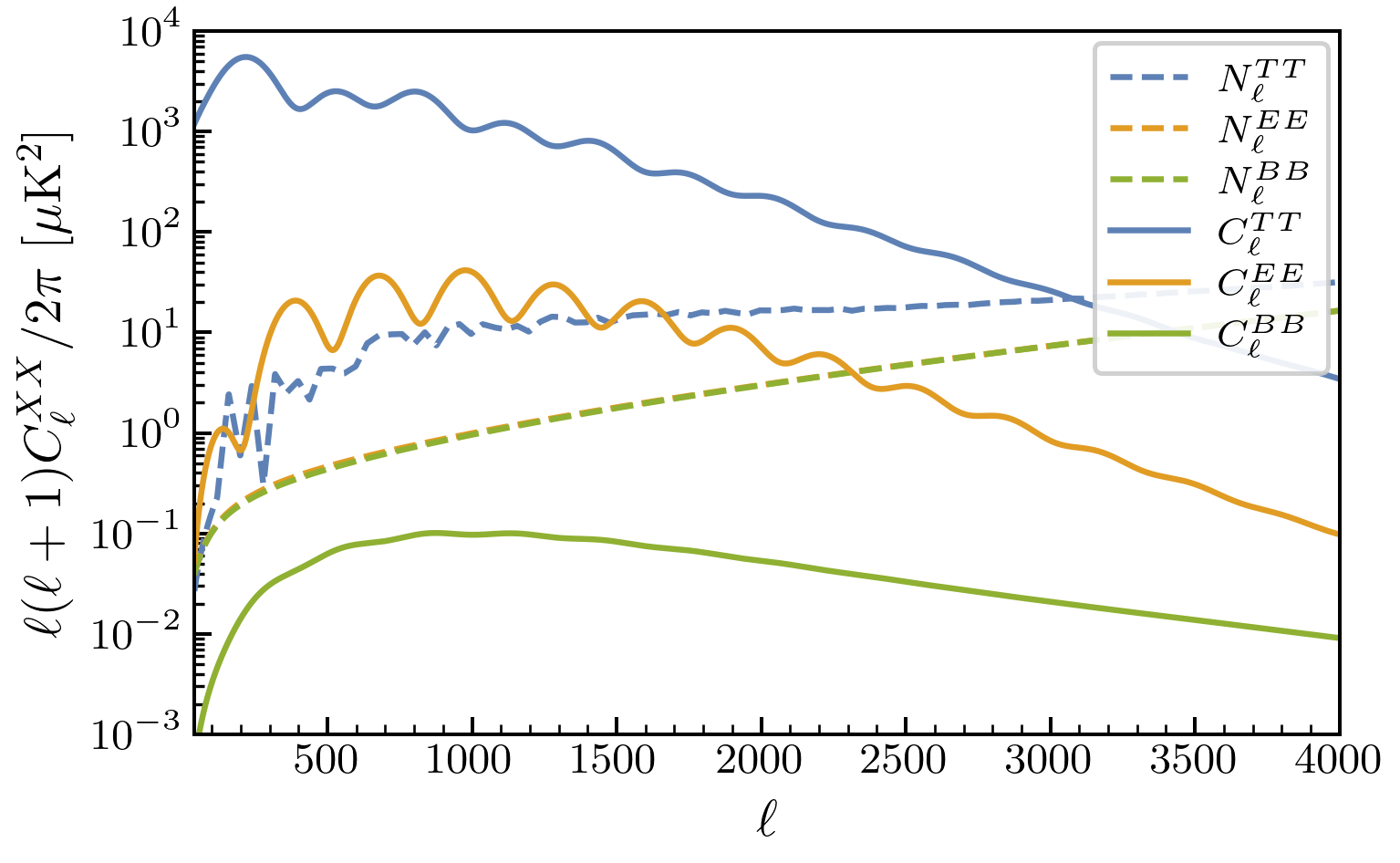}
    \caption{Lensed CMB temperature (TT) and polarization (EE/BB) signal and noise power spectra for the expected nominal SO observations~\cite{SimonsObservatory:2018koc} after component separation.}
    \label{fig:cmb_cl}
\end{figure}

We convert the $\bar{E}$/$\bar{B}$ fields to $\bar{Q}$/$\bar{U}$ by rotating in Fourier space, and we then apply the standard lensing remapping to simulate the effect of gravitational deflection. Given a convergence field $\kappa$ from {\tt MassiveNuS}, the lensed CMB fields are constructed as
\begin{align}\label{eq:lensing_remap}
    T(\hat{\mathbf{n}}) &= \bar{T}(\hat{\mathbf{n}} + \boldsymbol{\alpha})\,, \nonumber\\
    Q(\hat{\mathbf{n}}) &= \bar{Q}(\hat{\mathbf{n}} + \boldsymbol{\alpha})\,, \nonumber\\
    U(\hat{\mathbf{n}}) &= \bar{U}(\hat{\mathbf{n}} + \boldsymbol{\alpha})\,,
\end{align}
where $\boldsymbol{\alpha}$ is the deflection field, defined as $\boldsymbol{\alpha} = \nabla(\nabla^{-2}\kappa) = \nabla\phi$, with $\phi$ the lensing potential. Finally, we add noise realizations generated from the expected Simons Observatory (SO) noise power spectra after component separation, yielding the observed fields $\tilde{T}$, $\tilde{Q}$, and $\tilde{U}$. Figure~\ref{fig:cmb_cl} shows the signal and noise power spectra corresponding to the nominal SO configuration~\cite{SimonsObservatory:2018koc},\footnote{Note that we do not include anticipated observations from the enhanced SO LAT survey described in Ref.~\cite{SimonsObservatory:2025wwn}.} obtained using a standard internal linear combination (ILC) pipeline for component separation~\cite{SimonsObservatory:2018koc}\footnote{The noise spectra are publicly available at \url{https://github.com/simonsobs/so_noise_models/tree/master/LAT_comp_sep_noise/v3.1.0}}. We use the publicly available package {\tt pixell}\footnote{\url{https://github.com/simonsobs/pixell}} to implement the above pipeline.  Throughout, we use the flat-sky approximation.

\subsection{Lensing Reconstruction}
Once the simulated observed CMB maps are generated, we perform lensing reconstruction using the quadratic estimator~\cite{Hu:2001kj,Madhavacheril:2020ido}. At leading order, when the lensing potential $\phi$ is treated as fixed, the remapping in Eq.~\eqref{eq:lensing_remap} induces off-diagonal correlations between previously independent Fourier modes. These mode couplings are linearly proportional to $\phi(\mathbf{L})$, and by appropriately weighting and averaging the cross-mode correlations, we can construct an unbiased estimate of the lensing potential. The quadratic estimator for a given pair of fields $XY \in \{TT,\,TE,\,EE,\,TB,\,EB\}$ takes the form
\begin{align}
    &\hat{\phi}_{XY}(\boldsymbol{L})=A_{XY}(\boldsymbol{L}) \int_{\boldsymbol{\ell_1+\ell_2=L}}g_{XY}(\boldsymbol{\ell}_1,\boldsymbol{\ell}_2)\nonumber\\
    &\times\left[\frac{\tilde{X}}{C^{XX}+N^{XX}}\right](\boldsymbol{\ell}_1)\left[\frac{\tilde{Y}}{C^{YY}+N^{YY}}\right](\boldsymbol{\ell}_2)\,,
\end{align}
where $A_{XY}$ is the normalization for the estimator, and $g_{XY}$ is the weight defined as
\begin{align}
    &g_{XY}(\boldsymbol{\ell}_1,\boldsymbol{\ell}_2)\nonumber\\ 
    &=\begin{cases}
        \tfrac12\,f_{XY}(\boldsymbol{\ell}_1,\boldsymbol{\ell}_2), 
           \quad\text{if } XY=\mathrm{TT},\,\mathrm{EE}\\[4pt]
        f_{XY}(\boldsymbol{\ell}_1,\boldsymbol{\ell}_2), 
           \quad\text{if } XY=\mathrm{EB},\,\mathrm{TB}\\[4pt]
        f_{XY}(\boldsymbol{\ell}_1,\boldsymbol{\ell}_2)
         - f_{XY}(\boldsymbol{\ell}_2,\boldsymbol{\ell}_1)\,
         \dfrac{C^{TE}_{\ell_1}}{C^{EE}_{\ell_1}}\,
         \dfrac{C^{TE}_{\ell_2}}{C^{TT}_{\ell_2}},
           \quad\text{if } XY=\mathrm{TE}\,.
    \end{cases}
\end{align}
Here
\begin{align}
    f_{XY}(\boldsymbol{\ell}_1,\boldsymbol{\ell}_2) = \frac{\partial}{\partial\phi(\boldsymbol{\ell}_1+\boldsymbol{\ell}_2)}\langle X(\boldsymbol{\ell}_1)Y(\boldsymbol{\ell}_2) \rangle_{\phi_\text{fixed}}
\end{align}
encodes the response of the two-point function to lensing. The corresponding normalization is
\begin{align}
    A_{XY}(\boldsymbol{L})
    = L^{2}\Bigg[
    &\int_{\boldsymbol{\ell}_1 + \boldsymbol{\ell}_2 = \boldsymbol{L}}
    g_{XY}(\boldsymbol{\ell}_1, \boldsymbol{\ell}_2)\,
    \frac{1}{C_{\ell_1}^{XX} + N_{\ell_1}^{XX}}
    \notag\\[-2pt]
    &\times
    \frac{1}{C_{\ell_2}^{YY} + N_{\ell_2}^{YY}}\,
    f_{XY}(\boldsymbol{\ell}_1, \boldsymbol{\ell}_2)
    \Bigg]^{-1}.
\end{align}
With this normalization, $\hat{\phi}$ is unbiased provided that the fiducial cosmology used in the weights matches that of the data; techniques to avoid biases that could arise from a small mismatch have been developed~\cite{Namikawa:2012pe}.

For each quadratic estimator pair, we can compute the power spectrum $C_L^{\hat{\phi}_{XY},\hat{\phi}_{UV}}$. However, since this power spectrum is effectively the trispectrum of the field(s), it has a non-zero expectation value even when the field is purely Gaussian. The leading term of this disconnected contribution $N^{(0)}$ to the estimation of the power spectrum in an isotropic case can be written down explicitly as
\begin{align}
    &N^{(0)}_{XY,\,UV}(\boldsymbol{L})
    = A_{XY}(\boldsymbol{L})A_{UV}(\boldsymbol{L})\nonumber\\
    &\times\Big[\int_{\boldsymbol{\ell}_1+\boldsymbol{\ell}_2=\boldsymbol{L}}
    g_{XY}(\boldsymbol{\ell}_1,\boldsymbol{\ell}_2)
    g_{UV}(-\boldsymbol{\ell}_1,-\boldsymbol{\ell}_2)
    S^{XU}_{\ell_1}S^{YV}_{\ell_2}
    \nonumber\\
    &+
    \int_{\boldsymbol{\ell}_1+\boldsymbol{\ell}_2=\boldsymbol{L}}
    g_{XY}(\boldsymbol{\ell}_1,\boldsymbol{\ell}_2)
    g_{UV}(-\boldsymbol{\ell}_2,-\boldsymbol{\ell}_1)
    S^{XV}_{\ell_1}S^{YU}_{\ell_2}
    \Big]\,,
\end{align}
where 
\begin{align}
    S^{XU}_{\ell}= \frac{\tilde{C}^{\mathrm{obs},\,XU}_{\ell}}{(C^{XX}_{\ell} + N^{XX}_{\ell})(C^{UU}_{\ell} + N^{UU}_{\ell})}\,.
\end{align}
We use this $N^{(0)}$ term in the Wiener filtering of the reconstructed convergence field, as it is also the effective noise power spectrum of the reconstruction.  We neglect higher-order bias terms such as $N^{(1)}$, $N^{(2)}$, and $N^{(3/2)}$, whose details can be found in Refs.~\cite{Madhavacheril:2020ido, Fabbian:2019tik, Kesden:2003cc, Hanson:2010rp, Bohm:2016gzt, Bohm:2018omn, Beck:2018wud}.

We fix the cosmology of all unlensed CMB power spectra that enter the quadratic-estimator weights. To compute the lensed spectra that enter the estimator weights, we likewise hold the unlensed CMB cosmology fixed, but we vary the lensing convergence power spectrum $C_\ell^{\kappa\kappa}$ according to the cosmology associated with each simulated $\kappa$ map. This updated $C_\ell^{\kappa\kappa}$ is then used to lens the fiducial unlensed CMB spectra. Our goal in doing so is to ensure that the lensing reconstruction noise depends only weakly on cosmology, i.e., through higher-order terms, allowing our forecast to isolate the information coming specifically from the CMB lensing convergence field. For the reconstruction itself, we use temperature and polarization multipoles in the range $\ell \in [200, 4000]$.

\begin{figure}[t]
    \centering
    \includegraphics[width=0.9\linewidth]{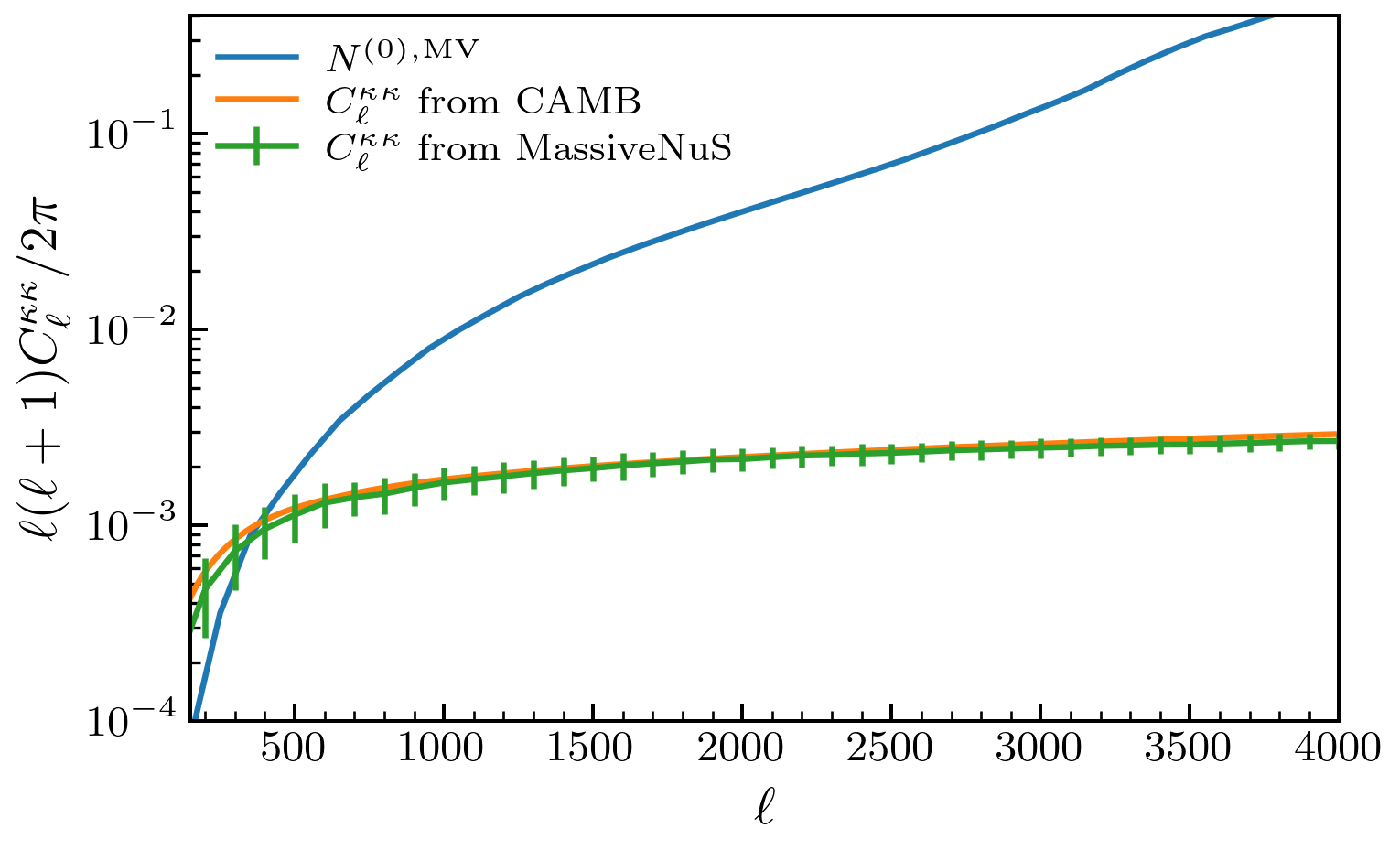}
    \caption{Fiducial $C_\ell^{\kappa\kappa}$ (orange) and the lensing reconstruction noise $N^{(0)}$ (blue) estimated from the minimum variance combination expected for an SO-nominal-like experiment. The green line indicates the mean and standard deviation of $C_{\ell^{\kappa\kappa}}$ from 10,000 realizations of {\tt MassiveNuS} at the fiducial cosmology. We can see that the differences between CAMB and {\tt MassiveNuS} begin to increase beyond $\ell\sim2000$ due to the finite resolution of the simulation.}
    \label{fig:kappa_cl}
\end{figure}

We also construct a smooth, compactly supported window function based on a normalized bump cumulative distribution.
The bump function is defined as
\begin{align}
    b(x) &=
    \begin{cases}
    \exp\!\left[-\dfrac{1}{x(1-x)}\right], & 0 < x < 1, \\[6pt]
    0, & \text{otherwise}.
    \end{cases}
\end{align}
It is $C^\infty$, i.e., for every integer $n\geq0$ the $n$-th derivative exists and is continuous, and moreover all derivatives vanish at the endpoints. Its cumulative distribution is given by
\begin{align}
    S(x) =
    \begin{cases}
    0, & x \le 0, \\[4pt]
    \dfrac{1}{B} \displaystyle\int_0^x b(t)\, dt, & 0 < x < 1, \\[10pt]
    1, & x \ge 1.
    \end{cases}
\end{align}
with $B = \int_0^1 b(t)\, dt$. This construction yields a $C^\infty$ window that smoothly transitions from 0 to 1 over the low-$\ell$ ramp, remains flat on the plateau, and decays smoothly back to 0 over the high-$\ell$ ramp, with all derivatives vanishing at the boundaries. We apply this bump function to the ranges $[\ell_{\rm min},\,1.1\ell_{\rm min}]$ and $[0.9\ell_{\rm max},\,\ell_{\rm max}]$ to ensure smooth transitions to zero so that we do not see artificial wiggles in real space.

We implement the above pipeline with the publicly available code {\tt symlens}\footnote{\url{https://github.com/simonsobs/symlens}}. Figure~\ref{fig:kappa_cl} shows $C_\ell^{\kappa\kappa}$ computed with our fiducial cosmology using CAMB and the corresponding minimum variance $N^{(0)}$ lensing reconstruction noise. We can see that the $S/N=1$ at roughly $\ell\approx400$, indicating a great amount of information can be extracted from the field even though the area of our simulation is small. 

\begin{figure}[t]
    \centering
    \includegraphics[width=0.9\linewidth]{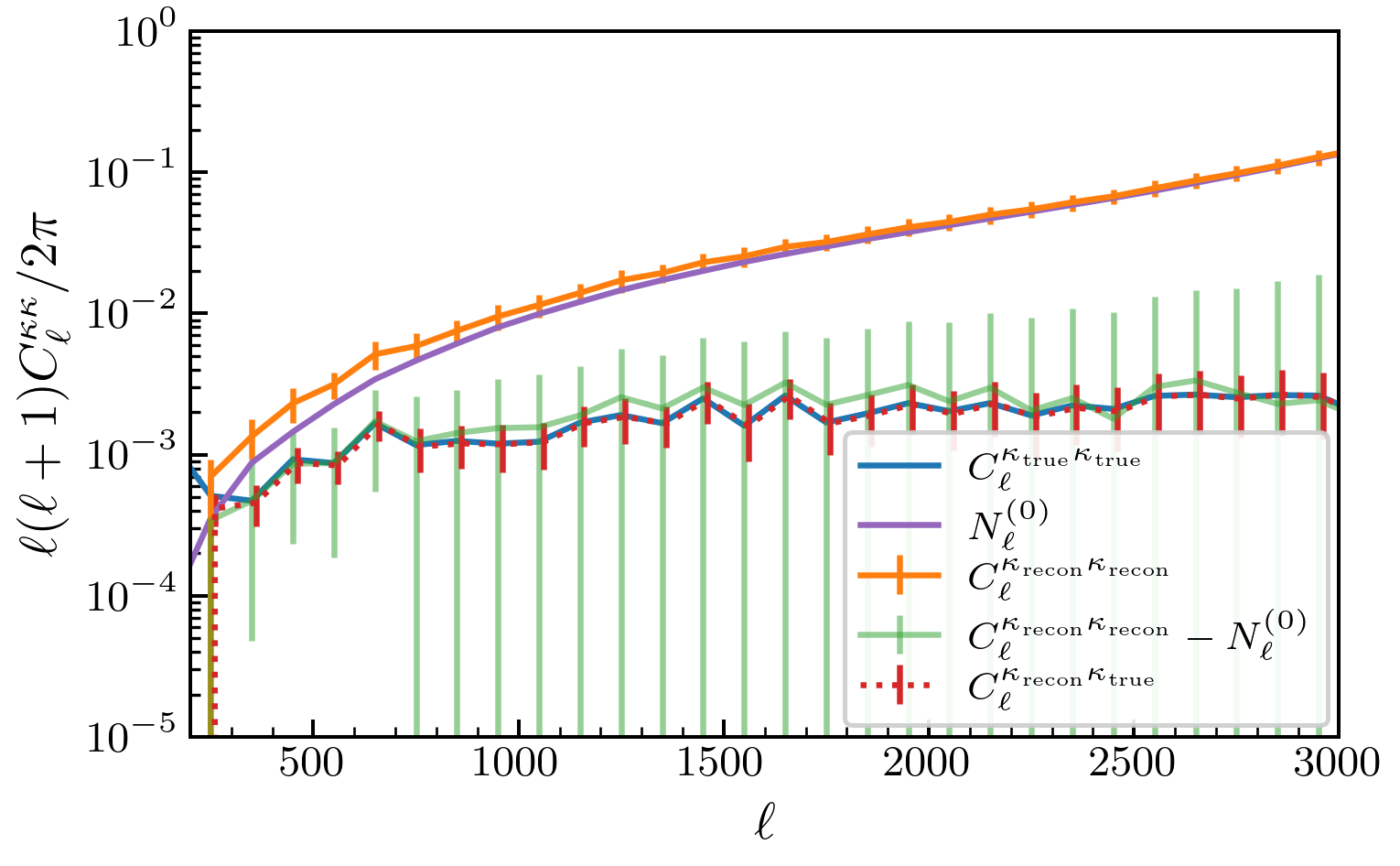}
    \caption{Lensing reconstruction performance for a single {\tt MassiveNuS} convergence map, averaged over 1,000 different realizations of unlensed CMB maps using the minimum variance quadratic estimator. The reconstructed $\kappa$ power spectrum (orange) and the noise bias $N^{(0)}$ (purple) are shown along with their difference (green) compared to the true input spectrum (blue). The cross-spectrum between the reconstructed and true $\kappa$ fields (dotted red) demonstrates excellent consistency, validating the reconstruction pipeline.}
    \label{fig:recon_kappa_cl}
\end{figure}

For a fixed convergence $\kappa$ map from the {\tt MassiveNuS} simulation suite, Figure~\ref{fig:recon_kappa_cl} demonstrates the performance of our lensing reconstruction pipeline. We generate 1,000 independent realizations of unlensed CMB $\bar{T}/\bar{Q}/\bar{U}$ maps to estimate the mean and variance of the reconstructed power spectra. The orange and purple curves show the reconstructed $\kappa$ power spectrum and the corresponding $N^{(0)}$ noise bias, respectively. Their difference, shown in green, tracks the true input power spectrum (blue) remarkably well, with only small residual deviations attributable to higher-order bias terms. The cross-spectrum between the reconstructed and true $\kappa$ fields also exhibits excellent agreement, further validating the accuracy and robustness of our reconstruction procedure.

\begin{figure}[!t]
    \centering
    \includegraphics[width=1.0\linewidth]{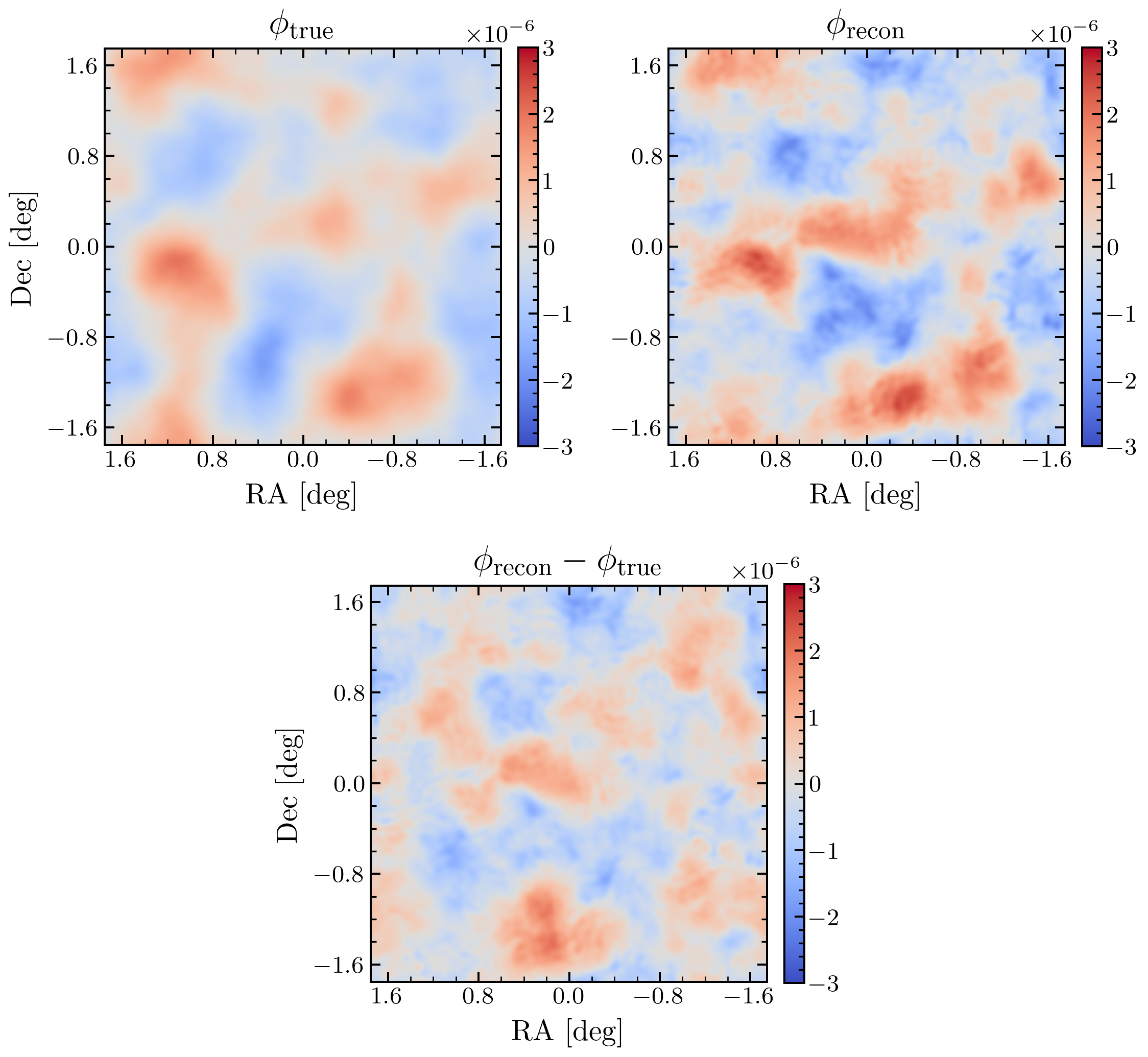}
    \caption{True ($\phi_{\rm true}$) and reconstructed ($\phi_{\rm recon}$) CMB lensing potential maps from one {\tt MassiveNuS} realization, shown at the map-level, together with their difference ($\phi_{\rm recon}-\phi_{\rm true}$). The reconstructed map recovers the main structures of the true potential, demonstrating the performance of our lensing reconstruction pipeline. Note that we apply a simple top-hat filter in Fourier space for both fields with $\ell_{\rm min}=200$ and $\ell_{\rm max}=4000$ here; thus, some small-scale noise clearly remains in the reconstructed map. Also, note that the small-scale smoothness of $\phi_{\rm true}$ is due to the resolution of the simulation, as shown in Figure~\ref{fig:kappa_cl}.}
    \label{fig:recon_kappa_map}
\end{figure}

Figure~\ref{fig:recon_kappa_map} provides a map-level comparison between the true and reconstructed lensing potential $\phi$ for a single realization of the unlensed CMB. For visualization, we apply a top-hat filter to retain modes within $[\ell_{\min},\ell_{\max}] = [200, 4000]$, the range relevant to our analysis. The large-scale structure of the reconstructed map closely matches that of the true field, illustrating the fidelity of the reconstruction in the regime most important for our forecasts.

\section{Higher-Order Statistics}\label{sec:higherOrderStatistics}
In this section, we describe the statistics used in our forecast: the power spectrum, Minkowski functionals, peak and minima counts, moments, and the wavelet scattering transform. Before computing any of these statistics from the reconstructed $\kappa$ maps, we apply the Wiener filter $C_\ell^{\kappa\kappa}/(C_\ell^{\kappa\kappa}+N_\ell^{(0)})$, where $C_\ell^{\kappa\kappa}$ is evaluated at the fiducial cosmology and $N_\ell^{(0)}$ is the zeroth-order bias from the lensing reconstruction. We do not include the bispectrum amongst our statistics because the number of simulations is not sufficient for our emulator to converge reliably.

\subsection{Power spectrum}
Under the flat-sky approximation, the power spectrum is estimated with the binned power spectrum estimator:
\begin{align}
    \hat{C}_{\ell_i} = \frac{1}{N_i}\sum_{|\boldsymbol{\ell}|\in[\ell_i,\ell_{i+1}]}\kappa(\boldsymbol{\ell})\kappa^*(\boldsymbol{\ell})\,,
\end{align}
where $\ell_i$s are the edges of each $\ell$-bin, $N_i$ is the number of modes within the range $\ell_i\leq|\boldsymbol{\ell}|\leq\ell_{i+1}$, and $\kappa(\boldsymbol{\ell})$ is the Fourier transform of the lensing convergence map. In our main analysis, we consider $[\ell_{\rm min},\ell_{\rm max}]=[300,3000]$ with $\Delta\ell=100$, which is slightly narrower than the range $[200,4000]$ used for the lensing reconstruction.


\subsection{Minkowski Functionals}
Minkowski functionals (MFs) provide a set of measures that characterize the global morphological properties of a field~\cite{Minkowski1903}. For a two-dimensional field, there are three functionals, $V_0$, $V_1$, and $V_2$, defined as
\begin{align}
    V_0(\nu) &= \frac{1}{A}\int_{\Sigma(\nu)} \di a\nonumber\\
    V_1(\nu) &= \frac{1}{4A}\int_{\partial\Sigma(\nu)} \di\ell\nonumber\\
    V_2(\nu) &= \frac{1}{4\pi A}\int_{\partial\Sigma(\nu)}\mathcal{K}\di\ell
\end{align}
where $\Sigma(\nu)$ is defined as the excursion set $\{y:y>\nu\}$, $\partial\Sigma(\nu)$ denotes the boundary of it, and $\mathcal{K}$ is the curvature at the point. As a result, $V_0$ measures the area of the excursion set under different thresholds; $V_1$ measures the total perimeter of each excursion set; and $V_2$ measure the perimeter weighted by the curvature at each point. 

If the field is a pure Gaussian random field, the MFs have the expectation values~\cite{Matsubara_2003}:
\begin{align}
    V_0^{\rm G}(\nu) &= \frac{1}{2}\mathrm{erfc}\left(\frac{\nu}{\sqrt{2}}\right), \nonumber\\
    V_1^{\rm G}(\nu) &= \frac{1}{8}\frac{\sigma_{G,1}}{\sigma_{G,0}} e^{-\nu^2/2}, \nonumber\\
    V_2^{\rm G}(\nu) &= \frac{1}{2\pi} \left(\frac{\sigma_{G,1}}{\sigma_{G,0}}\right)^2 \nu e^{-\nu^2/2}\,,
\end{align}
where
\begin{align}
    \sigma_{G,0}^2 \equiv \int \frac{\ell \di\ell}{2\pi}C_\ell^{\kappa\kappa}\text{   and   }
    \sigma_{G,1}^2 \equiv \int \frac{\ell \di\ell}{2\pi}\ell^2 C_\ell^{\kappa\kappa}\,.
\end{align}

Our binning scheme is constructed by having 20 equally spaced bins within the range $[-3\sigma_0, 3\sigma_0]$, resulting in a total of $3\times20$ bins for the three MFs. Here $\sigma_0$ is chosen to be the standard deviation of the overall pixel values for the simulated observed lensing convergence maps at the fiducial cosmology with Wiener filtering. 

\subsection{Peak/Minima Counts}
The peak and minima counts, $\mathrm{PC}(\nu)$ and $\mathrm{MC}(\nu)$, are defined as the number of pixels that are local maxima or minima with values within the range of the given bin size~\cite{1987MNRAS.226..655B,Bardeen:1985tr}. In practice, we identify a pixel as a peak or a minimum if its value is larger or smaller, respectively, than those of its eight neighboring pixels.

For the binning scheme, we use 13 equally spaced bins. For peak counts, the bin edges span the range $[-1\sigma_0,\,3\sigma_0]$, while for minima counts they span $[-3\sigma_0,\,1\sigma_0]$. These asymmetric ranges ensure good sampling of both the high-density and low-density excursion sets, where the distributions of peaks and minima are most sensitive to non-Gaussian features. Here $\sigma_0$ is the same variance parameter defined in the context of the MFs.

\subsection{Moments}
For the moments (MOM), we consider the following four lower-order moments: $\sigma_0$, $\sigma_1$, $S_1$, and $K_1$, which are defined as
\begin{align}
    \hat{\sigma}_0 &\equiv \sqrt{\langle\kappa^2\rangle}\,,\quad \hat{\sigma}_1 \equiv \sqrt{\langle|\nabla\kappa|^2\rangle}\,,\nonumber\\
    \hat{S}_1 &\equiv \langle\kappa^2\nabla^2\kappa\rangle\,,\quad \hat{K}_1 \equiv \langle\kappa^3\nabla^2\kappa\rangle \,.
\end{align}
These moments probe complementary aspects of the field: $\sigma_0$ and $\sigma_1$ capture the variance of the field and its gradient, while $S_1$ and $K_1$ quantify departures from Gaussianity through skewness- and kurtosis-like combinations involving spatial derivatives. They are also motivated by the expansion of MFs beyond the Gaussian approximation~\cite{Petri:2013ffb,Matsubara:2000mw,Matsubara:2010te}. Together with MFs and PC/MC, these statistics are computed with {\tt Lenstools}.

\subsection{Wavelet Scattering Transform}
The wavelet scattering transform (WST) is an emerging method that provides a translation-invariant, multiscale representation capable of capturing non-Gaussian information beyond the power spectrum~\cite{2011arXiv1101.2286M,Cheng:2021xdw}. It has been successfully applied to the diffuse interstellar medium~\cite{Regaldo-SaintBlancard:2020dlb}, weak lensing~\cite{Cheng:2020qbx,Boone:2025obt}, and redshift-space galaxy overdensity fields~\cite{Valogiannis:2022xwu,Valogiannis:2023mxf}, demonstrating strong sensitivity to higher-order statistics and robustness against noise. Conceptually, the WST shares similarities with convolutional neural networks (CNNs), in that it performs local convolutions across the field; however, unlike CNNs, the filters in the WST are fixed analytic wavelets rather than trainable parameters, allowing the method to retain interpretability and stability while still extracting non-Gaussian signatures.

Given an initial field $I_0$ and a set of filters $\psi_{j_n l_n}$, we construct a sequence of fields iteratively as
\begin{align}
    I_{n+1} = \psi_{j_n l_n}\star I_n\,,
\end{align}
where $\star$ denotes convolution. The corresponding wavelet scattering coefficients are defined as spatial averages of the modulus of these filtered fields: 
\begin{align}
    S_0 &= \langle I_0 \rangle \nonumber\\
    S_1^{j_1l_1} &= \langle |\psi_{j_1l_1}\star I_0| \rangle \nonumber\\
    S_2^{j_1l_1,j_2l_2} &= \langle |\psi_{j_2l_2}\star(\psi_{j_1l_1}\star I_0)| \rangle\,,
\end{align}
where $\langle\cdot\rangle$ denotes a spatial average over all pixels. For the wavelets, we use the Morlet wavelets of the form:
\begin{align}
    \psi_{j,l}(\vec{x}) = 
    \frac{1}{\sigma}
    \exp\!\left(-\frac{x^2}{2\sigma^2}\right)
    \left[\exp\!\left(i\vec{k}_0 \cdot \vec{x}\right) - \beta\right]\,,
\end{align}
where $\sigma = 0.8\times 2^j$, $k_0 = 0.7\pi \times 2^{-j}$, and $\arg(k_0) = (L/2 - 1 - l)\pi/L$. Lower values of $j$ correspond to smaller spatial scales, which in turn lead to scattering coefficients that are more strongly affected by noise. For the second-order coefficients $S_2^{j_1 l_1, j_2 l_2}$, we retain only the cases with $j_1 > j_2$, since filtering with a large-scale wavelet after a small-scale one does not yield additional information. Because our maps are isotropic, the expectation values of the WST coefficients are identical across different $l$ values. The maximum accessible scale index $J$ is set by the map resolution, $J = \log_2(N_{\rm pix}) - 1$. For our $512^2$-pixel maps, we use $J = 8$, which yields $1 + 8 + 28 = 37$ coefficients when restricting to a single orientation, i.e., $L = 1$. We further remove pairs that have correlations with absolute values greater than 0.99, which reduces the number of coefficients to only 22 in our forecast. We implement the WST with the public package {\tt PyWST}\footnote{\url{https://github.com/bregaldo/pywst}}~\cite{Regaldo-SaintBlancard:2020dlb}.

\section{Emulator}\label{sec:emulator}
\subsection{Model Specification}
We construct emulators for the statistics described in Section~\ref{sec:higherOrderStatistics} using Gaussian Processes (GP)~\cite{2006gpml.book.....R}. For each of the 101 cosmologies in {\tt MassiveNuS}, we compute the mean of each statistic from 10,000 pseudo-independent realizations. Because the random seeds are fixed across cosmologies in {\tt MassiveNuS}, we also fix the seeds used to generate the unlensed CMB $\bar{T}/\bar{E}/\bar{B}$ maps and the noise realizations, allowing us to isolate how the statistics change solely as a function of cosmology. 

For the GP model, we use a Matérn kernel~\cite{matern2013spatial}, defined for two points separated by distance $r = |x - x'|$ as
\begin{align}
    k_\nu(r) = \sigma_f^2 \frac{2^{1-\nu}}{\Gamma(\nu)}\left(\frac{\sqrt{2\nu}r}{\ell_G}\right)^\nu K_{\nu}\left(\frac{\sqrt{2\nu}r}{\ell_G}\right)\,,
\end{align}
where $\sigma_f$ is the signal variance inferred from the training data, $K_\nu$ is the modified Bessel function of the second kind, $\ell_G$ is the correlation length, and $\nu>0$ controls the smoothness of the resulting function. We adopt $\nu = 0.5$, for which the kernel reduces to the exponential form $k_{0.5}(r) \propto \exp(-r/\ell_G)$, yielding a process that is continuous but nowhere differentiable. This choice is motivated by the rapid fluctuations observed in the data across cosmologies once observational noise is included, combined with the sparsity of sampling in parameter space. Larger values of $\nu$ tend to underfit the data by enforcing overly smooth behavior, and we find that using $\nu>0.5$ often leads to multimodal posterior distributions for the cosmological parameters of interest. 

To specify the noise variance for the GP, we estimate the sample variance from the simulations and scale it by the square root of the number of realizations using the central limit theorem. The correlation length $\ell_G$ is optimized within the interval $[10^{-6},\,10^6]$ using the L-BFGS algorithm~\cite{liu1989limited}, a quasi-Newton method for function optimization. Finally, before training the Gaussian Process, we standardize the inputs and targets to improve numerical stability. The pipeline here is implemented with {\tt scikit-learn}\footnote{\url{https://scikit-learn.org/stable/}}.

\subsection{Leave-One-Out Cross-Validation}
\begin{figure*}[t]
    \centering
    \includegraphics[width=0.3\textwidth]{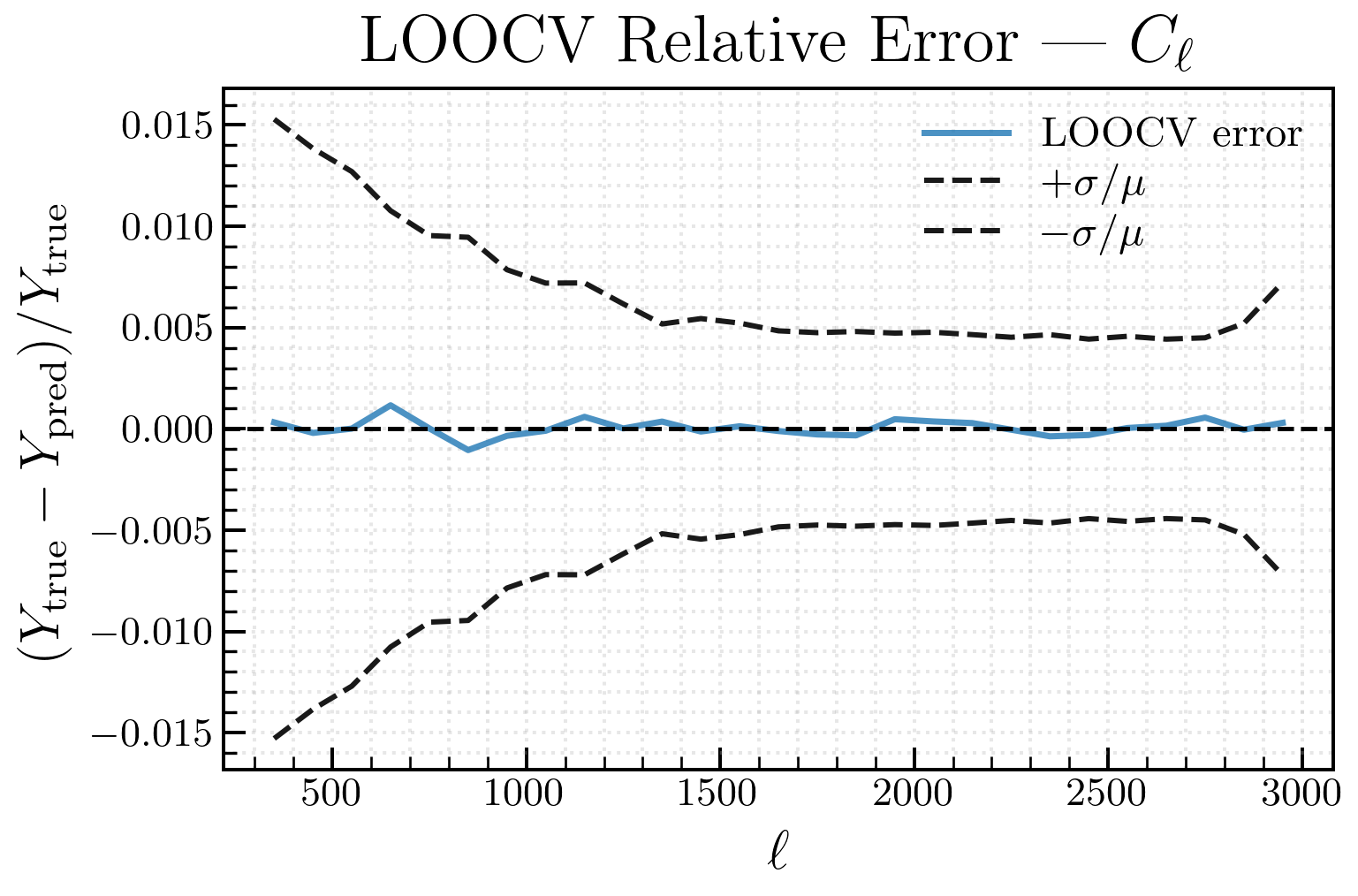}
    \includegraphics[width=0.3\textwidth]{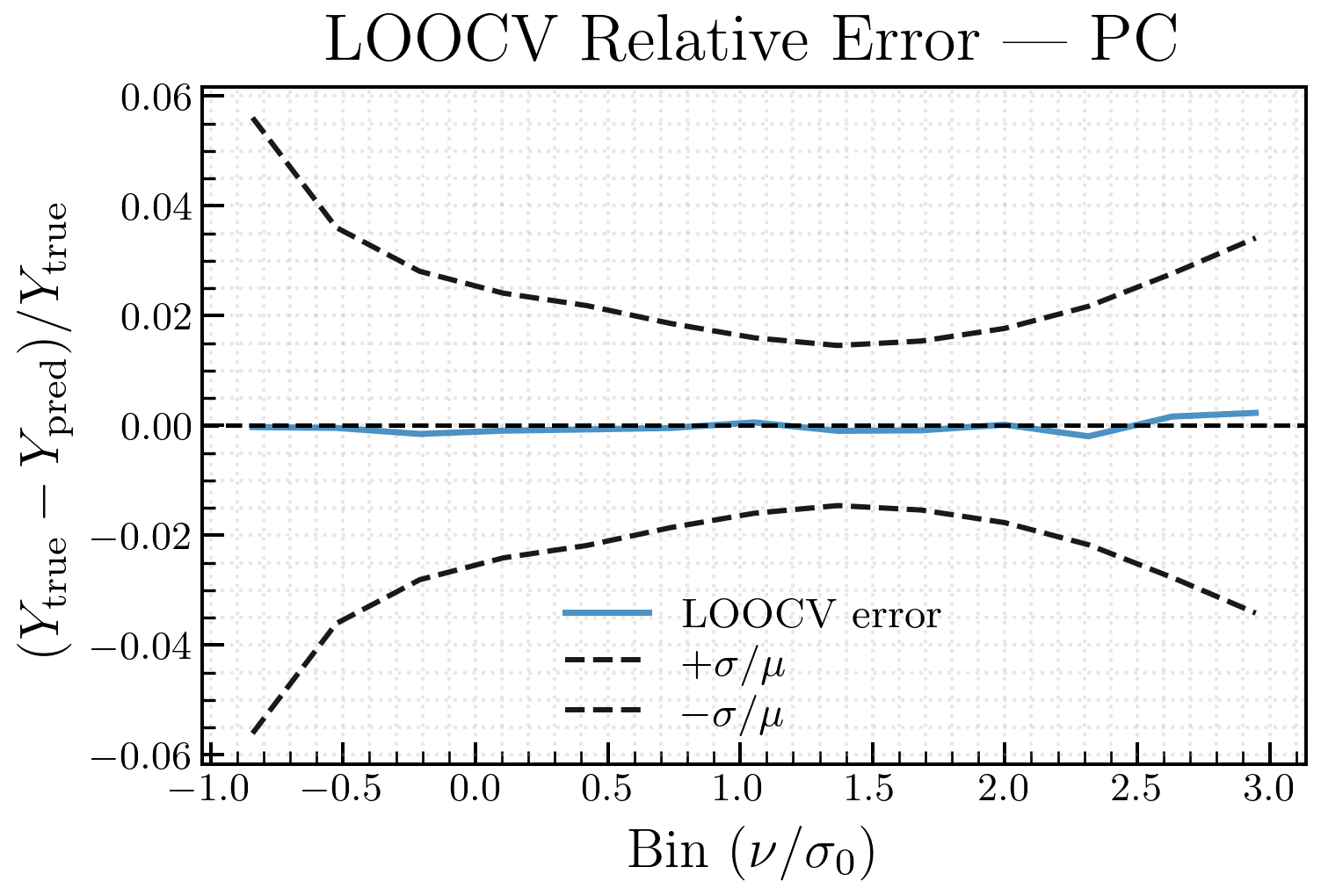}
    \includegraphics[width=0.3\textwidth]{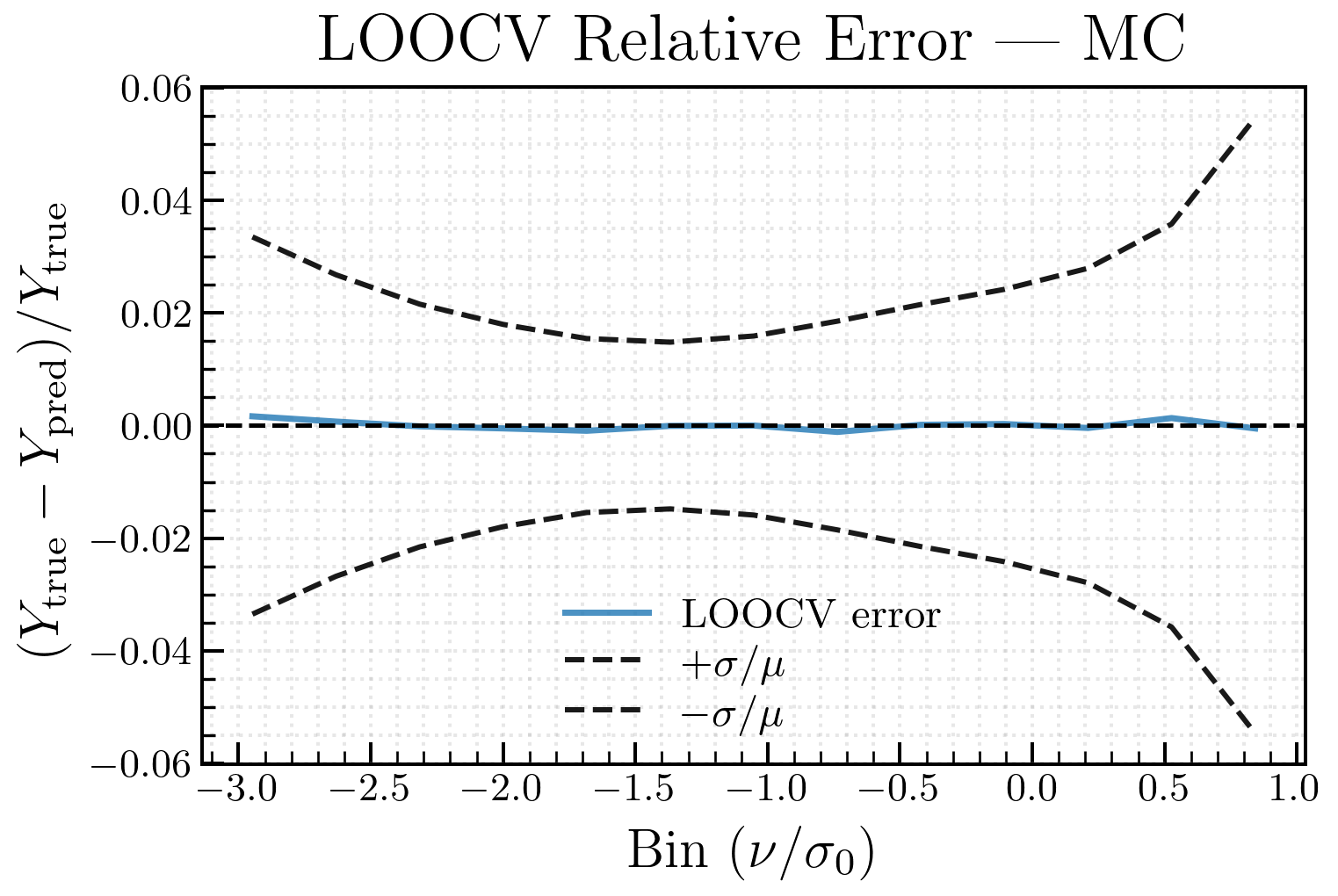}\\[8pt]
    \includegraphics[width=0.3\textwidth]{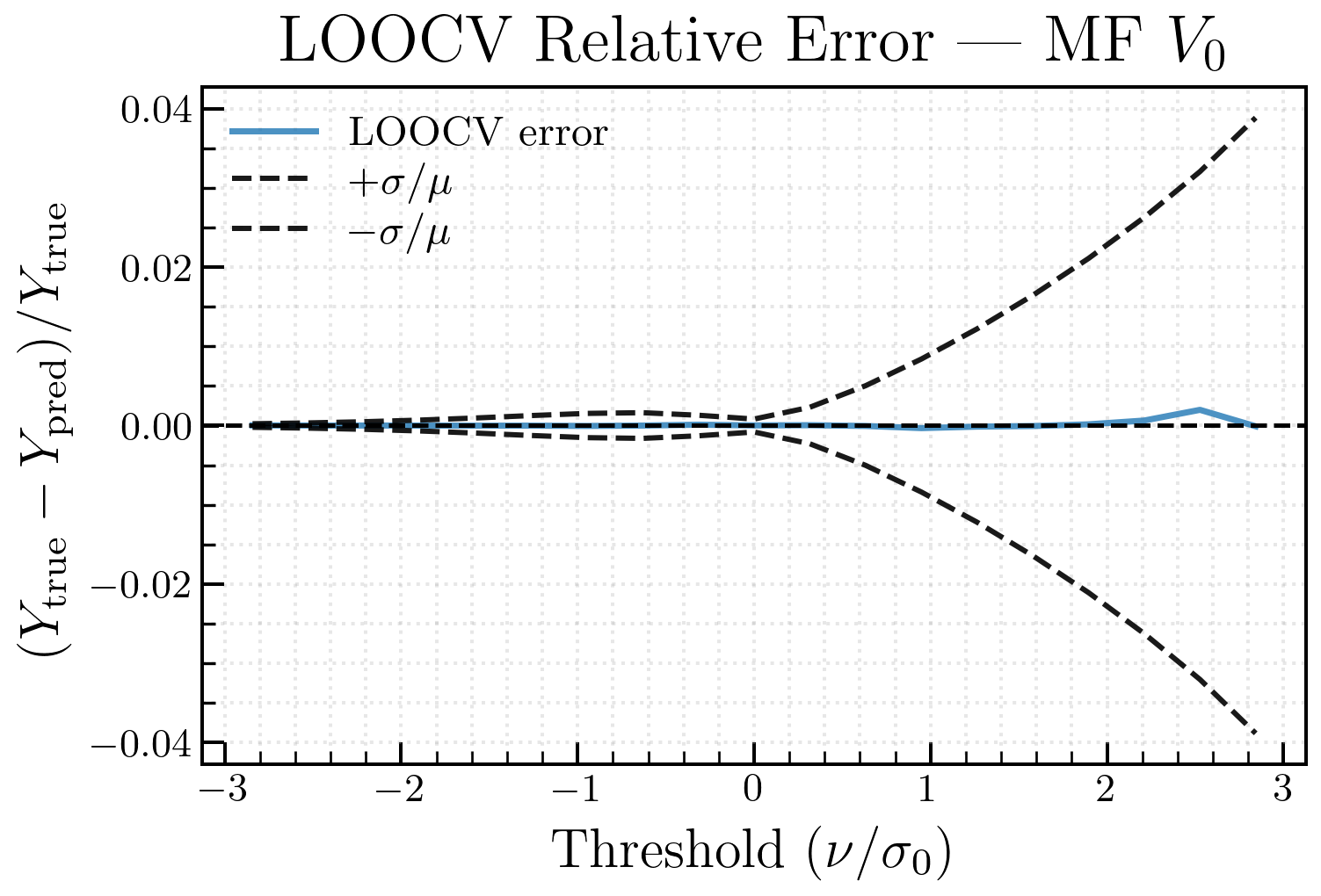}
    \includegraphics[width=0.3\textwidth]{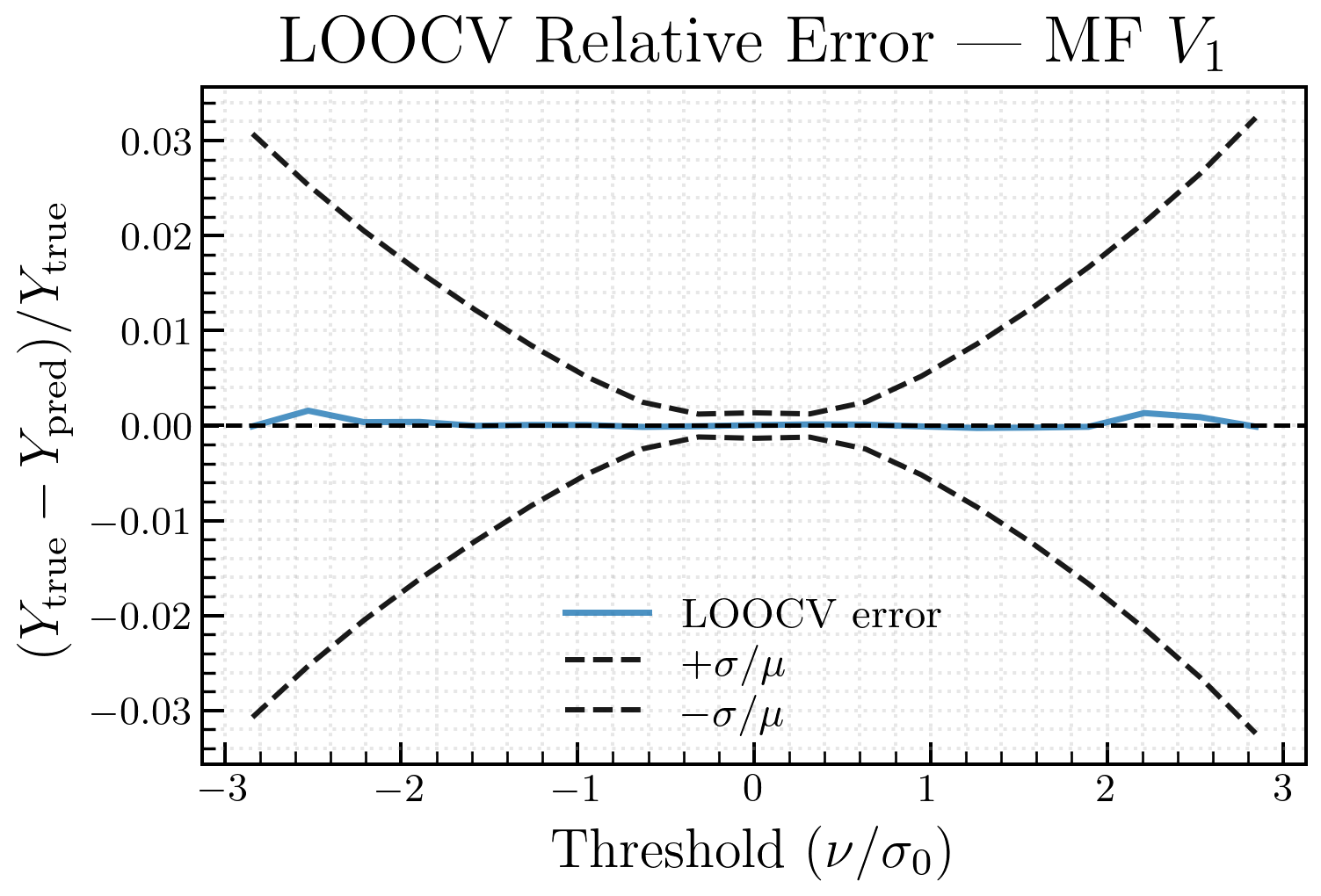}
    \includegraphics[width=0.3\textwidth]{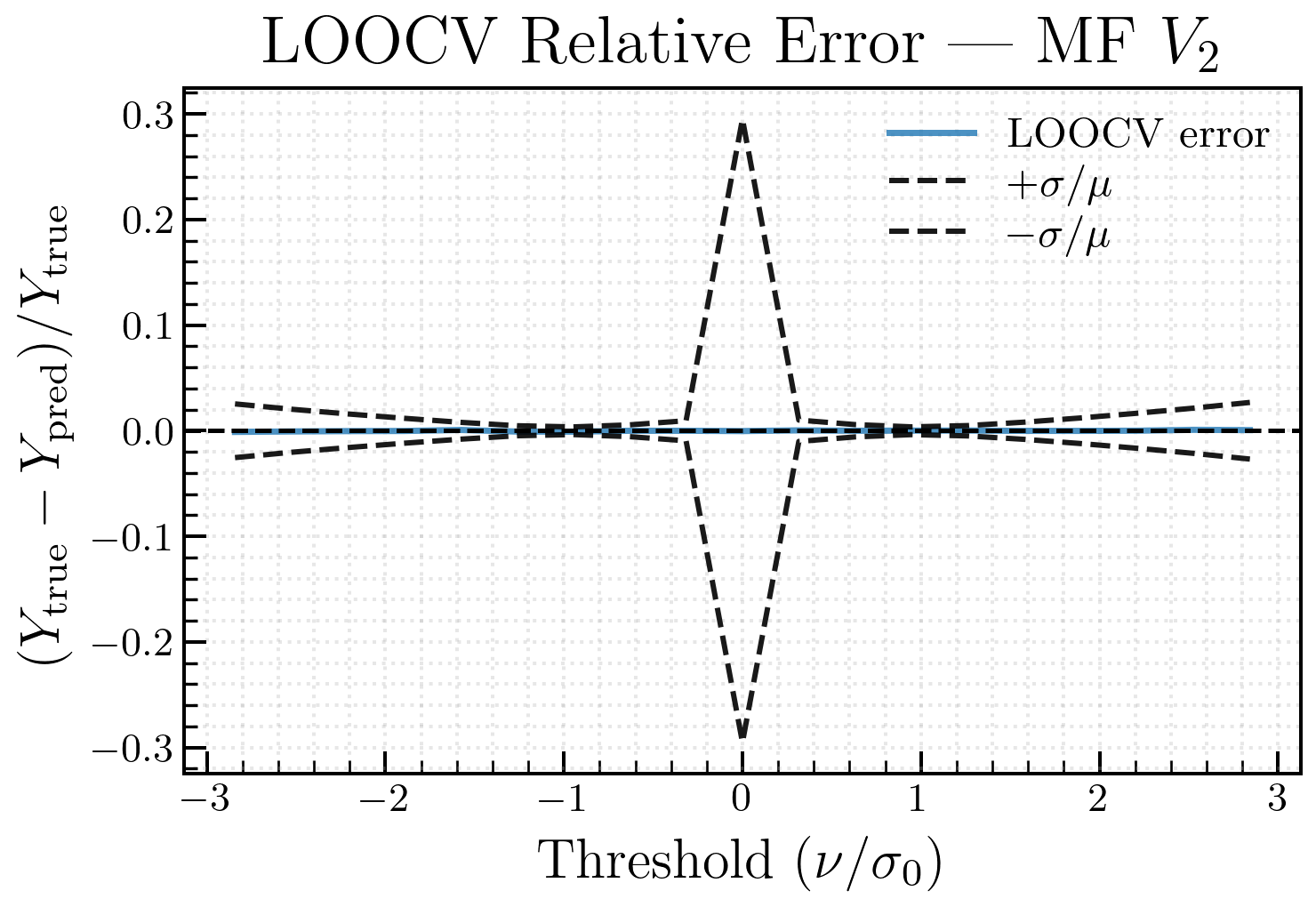}\\[8pt]
    \includegraphics[width=0.3\textwidth]{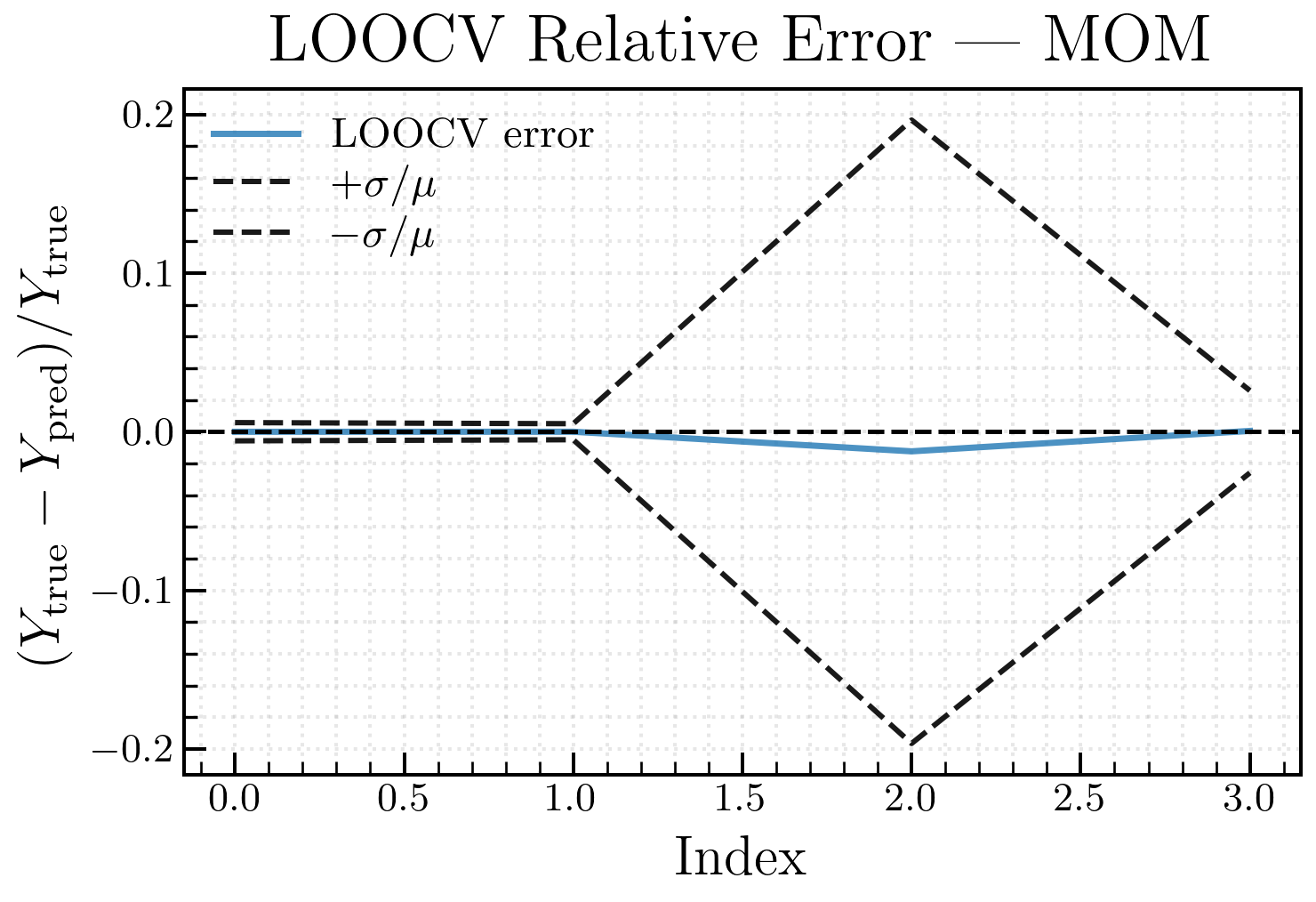}
    \includegraphics[width=0.3\textwidth]{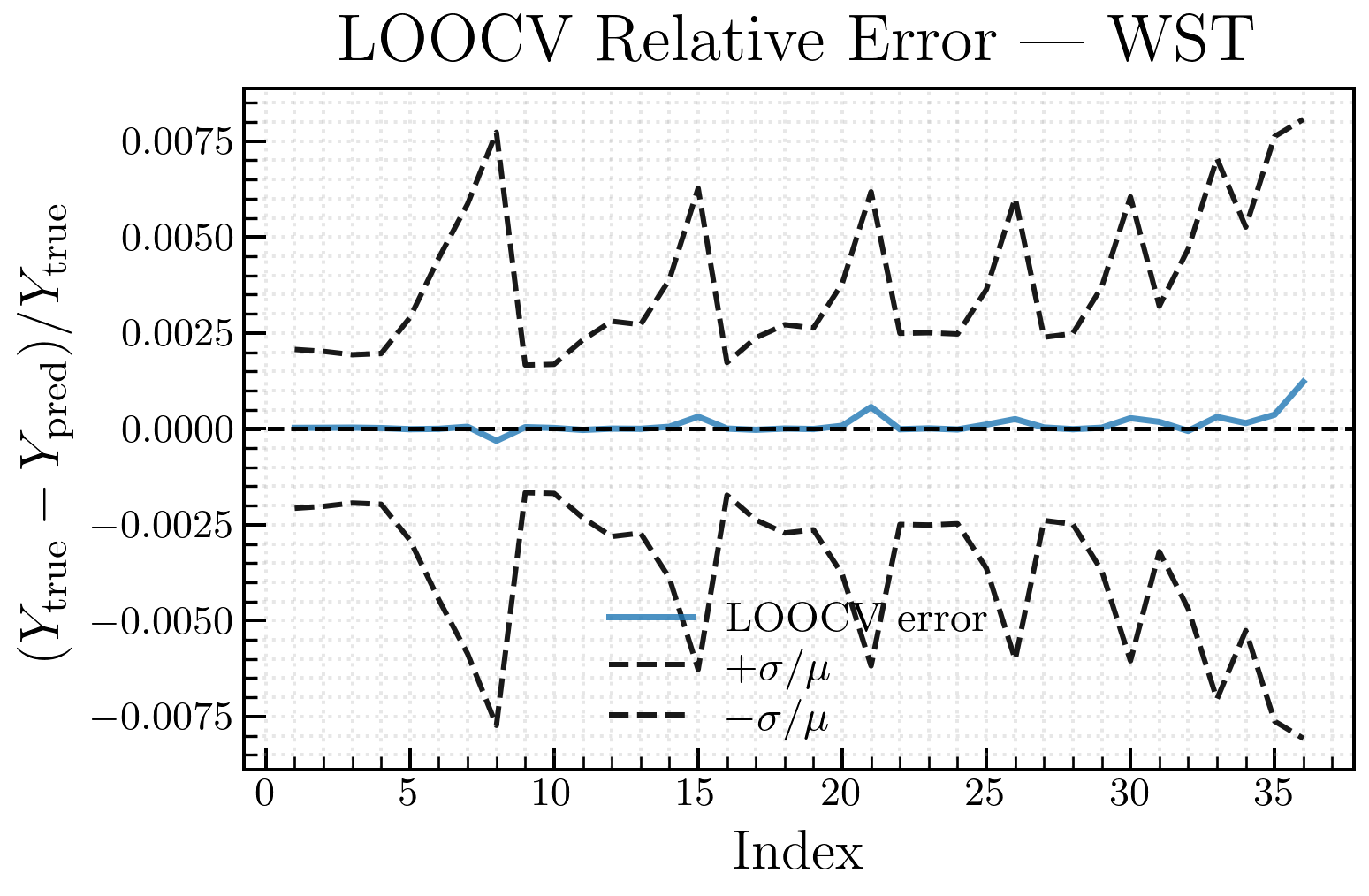}
    \caption{Leave-one-out cross-validation (LOOCV) performance of our emulator across different summary statistics: the angular power spectrum $C_\ell$, peak counts (PC), minima counts (MC), three Minkowski functionals (MFs), moments (MOM), and wavelet scattering transform coefficients (WST). Each panel shows the fractional error between the true statistic at the fiducial cosmology and the emulator prediction, with black dashed lines indicating the expected SO-nominal uncertainty.}
    \label{fig:loocv}
\end{figure*}

To assess the performance of our emulators, we perform a leave-one-out cross-validation (LOOCV) analysis across multiple summary statistics. Figure~\ref{fig:loocv} presents LOOCV results for the angular power spectrum $C_\ell$, peak/minima counts (PC/MC), and Minkowski functionals (MFs). For each statistic, the upper panel shows the fractional difference between the true measurement at the fiducial cosmology (which is excluded here from the GP fitting) and the emulator prediction trained on the remaining cosmologies. The blue curve denotes the fractional error, while the black dashed lines indicate the expected observational uncertainty from the nominal SO forecast (see Section~\ref{subsec:cov}). In all six cases, the emulator errors remain well below $0.1\%$ and lie comfortably within the SO uncertainty band, demonstrating that the emulator reproduces these statistics with accuracy far exceeding the level of observational noise.  The relatively broad $1\sigma$ ranges for PC and MC, compared to the narrower bands for the MFs, highlight the importance of properly accounting for sample variance in future analyses to avoid potential parameter biases.

We have also performed LOOCV on other cosmologies. In most cases, we obtain comparable fractional errors when the withheld cosmology lies near the center of the sampled parameter space. However, for cosmologies located near the boundary of the sampled region, the fractional error can increase to the $\sim 5\%$ level or above --- larger than the expected observational noise as this becomes an extrapolation. We will introduce an additional prior on the cosmological parameters to reduce this error later in our analysis.

\subsection{Cosmology Dependence of Higher-Order Statistics}
\begin{figure*}[t]
    \centering
    \includegraphics[width=\linewidth]{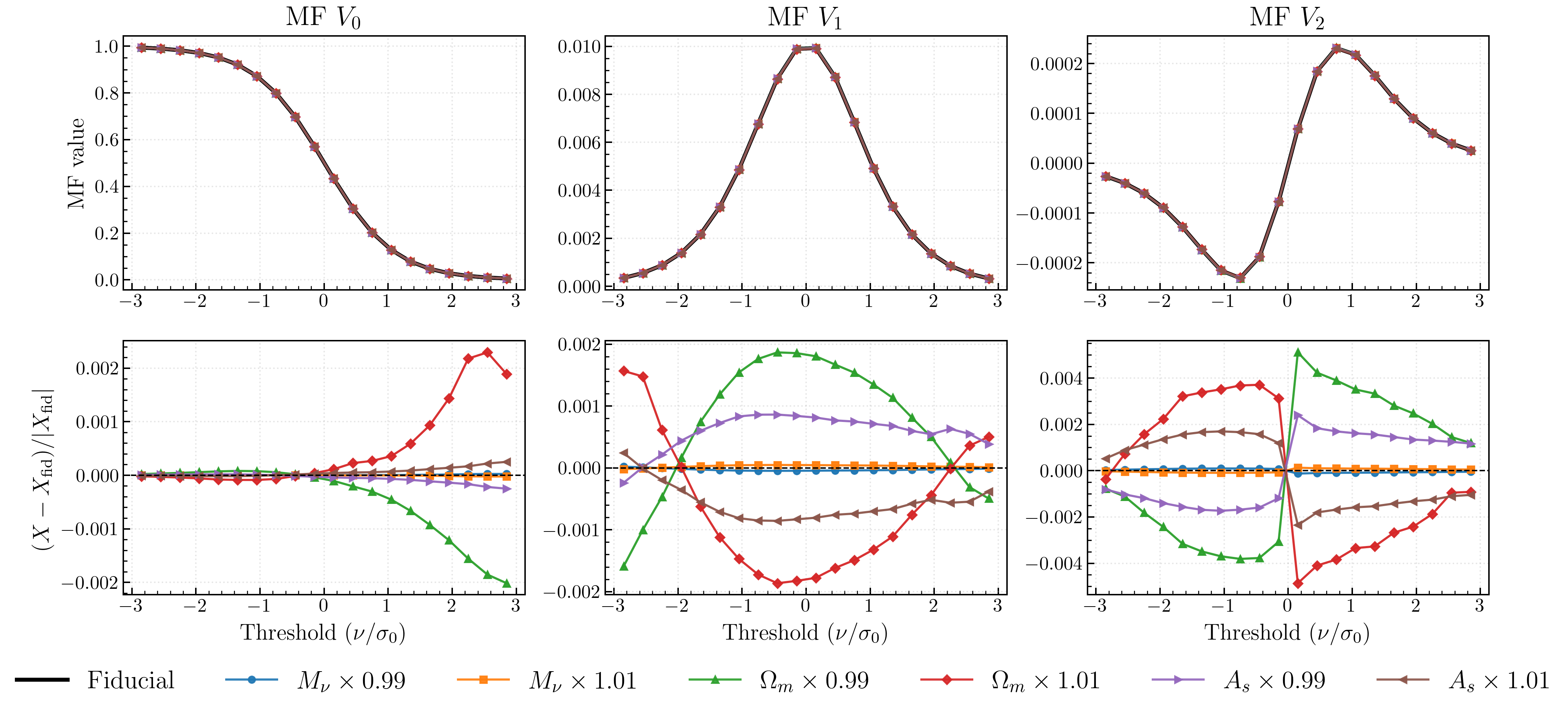}
    \caption{Dependence of the Minkowski functionals (MFs) on cosmological parameters. The upper panels show the MFs as a function of the field threshold, while the lower panels display the corresponding fractional differences with respect to the fiducial cosmology. Increasing $\Omega_m$ or $A_s$ enhances nonlinear structure growth and produces distinctive non-monotonic features in the MFs, whereas decreasing $M_\nu$ has a similar effect to increasing $\Omega_m$, consistent with the damping of small-scale structure by massive neutrinos.}
    \label{fig:vary_mf}
\end{figure*}

\begin{figure*}[t]
    \centering
    \includegraphics[width=\linewidth]{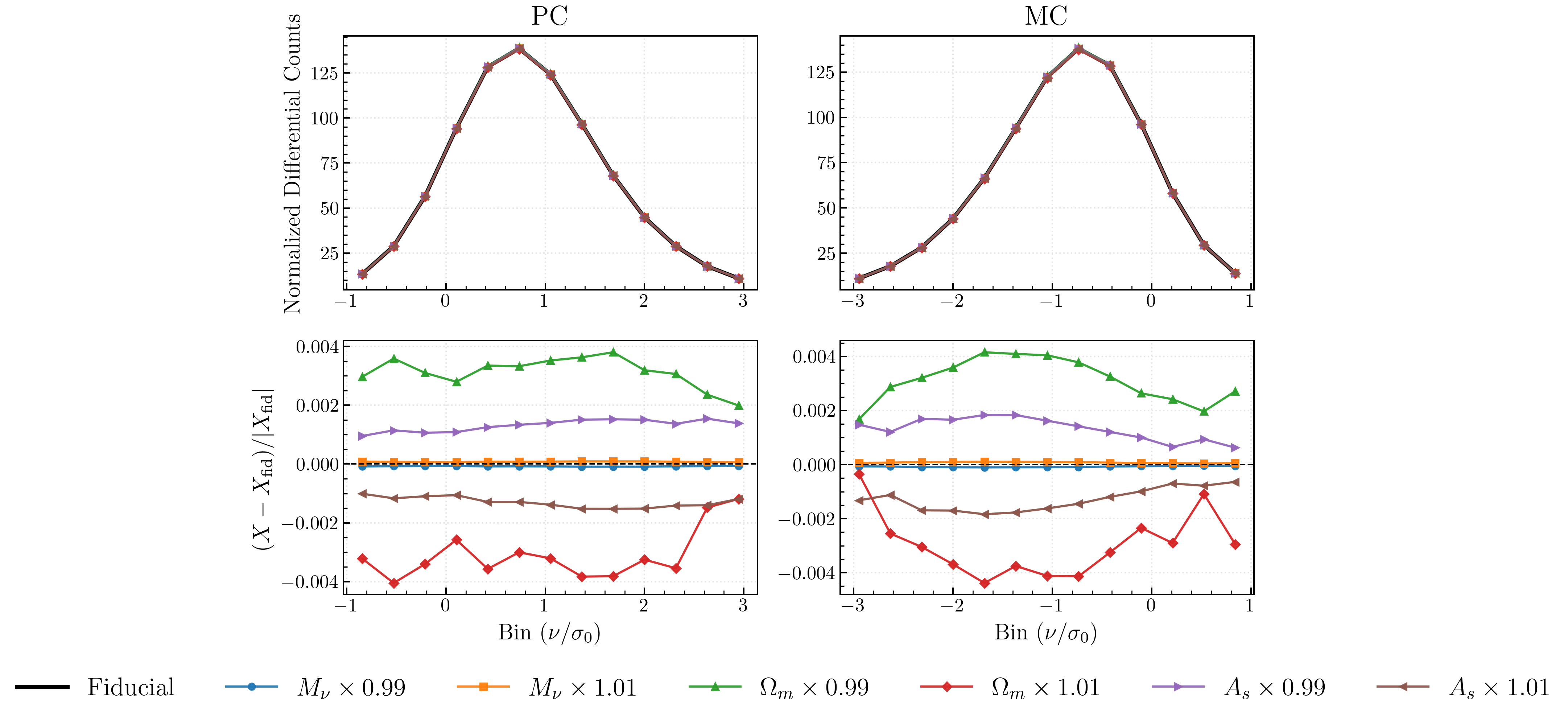}
    \caption{Dependence of the peak counts (PC) and minima counts (MC) on cosmological parameters.
        The upper panels show the absolute counts as a function of threshold, while the lower panels present the corresponding fractional differences relative to the fiducial cosmology. Both statistics are sensitive to changes in $\Omega_m$, $A_s$, and $M_\nu$, reflecting their dependence on the abundance of nonlinear structures. We observe similar behavior on how each of them responds to the change in the cosmological parameters as in Refs.~\cite{Li:2018owg,Coulton:2019enn}.  
        }
    \label{fig:vary_pcmc}
\end{figure*}

\begin{figure*}
    \centering
    \includegraphics[width=\linewidth]{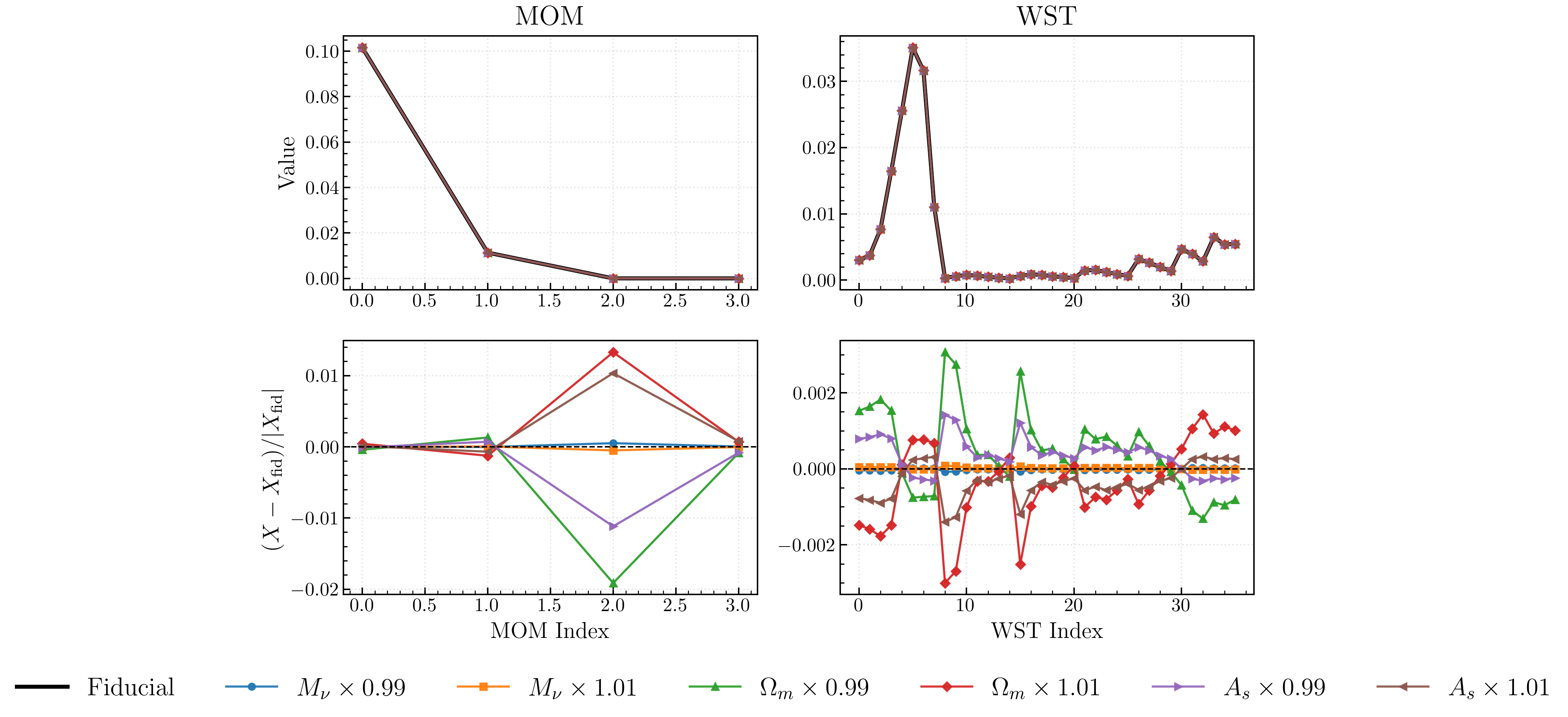}
    \caption{Dependence of the moments (MOM) and wavelet scattering coefficients (WST) on cosmological parameters.  The upper panels show the absolute values of the statistics, while the lower panels present the corresponding fractional differences relative to the fiducial cosmology.}
    \label{fig:vary_momwst}
\end{figure*}

Once the emulator is trained, we examine how each summary statistic responds to variations in the cosmological parameters. Figures~\ref{fig:vary_mf}–\ref{fig:vary_momwst} show the fractional differences relative to the fiducial cosmology for the Minkowski functionals (MFs), peak and minima counts (PC/MC), the low-order moments, and the wavelet scattering transform (WST).
Figure~\ref{fig:vary_mf} shows the response of the three MFs: $V_0$, $V_1$, and $V_2$. The MF curves remain smooth across all thresholds, and the fractional differences display clear and systematic trends. Increasing $\Omega_m$ produces positive deviations at high thresholds and negative deviations at low thresholds, while decreasing $\Omega_m$ reverses this behavior. It also generates a peak at high $\nu$. Variations in $A_s$ generate similar antisymmetric patterns, but with smaller amplitudes and without peaks. In contrast, the sensitivity to $M_\nu$ is much weaker across all thresholds. These responses reflect how $\Omega_m$ and $A_s$ affect nonlinear structure formation and the geometry of excursion sets, whereas changes in $M_\nu$ leave comparatively smaller imprints on the reconstructed field.

The peak and minima counts in Figure~\ref{fig:vary_pcmc} show smooth, monotonic variations with threshold. Similar to what was found in Refs.~\cite{Li:2018owg,Coulton:2019enn}, decreasing $A_s$ and $\Omega_m$, or increasing $M_\nu$, suppresses the number of significant peaks and minima ($|\nu/\sigma_0| > 3$), as most of them originate from rare, massive halos. Conversely, at moderate thresholds (which is the range shown in the figure), the counts are boosted. We exclude the high-significance peaks and minima from our analysis as their mean does not converge sufficiently with the number of simulations available, leading to an unreliable emulator in this regime. 

Figure~\ref{fig:vary_momwst} shows how MOM and WST coefficients vary with cosmology. For the MOM, only a few bins respond noticeably: the variance-like quantities $\sigma_0$ and $\sigma_1$ show small fractional shifts, while the higher-order combinations ($S_1$ and $K_1$) exhibit somewhat larger deviations. Increasing $\Omega_m$ or $A_s$ broadly increases the magnitude of these statistics, while the effects of varying $M_\nu$ remain comparatively modest.

The WST coefficients show a different pattern. The first-order coefficients vary smoothly with cosmology, while the second-order coefficients display oscillatory behavior across bins. The magnitude of these differences is generally small, and the responses to changes in all three cosmological parameters are relatively subtle compared to those seen in the other summary statistics. These trends simply reflect how the multiscale filtering used in the WST responds to shifts in the underlying field, without indicating strong discriminating power.

Across all statistics, we observe consistent qualitative trends: $\Omega_m$ induces the largest variations, $A_s$ produces similar but smaller responses, and $M_\nu$ shows the weakest dependence. These results summarize how each statistic changes with cosmology before incorporating observational uncertainties or forecast parameter sensitivities. In Appendix~\ref{appendix:noiseOnStatistics}, we also show how Wiener-filtering and noise can affect the shape of MFs and PC/MC.

\section{Likelihood function}\label{sec:likelihood}
\subsection{Covariance Matrix}\label{subsec:cov}
We assume a Gaussian likelihood function in our analysis. The covariance matrix is obtained directly from the simulation suite {\tt MassiveNuS}. We construct our covariance matrix at the fiducial cosmology: $\{\Omega_m=0.3,\,M_\nu=0.1\,\text{eV},\,10^9A_s=2.1\}$. We use $N_{\rm sim}=10,000$ simulations in total to construct the covariance matrix:
\begin{align}
    \hat{\boldsymbol{C}}_{\rm sim} = \frac{r_{\rm sky}}{N_{\rm sim}-p-1}\sum_{i=0}^{N_{\rm sim}}(\boldsymbol{d}_i-\bar{\boldsymbol{d}})(\boldsymbol{d}_i-\bar{\boldsymbol{d}})^{T}\,,
\end{align}
where $\boldsymbol{d}_i$ denotes the data vector for each simulation that contains the summary statistics from each map and $\bar{\boldsymbol{d}}=\sum_{i=0}^{N_{\rm sim}}\boldsymbol{d}_i$. Here $p$ is the length of the data vector $\boldsymbol{d}_i$ and the denominator is the Hartlap correction~\cite{Hartlap:2006kj} used to ensure an unbiased estimation of the covariance matrix from a limited number of simulations. In our case, the ratio $N_{\rm sim}-p-1/(N_{\rm sim}-1) \approx 0.98-0.99$, which is very close to unity because the length of our data vector is roughly $100$. Lastly, $r_{\rm sky}$ is a parameter used to rescale the covariance matrix in order to match the sky coverage for a specific experiment given the small sky coverage of our simulated maps. In our case, we assume the fraction of sky $f_{\rm sky} \simeq 0.4$ is covered by the nominal SO survey~\cite{SimonsObservatory:2018koc}.\footnote{Updated (wider) SO survey strategies are described in Ref.~\cite{SimonsObservatory:2025wwn}, but we do not consider these here.} As a result, $r_{\rm sky}$ in our case is $2.97\times10^{-4}/0.4$, where the numerator comes from the sky area $(3.5\,\text{deg})^2$ of {\tt MassiveNuS}. 

\begin{figure*}
    \centering
    \includegraphics[width=0.3\linewidth]{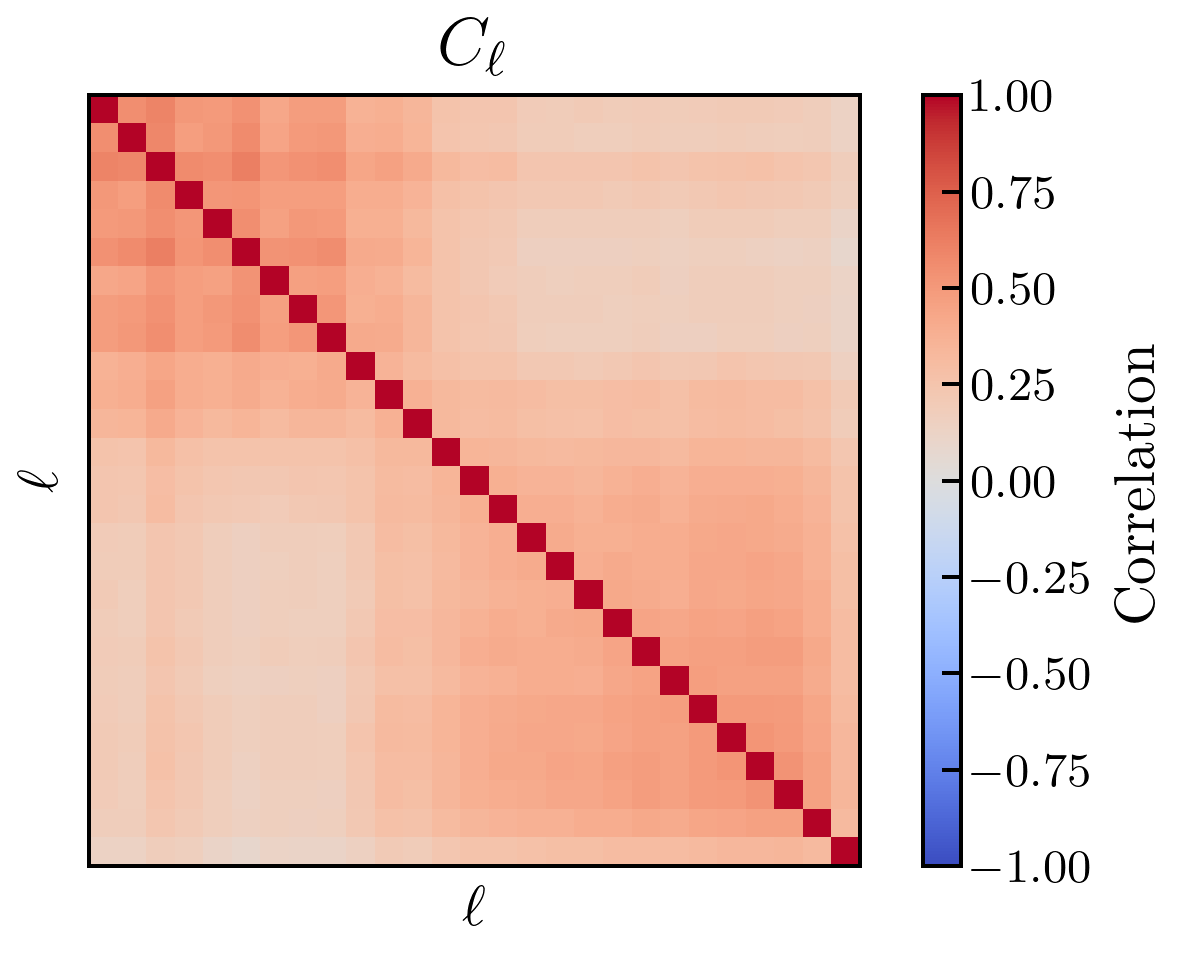}
    \includegraphics[width=0.3\linewidth]{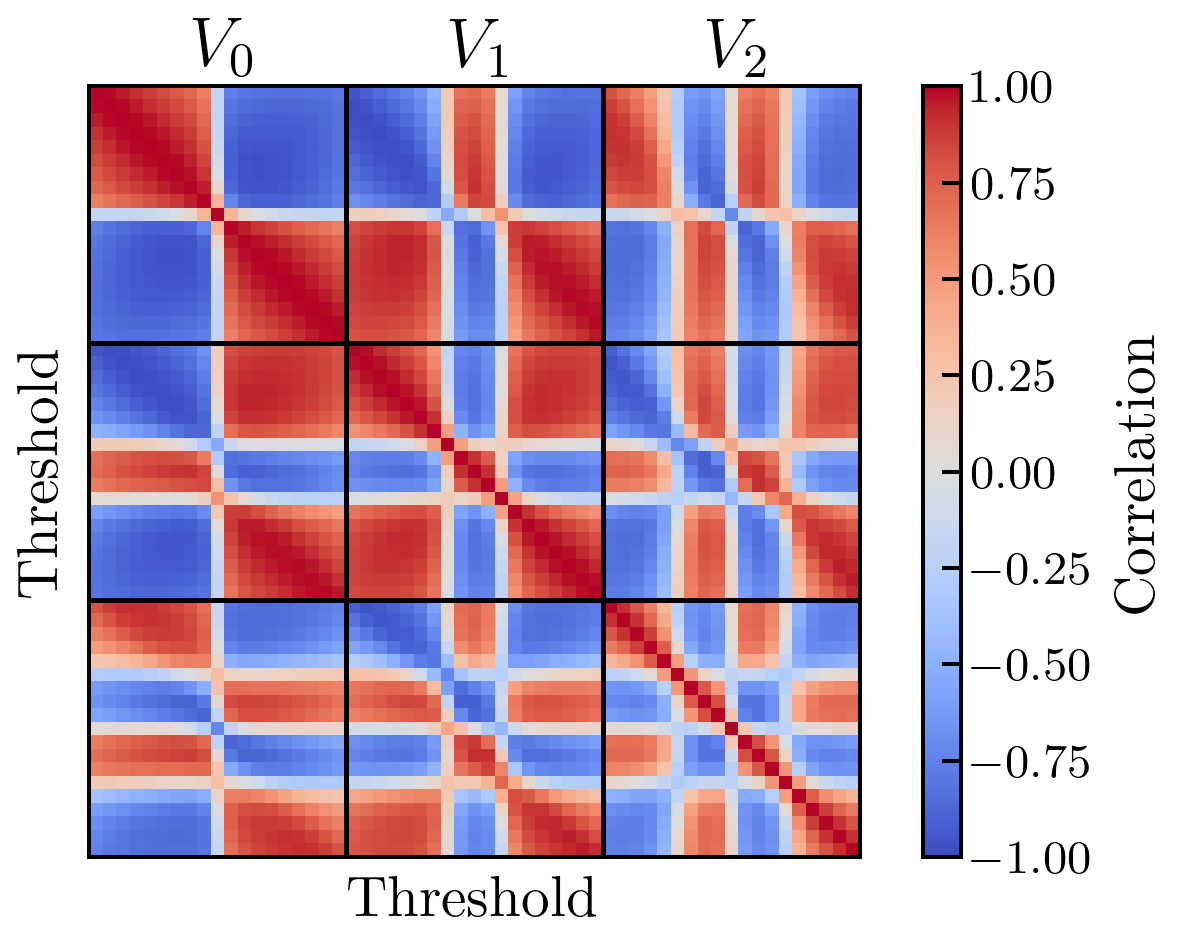}
    \includegraphics[width=0.3\linewidth]{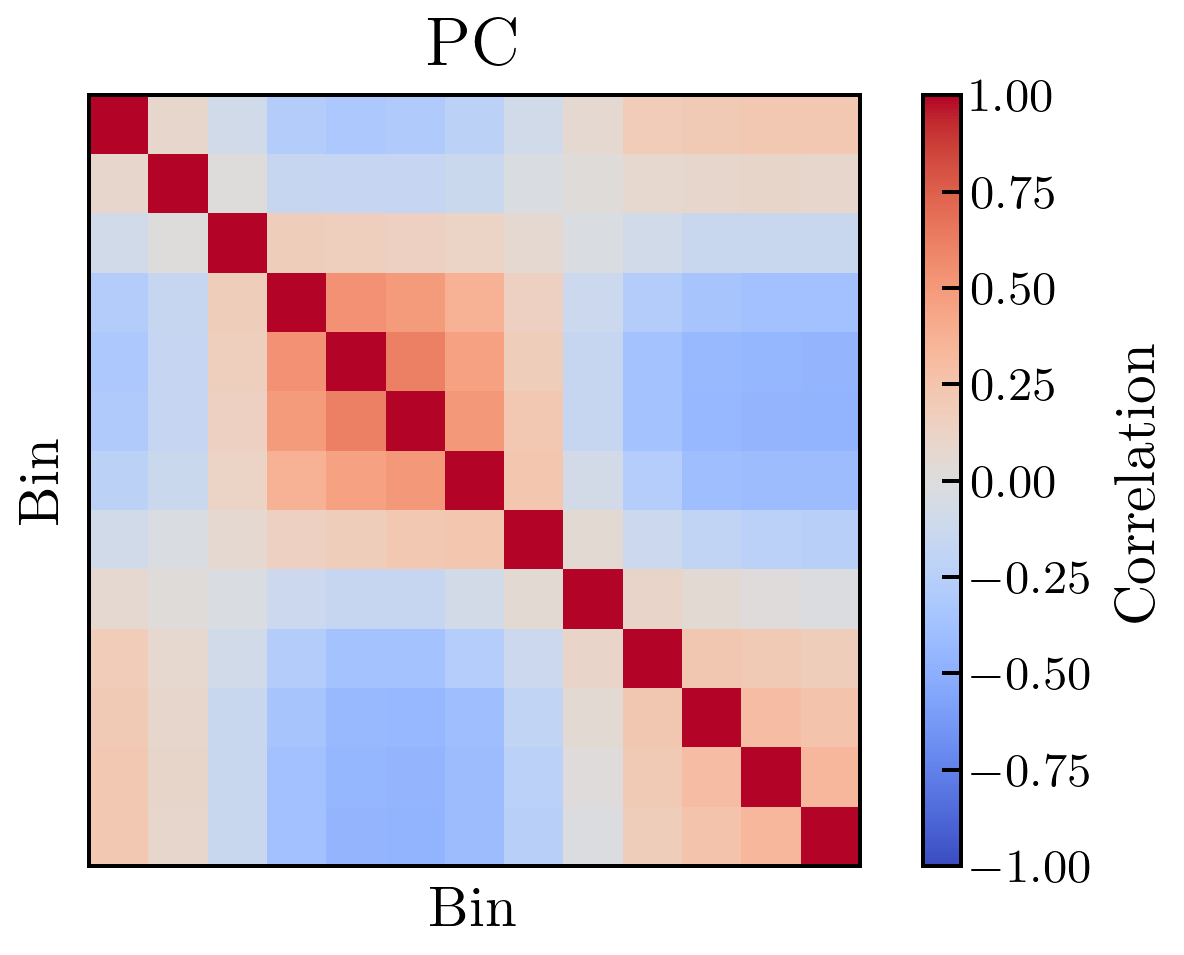} \\
    \includegraphics[width=0.3\linewidth]{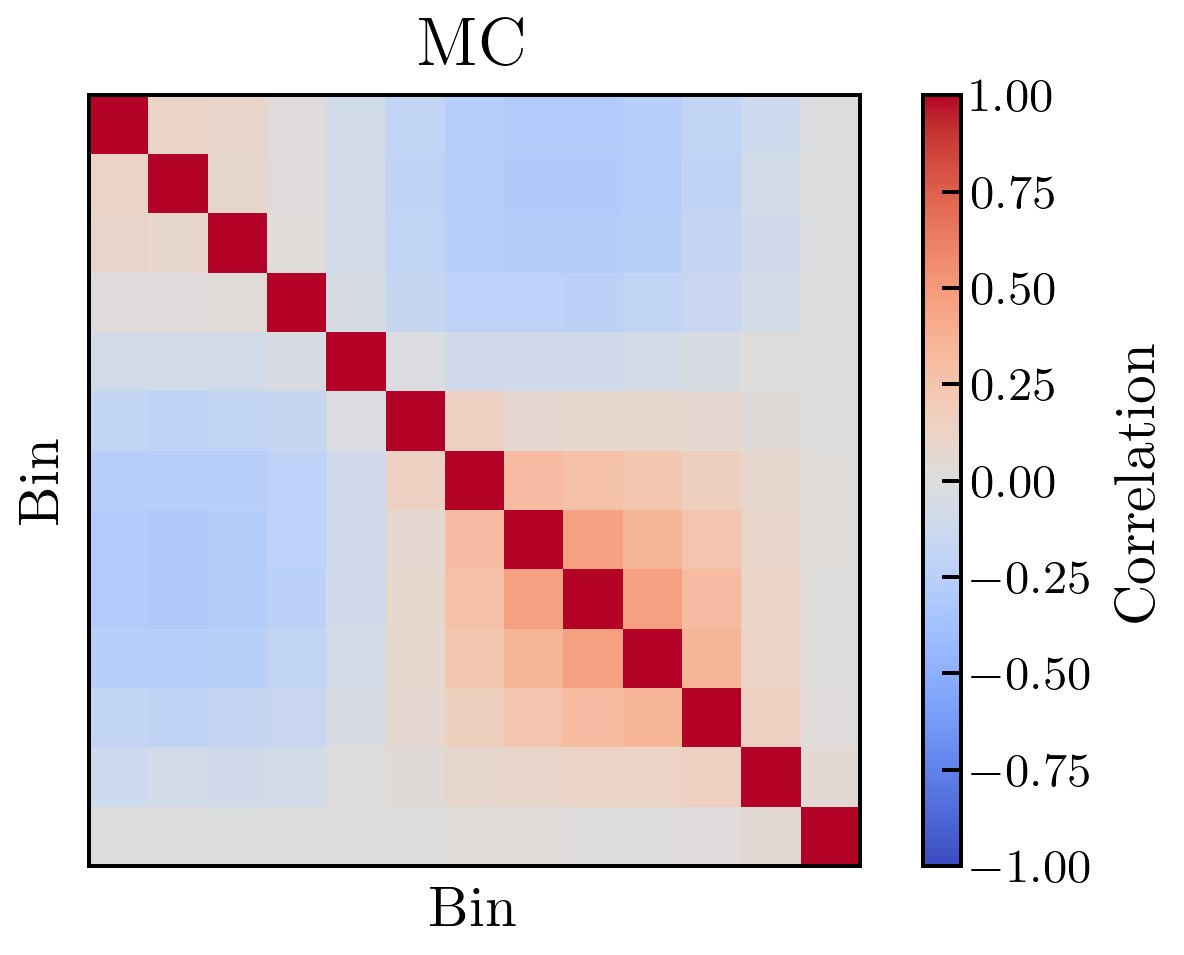}
    \includegraphics[width=0.3\linewidth]{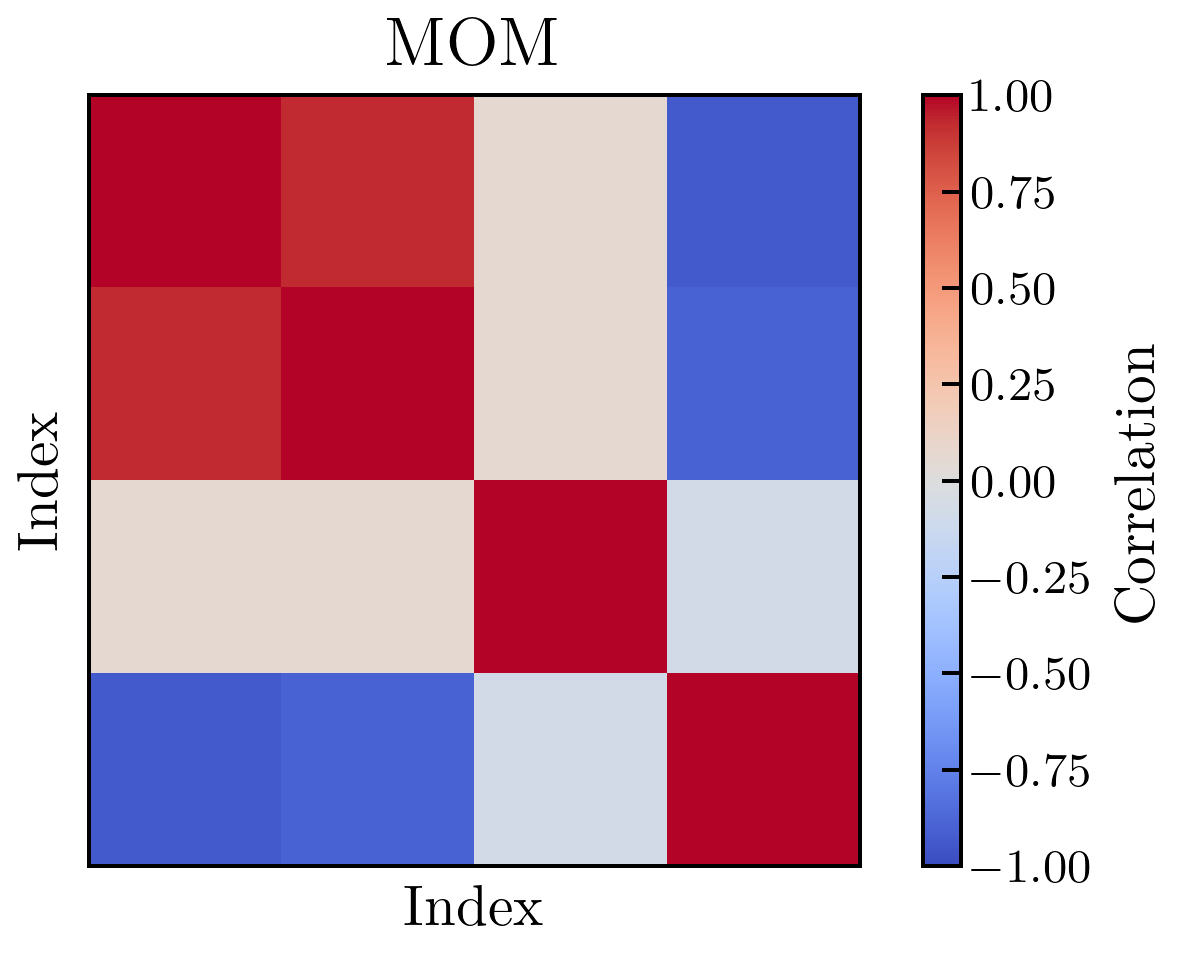}
    \includegraphics[width=0.3\linewidth]{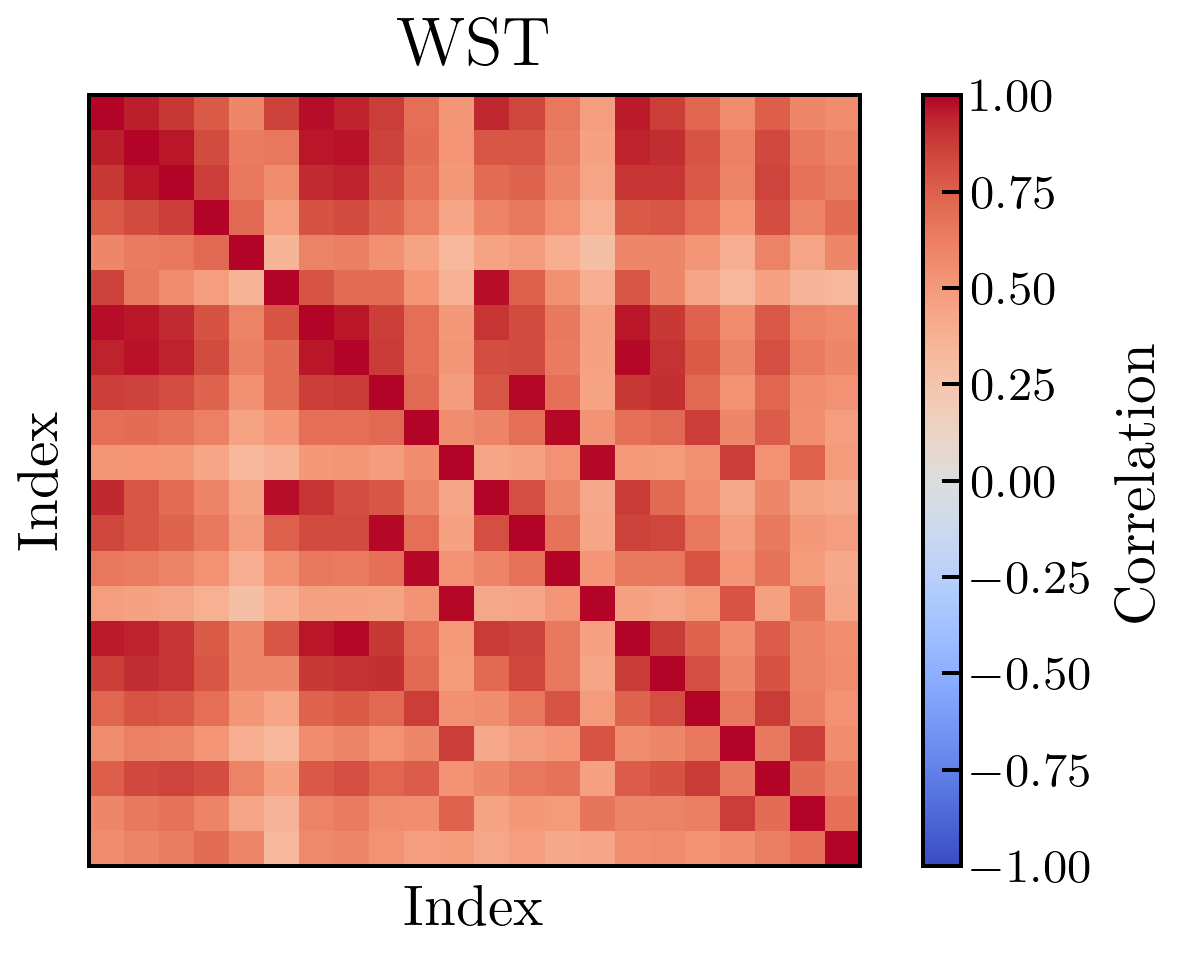} \\
    \includegraphics[width=0.75\linewidth]{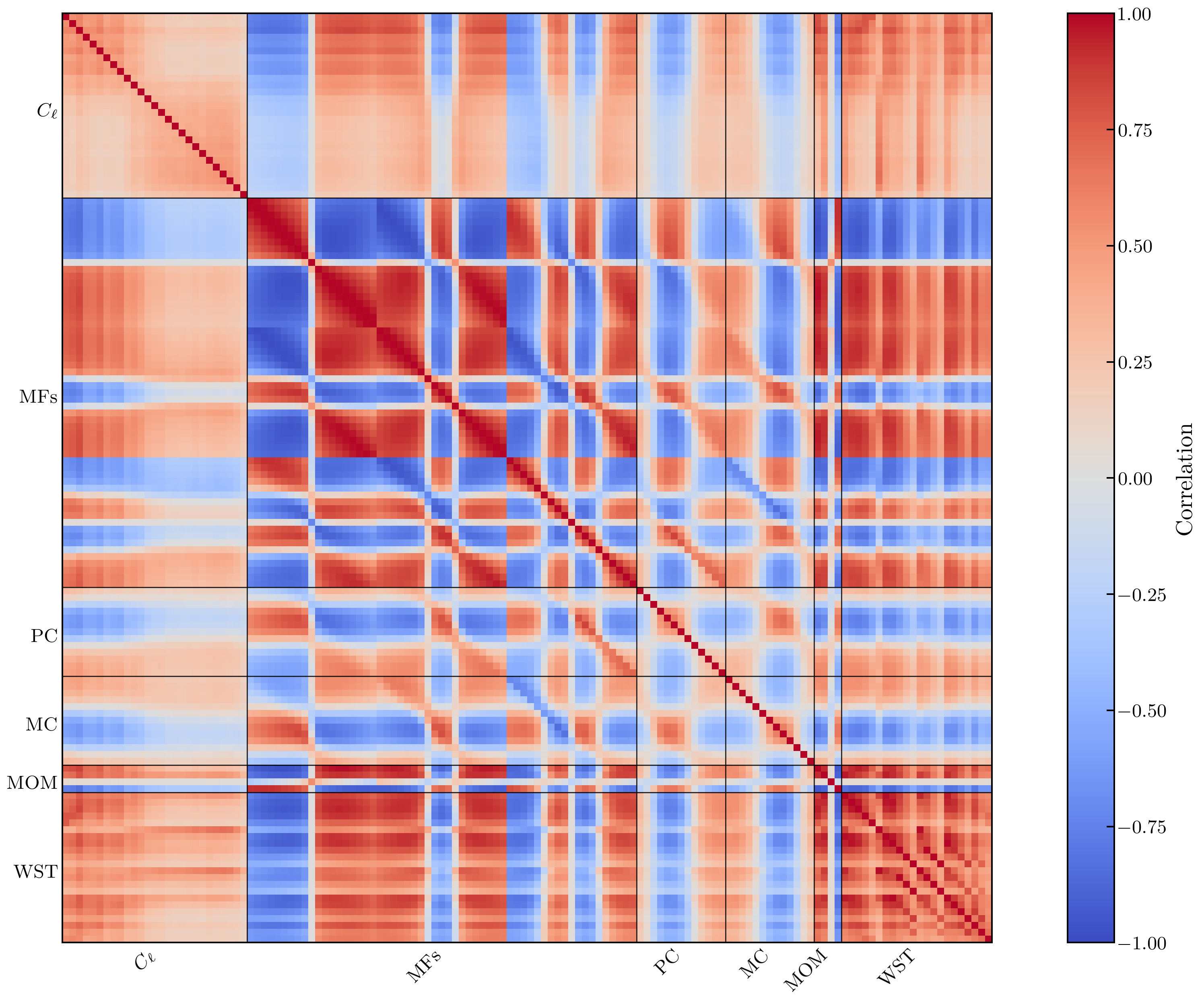}
    \caption{Correlation matrices for each statistic we consider in our analysis. This includes the power spectrum $C_\ell$, the three Minkowski functionals (MFs) $V_0$, $V_1$, and $V_2$, the peak/minima counts (PC/MC), moments (MOM), and the wavelet scattering transforms coefficients (WST). The bottom panel shows the full correlation matrix formed by combining all statistics, with block boundaries indicating the individual contributions from each summary statistic. The correlation matrices for $C_\ell$, PC, and MC are almost diagonal, while the rest show strong correlations across different modes.}
    \label{fig:cov}
\end{figure*}

Figure~\ref{fig:cov} shows the correlation matrix for each statistic. For statistics like $C_\ell$, PC, and MC, the correlation matrix is nearly diagonal. The MFs and MOM show moderate correlations with each other, while the WST shows the highest correlation coefficient even though we have already removed pairs of WST coefficients that have correlation $|r|$ greater than 0.99. The highly correlated WST coefficients will lead to only marginal improvements on the parameter constraints which will be shown in Section~\ref{sec:forecast}. In addition to the individual statistics, we show the full correlation matrix including cross-correlations between different statistics in the lower panel of Figure~\ref{fig:cov}. The absolute values of the cross-correlation coefficients are generally modest, with typical values around $\sim 0.25$, likely reflecting the impact of noise and the fact that these statistics probe largely distinct aspects of the field. We find that WST and MOM exhibit relatively stronger correlations with the MFs, suggesting that a significant fraction of the information captured by these statistics overlaps with that encoded by the MFs. In contrast, PC/MC remain largely uncorrelated with the MFs, indicating that they provide complementary information. The corresponding convergence test of our covariance matrix can be found in Appendix~\ref{appendix:convergenceCov}.

In addition to the covariance $\hat{\boldsymbol{C}}_{\rm sim}$ constructed directly from the simulation, we include the emulator uncertainty covariance $\hat{\boldsymbol{C}}_{\rm emu}$, which accounts for the intrinsic error of the GP model. For a test parameter point $\boldsymbol{\theta}_*$, a trained GP defines a predictive posterior distribution
\begin{align}
    p(y_* \,|\, \boldsymbol{\theta}_*, X, y)
    &= \mathcal{N}\!\left(y(\boldsymbol{\theta}_*),\, \sigma_y^2(\boldsymbol{\theta}_*)\right),
\end{align}
where $y(\boldsymbol{\theta}_*)$ is the GP predictive mean and $\sigma_y(\boldsymbol{\theta}_*)$ is the predictive standard deviation.

The predictive mean $y(\boldsymbol{\theta}_*)$ serves as our theoretical prediction for the summary statistics  when performing posterior inference, while the predictive uncertainty $\sigma_y(\boldsymbol{\theta}_*)$ quantifies the emulator’s intrinsic interpolation error. Following the standard GP formalism with a Matérn kernel, the predictive variance takes the analytical form
\begin{align}
    \sigma_y^2(\boldsymbol{\theta}_*) 
    &= k_{\nu}(\boldsymbol{\theta}_*, \boldsymbol{\theta}_*) 
    - \boldsymbol{k}_{\nu}^{\top}(\boldsymbol{\theta}_*) 
    \left[ \boldsymbol{K}_{\nu} + \sigma_n^2 \boldsymbol{I} \right]^{-1} 
    \boldsymbol{k}_{\nu}(\boldsymbol{\theta}_*),
\end{align}
where $k_{\nu}$ denotes the Matérn kernel with smoothness parameter $\nu$, $\boldsymbol{K}_{\nu}$ is the kernel matrix among the training points, and $\boldsymbol{k}_{\nu}(\boldsymbol{\theta}_*)$ is the covariance vector between $\boldsymbol{\theta}_*$ and the training data. 

The resulting emulator covariance matrix is incorporated into the total covariance used for the likelihood evaluation:
\begin{align}
    \hat{\boldsymbol{C}}_{\rm tot} 
    &= \hat{\boldsymbol{C}}_{\rm sim} + \hat{\boldsymbol{C}}_{\rm emu}.
\end{align}
This formulation ensures that both the stochastic variance from finite simulation sampling and the interpolation uncertainty of the emulator are consistently propagated into our posterior inference. As a conservative choice, when we get $\hat{\boldsymbol{C}}_{\rm emu}$ at the fiducial cosmology, we exclude the statistics at the fiducial cosmology in the training data set to avoid a narrow variance prediction from the emulator at that point. Note that we still have that information in our emulator used for posterior sampling. We also fix the uncertainties within the emulator which might not be applicable when we sample the emulator in regions with fewer number of simulations. Attempting to vary it together with the cosmological parameters leads to a very localized and multi-modal posterior distribution. Lastly, we also leave the inclusion of the contribution from the sample variance of the random seed used for future work, because we currently use only one random seed to generate the initial conditions. 

\subsection{Gaussian Likelihood Function}
After we build our covariance matrix, the log Gaussian likelihood function up to a constant is constructed as follows:
\begin{align}
    -\ln \mathcal{L}(\boldsymbol{\theta}) = \frac{1}{2}\left(\boldsymbol{y}(\boldsymbol{\theta})-\boldsymbol{y}_{\rm fid}\right)^{T}\hat{\boldsymbol{C}}_{\rm tot}^{-1}\left(\boldsymbol{y}(\boldsymbol{\theta})-\boldsymbol{y}_{\rm fid}\right)\,,
\end{align}
where $\boldsymbol{\theta}$ denotes the cosmological parameters we are going to sample, $\boldsymbol{y}(\boldsymbol{\theta})$ is the emulator trained for each statistic as a function of cosmological parameters, and $\boldsymbol{y}_{\rm fid}$ is the assumed observation at the fiducial cosmology. 

The posterior sampling is performed with {\tt emcee}~\cite{foreman2013emcee}\footnote{\url{https://emcee.readthedocs.io/en/stable/}}, an affine invariant Markov chain Monte Carlo (MCMC) ensemble sampler. We run 32 parallel chains and ensure that the length of the chain is at least 100 times longer compared to the auto-correlation time~\cite{2010CAMCS...5...65G}. 

\section{Forecast}\label{sec:forecast}
In this section, we explore two forecasting scenarios: one in which the total neutrino mass, $M_\nu$, is fixed to its fiducial value, and another in which it is treated as a free parameter. Our forecast includes only modes in each convergence map with $\ell\in[300,3000]$, a restriction imposed by the limited $3.5\text{ deg}^2$ sky area. Since this cut removes a substantial number of large-scale modes, we focus primarily on the relative gains in information rather than the absolute constraining power.\footnote{Note that the vast majority of the cosmological information in modes with $\ell < 300$ is contained in the standard angular power spectrum, since these modes are highly linear~\cite{Lewis:2006fu}.}

\subsection{Validation with Analytical Method}\label{subsec:validation}
\begin{figure}[t]
    \centering
    \includegraphics[width=0.85\linewidth]{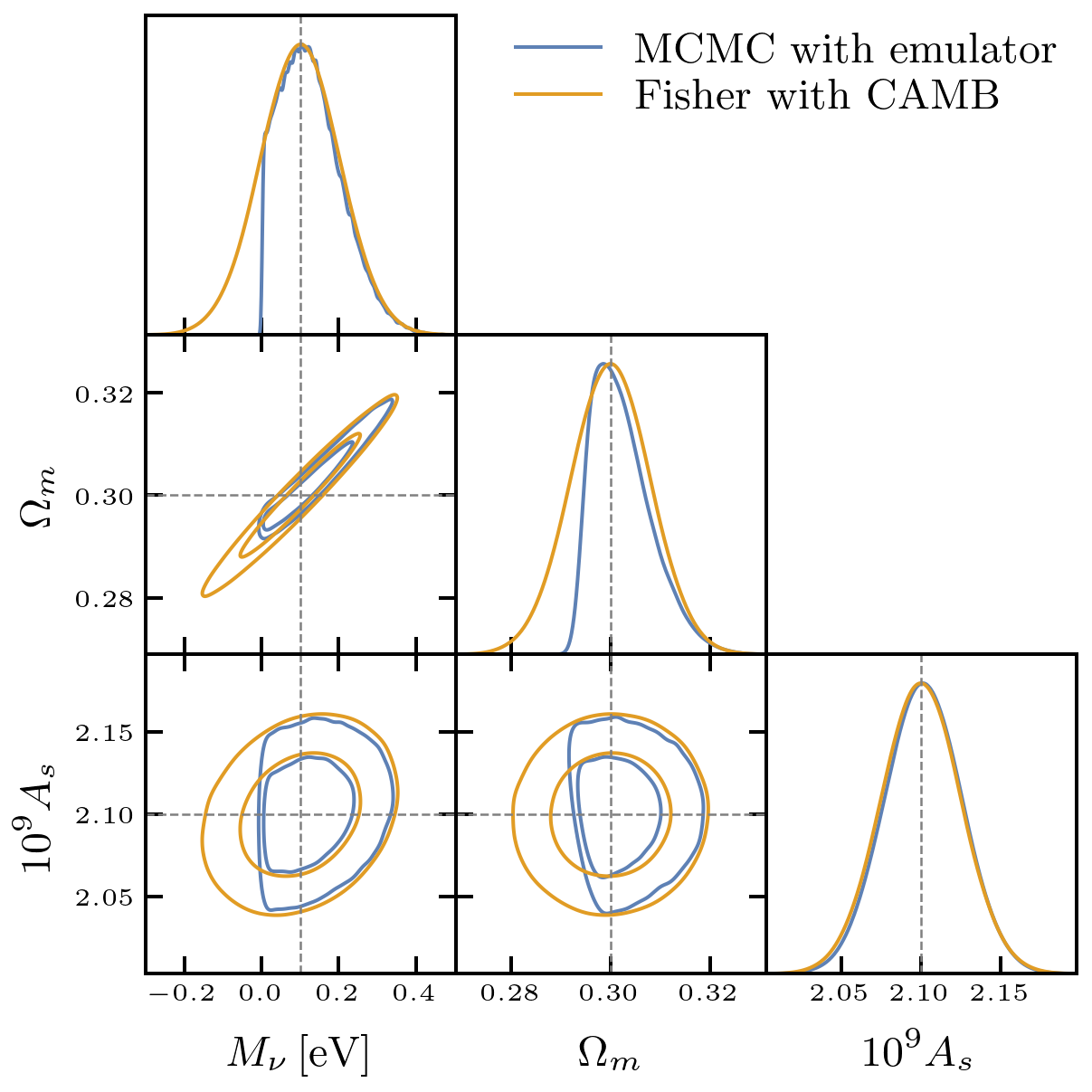}
    \caption{Comparison of parameter constraints between a Fisher forecast with CAMB (blue) and an MCMC with our emulator (yellow) using only $C_{\ell}^{\kappa\kappa}$. We put priors on both $\Omega_m$ and $A_s$ motivated by DESI DR2 BAO and Planck+ACT DR6 CMB data. We find excellent agreement between the two forecasts (after accounting for the truncation of the MCMC results due to the emulator training range being restricted to the physical region $M_\nu \geq 0$), showing that our forecast with the emulator is both robust and accurate.}
    \label{fig:validation}
\end{figure}

Before assessing the information content of the full set of statistics, we first validate our methodology at the level of the power spectrum alone by comparing constraints from a Fisher analysis using CAMB to those obtained from an MCMC forecast with our emulator. Figure~\ref{fig:validation} presents this comparison. For both approaches, we use the convergence power spectrum over $\ell\in[300,3000]$, following the setup described in Section~\ref{sec:higherOrderStatistics}. Note that the reason why we use MCMC instead of a Fisher forecast is because the number of simulated cosmologies is not sufficient to provide an accurate evaluation of the numerical derivative.

The Fisher matrix is computed as
\begin{align}
    \boldsymbol{F}_{\alpha\beta}
    = \sum_{\ell}\frac{f_{\rm sky}(2\ell+1)}{2(C_\ell^{\kappa\kappa}+N_\ell^{(0),\rm MV})^2}
    \frac{\partial C_\ell^{\kappa\kappa}}{\partial p_\alpha}
    \frac{\partial C_\ell^{\kappa\kappa}}{\partial p_\beta},,
\end{align}
where $C_\ell^{\kappa\kappa}$ is the CAMB prediction for the lensing convergence power spectrum, including the HaloFit prescription~\cite{Mead:2020vgs}, and $N_\ell^{(0),\rm MV}$ denotes the minimum-variance reconstruction noise described in Section~\ref{sec:simulation}. The parameter vector is $p_\alpha\in\{\Omega_m,\,A_s,\,M_\nu\}$, and derivatives are evaluated numerically by varying each parameter by $\pm1\%$ around the fiducial cosmology.

\begin{figure}[t]
    \centering
    \includegraphics[width=0.85\linewidth]{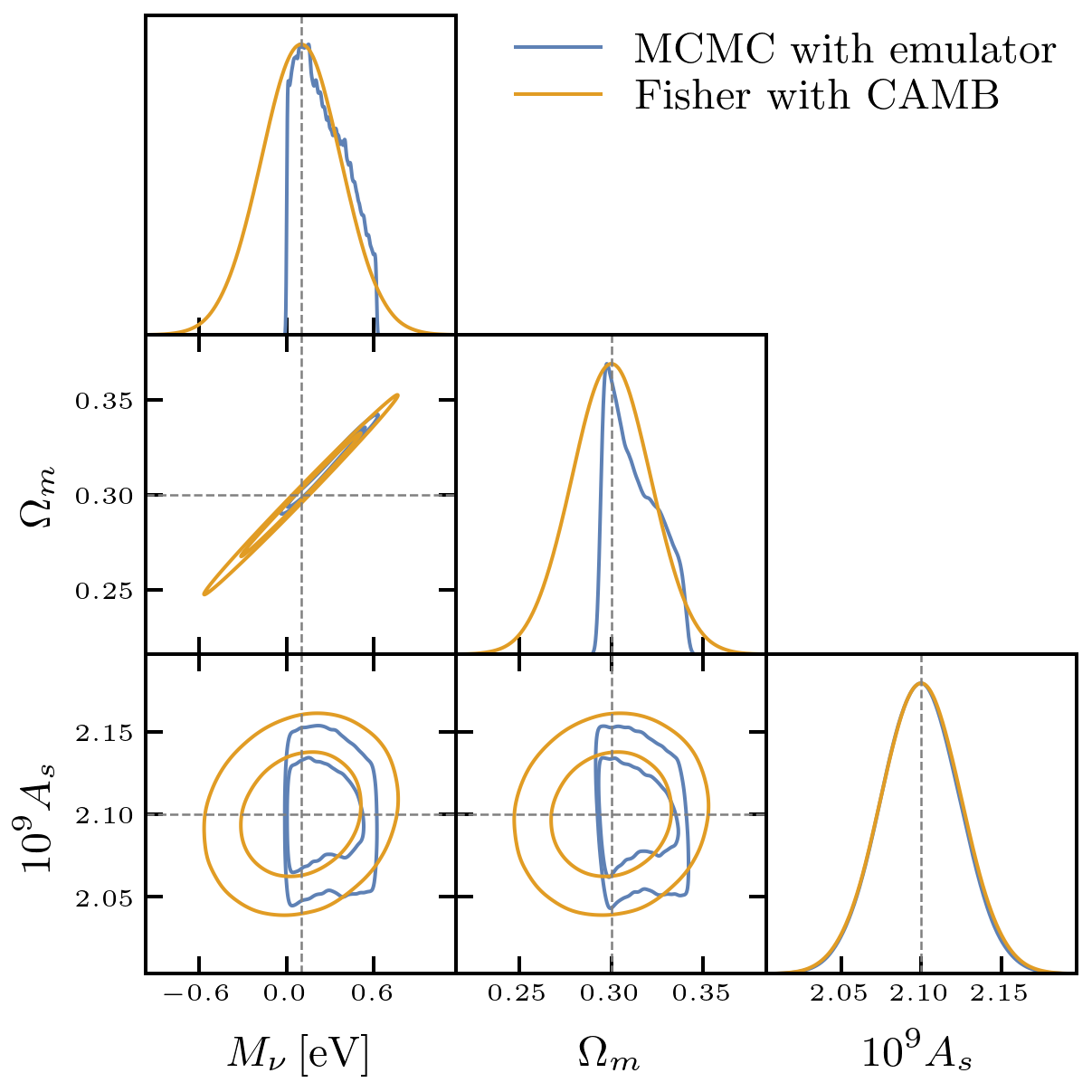}
    \caption{Comparison of parameter constraints between Fisher forecast with CAMB (blue) and an MCMC with our emulator (yellow) using only $C_{\ell}^{\kappa\kappa}$. We put a prior here only on $A_s$, motivated by Planck+ACT DR6 CMB data. We find that the agreement is good at the 68\% contour but not for the 95\% one, where the emulator hits the prior set by the available simulations, which can be seen in Figure~\ref{fig:massivenus}.}
    \label{fig:validation2}
\end{figure}

Because the lensing power spectrum does not independently constrain all three parameters, we impose Gaussian priors on $\Omega_m\sim\mathcal{N}(0.3,0.0086^2)$ and $10^9A_s\sim\mathcal{N}(2.1,0.025^2)$, motivated by DESI DR2 BAO-like~\cite{DESI:2025zgx} and Planck+ACT DR6 CMB-like~\cite{AtacamaCosmologyTelescope:2025blo} measurements. These priors prevent the MCMC sampler from entering regions of parameter space that are poorly represented in the emulator training set, as shown in Figure~\ref{fig:massivenus}. This effect is illustrated in Figure~\ref{fig:validation2}, where we apply a prior only on $A_s$: the 68\% contours from the MCMC and Fisher analyses remain in good agreement, but the 95\% contours diverge due to sampling challenges near the parameter boundaries imposed by the simulation suite. As a result, we can see sharp cutoffs at around 0.6 eV for the neutrino mass.

When priors on both $\Omega_m$ and $A_s$ are included, the agreement improves substantially. As shown in Figure~\ref{fig:validation}, the emulator-based constraints (blue contours) closely match the Fisher predictions. The sharp cutoff in the $M_\nu$ direction arises from the physical prior enforcing $M_\nu \geq 0$. Overall, this validation demonstrates that emulator-based forecasts reproduce the Fisher results with high fidelity. The close agreement between the two approaches also suggests that, for the power spectrum alone, the off-diagonal elements of the covariance matrix contribute minimally to the parameter constraints --- the diagonal approximation used in the Fisher analysis captures the bulk of the information. 

Lastly, we also run an independent forecast which mimics the same configuration as in Section~5.2 of Ref.~\cite{SimonsObservatory:2018koc}, which considers the combination of primary CMB TT, EE, TE, and CMB lensing spectra, and DESI BAO~\cite{DESI:2013agm}. We find good agreement between our results and those in Ref.~\cite{SimonsObservatory:2018koc} for the constraint on $M_\nu$ for two different assumed priors on $\tau$, thereby further indicating the validity of our pipeline.

\subsection{Baseline Analysis}
In our baseline analysis, we focus on the two-parameter space spanned by the matter density parameter $\Omega_m$ and the primordial amplitude $A_s$, fixing the total neutrino mass to its fiducial value of $M_\nu = 0.1$ eV. We do not impose external priors on either parameter, as the MCMC posterior remains well contained within the parameter ranges covered by the simulation suite.

\begin{figure}
    \centering
    \includegraphics[width=0.8\linewidth]{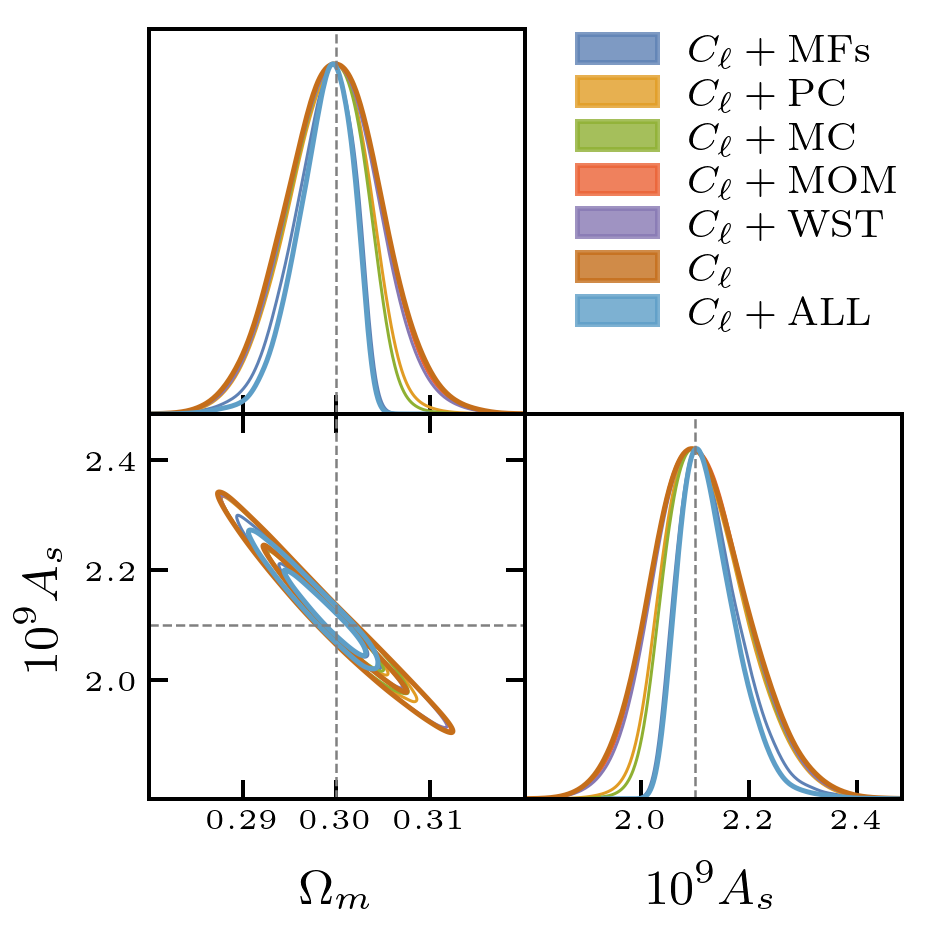}
    \caption{Marginalized 68\% and 95\% confidence contours for $\Omega_m$ and $A_s$, while holding $M_\nu = 0.1$~eV fixed. The baseline constraints from the angular power spectrum $C_\ell$ are shown together with results obtained by combining $C_\ell$ with each higher-order statistic. The inclusion of Minkowski functionals (MFs) yields the largest improvement, followed by peak counts (PC) and minima counts (MC), while moments (MOM) and the wavelet scattering transform (WST) provide negligible additional constraining power. The dashed lines mark the fiducial cosmology used to generate the simulations.}
    \label{fig:triangle_2param_clall}
\end{figure}

Figure~\ref{fig:triangle_2param_clall} summarizes our main findings and highlights how incorporating higher-order statistics sharpens cosmological constraints beyond what is achievable with the angular power spectrum alone. Each contour shows the joint posterior distribution obtained from combining $C_\ell$ with one of the higher-order statistics. The corresponding quantitative results --- including posterior means, marginal $1\sigma$ uncertainties, and the figure of merit (FoM) --- are provided in Table~\ref{tab:2param_clall}. We compute the FoM as $\mathrm{FoM} = 1/\sqrt{\det \mathrm{Cov}}$, where $\mathrm{Cov}$ is the $2\times2$ covariance matrix of $\{\Omega_m, A_s\}$.

Our results are broadly consistent with, and in fact slightly improve upon, those of Ref.~\cite{Liu:2016nfs}, with the improvements arising because we include polarization data (in addition to temperature) when reconstructing the lensing convergence field. Using the power spectrum alone, we obtain $\sigma(\Omega_m)=0.0051$, compared to their reported value of 0.0065. When all non-Gaussian statistics are combined, we achieve a constraint of $\sigma(\Omega_m)=0.0030$, surpassing their result of 0.0045 that used only the one-point PDF and PC. 

Overall, we find that incorporating non-Gaussian statistics yields $\sim$15–35\% improvements in the marginalized uncertainties on $\Omega_m$ and $A_s$, with the notable exceptions of MOM and WST. These two statistics perform poorly in the context of CMB lensing because the reconstruction noise is substantially higher than in galaxy weak-lensing or redshift-space analyses, and their reliance on real-space averaged quantities makes them especially susceptible to this noise. In the case of WST, removing highly correlated coefficients ($r > 0.99$) leaves only 22 usable coefficients for $J=8$, reducing the amount of independent information available. Together, the strong reconstruction noise and the resulting reduction in effective dimensionality explain the limited constraining power of MOM and WST in our analysis.

\begin{table}[!t]
    \centering
    \begin{tabular}{lccc}
    \hline\hline
    \textbf{Statistics} & $\sigma(\Omega_m)$ & $\sigma(A_s)$ & FoM \\
    \hline
    $C_\ell$ & $0.00512$ (--) & $0.0895$ (--) & 12370 (--) \\
    $C_\ell+\mathrm{MFs}$ & $0.00337$ (34.2\%) & $0.0613$ (31.5\%) & 20130 (62.7\%) \\
    $C_\ell+\mathrm{PC}$ & $0.00435$ (15.1\%) & $0.0775$ (13.4\%) & 15030 (21.5\%) \\
    $C_\ell+\mathrm{MC}$ & $0.00423$ (17.5\%) & $0.0761$ (15.0\%) & 15480 (25.1\%) \\
    $C_\ell+\mathrm{MOM}$ & $0.00500$ (2.4\%) & $0.0874$ (2.3\%) & 12750 (3.1\%) \\
    $C_\ell+\mathrm{WST}$ & $0.00492$ (3.9\%) & $0.0863$ (3.5\%) & 13160 (6.4\%) \\
    $C_\ell+\mathrm{ALL}$ & $0.00307$ (40.2\%) & $0.0555$ (38.0\%) & 23260 (88.0\%) \\
    \hline\hline
    \end{tabular}
    \caption{Marginalized $1\sigma$ constraints on $\Omega_m$ and $A_s$ from different statistics. Parentheses indicate the fractional reduction in uncertainty or the fractional improvement in the Figure of Merit (FoM) relative to the $C_\ell$-only case.}
    \label{tab:2param_clall}
\end{table}

Among the higher-order statistics examined, the MFs provide the strongest improvement in cosmological parameter constraints. This is, perhaps, unsurprising: MFs encode the full morphological and topological information of the field, capturing nonlinear structures such as filaments, clusters, and voids that are not represented in the two-point function. They are particularly sensitive to $\Omega_m$, since changes in the matter density alter the connectivity and geometry of large-scale structure, whereas variations in $A_s$ primarily rescale the field amplitude. Using MFs alone, we obtain 34.2\% and 31.5\% reductions in the marginalized uncertainties on $\Omega_m$ and $A_s$, respectively, along with a 62.7\% enhancement in the FoM. Similar conclusions were also reached by Ref.~\cite{Sabyr:2024kcu} when applying MFs to the Compton-$y$ field.

Peak counts and minima counts (PC/MC) also enhance constraining power by tracing the abundance of rare high- and low-density features generated by nonlinear gravitational evolution. Although they do not capture global morphology as comprehensively as the MFs, they remain complementary to the power spectrum. Each yields $\sim$15\% improvements in $\Omega_m$ and $A_s$, with corresponding FoM increases of 21\% (PC) and 25\% (MC), indicating that MC provides slightly stronger constraints. Combining both PC and MC further boosts parameter sensitivity.

By contrast, we find that MOM and WST contribute little additional information for CMB lensing. For MOM, this is likely due to its small data dimensionality, which causes its components to be highly correlated with one another. For WST, CMB lensing maps are noise-dominated at most of the scales, causing the WST coefficients to become highly correlated with one another and thus limiting its additional constraining power. 

When all non-Gaussian statistics are combined with the power spectrum, the gains are substantial. We obtain an 88\% improvement in the FoM relative to using $C_\ell$ alone, along with 40.2\% and 38.0\% reductions in the marginalized uncertainties of $\Omega_m$ and $A_s$, respectively. This improvement significantly exceeds that achieved by any individual non-Gaussian statistic. For example, the FoM gain from combining all statistics is more than twice that obtained from adding MFs (the best single type of statistic) alone, demonstrating that different non-Gaussian probes capture complementary information. In particular, MFs, PC/MC, WST, and MOM are sensitive to distinct aspects of the lensing field, and their combination leads to a marked enhancement in constraining power. Overall, these results highlight the rich cosmological information encoded in the non-Gaussian structure of the CMB lensing field: in terms of the FoM, the additional information provided by non-Gaussian statistics is comparable to roughly half of that captured by the power spectrum itself. We note that this comparison should be interpreted with care: our $C_\ell$ baseline only uses multipoles $\ell > 300$, and the lower multipoles in the power spectrum contain substantial additional constraining power. The relative contribution of non-Gaussian statistics would therefore likely appear more modest when compared against a full-range power spectrum analysis. However, we will see that the improvement is still not negligible when we consider large-scale multipoles $\ell<300$ in Section~\ref{subsec:rangeEllCl}. 

\subsection{Including the Neutrino Mass Sum $M_\nu$}
In this part, we study the information when we vary the sum of the neutrino massess, $M_\nu$, as an additional parameter. 

\begin{figure}
    \centering
    \includegraphics[width=0.95\linewidth]{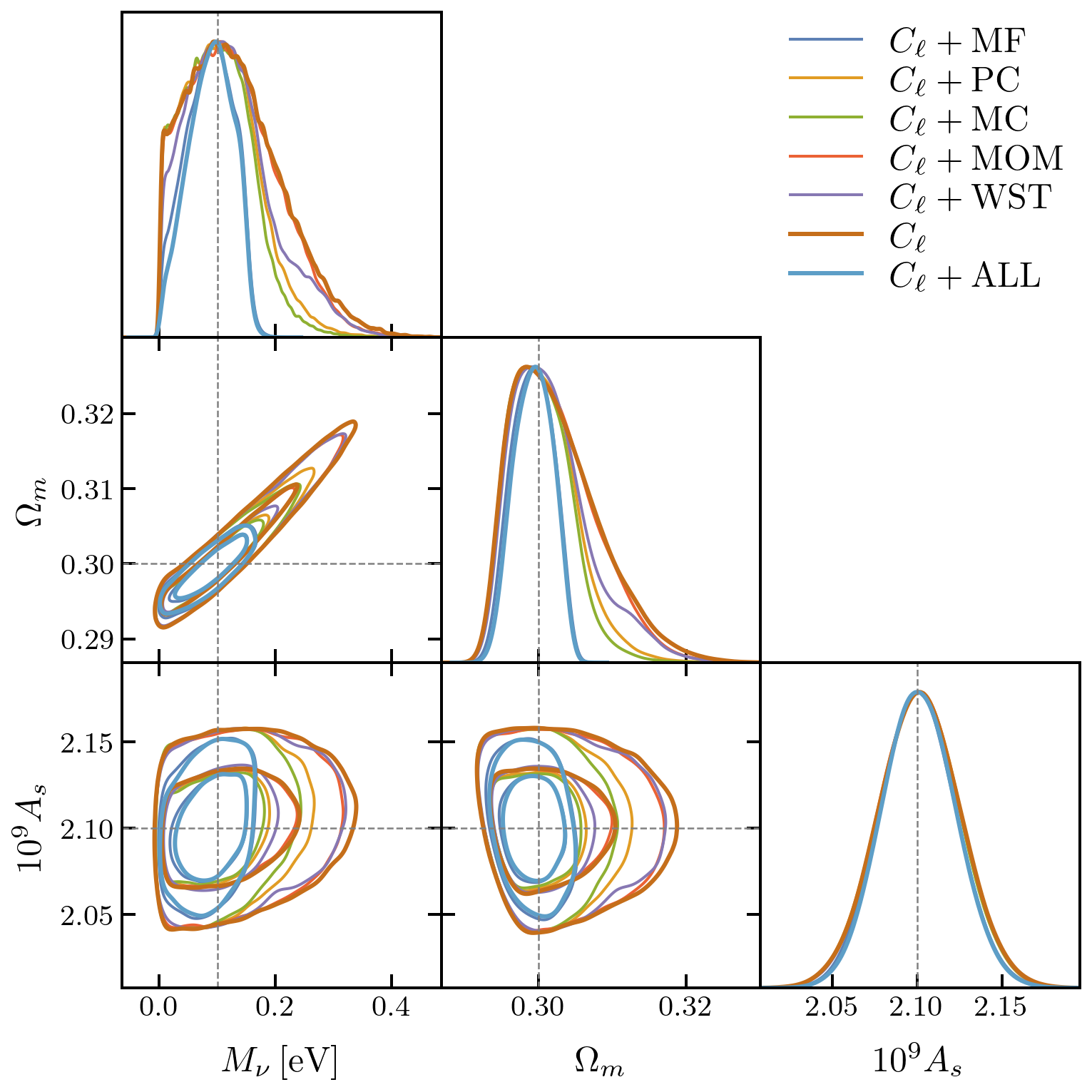}
    \caption{Marginalized 68\% and 95\% confidence contours for $M_\nu$, $\Omega_m$, and $A_s$. The baseline constraints from the angular power spectrum $C_\ell$ are shown together with results obtained by combining $C_\ell$ with each higher-order statistic. Similar to Figure~\ref{fig:triangle_2param_clall}, the inclusion of Minkowski functionals (MFs) yields the largest improvement, followed by peak counts (PC) and minima counts (MC), while moments (MOM) and the wavelet scattering transform (WST) provide negligible additional constraining power. Note that we impose external priors on both $A_s$ and $\Omega_m$ motivated by Planck + ACT DR6 CMB and DESI DR2 BAO data. The dashed lines mark the fiducial cosmology used to generate the simulations.}
    \label{fig:triangle_3param_clall}
\end{figure}

Figure~\ref{fig:triangle_3param_clall} shows the 68\% and 95\% confidence intervals for $M_\nu$, $\Omega_m$, and $A_s$, as inferred from $C_\ell^{\kappa\kappa}$ alone, $C_\ell^{\kappa\kappa}$ combined with each non-Gaussian statistic individually, and the combination of all statistics.  Note that in this analysis, we impose external priors on both $\Omega_m$ and $10^9 A_s$: $\Omega_m \sim \mathcal{N}(0.3, 0.0086^2)$ and $10^9 A_s \sim \mathcal{N}(2.1, 0.025^2)$, respectively. We present the corresponding quantitative constraints in Table~\ref{tab:3param_clall}, which includes the mean $\pm\,1\sigma$ for $\Omega_m$ and $A_s$, the 68\% and 95\% confidence intervals for $M_\nu$, and the FoM for the $(\Omega_m, A_s)$ plane. We also compute the one-sided standard deviation for $M_\nu$ as there is a hard cutoff at 0 eV. As a result, we want to focus on the tail at the high end. The one-sided standard deviation is computed as follows:
\begin{align}
    \sigma^{\text{one-sided}}_{M_\nu} = \sqrt{E[(X-0.1\text{ eV})^2 | X > \text{0.1 eV}]}\,,
\end{align}
where $X$ is the MCMC-sampled $M_\nu$ and 0.1 eV is the fiducial value of $M_\nu$ for our analysis. 

Similar to what we observed in the two-parameter case varying only $\Omega_m$ and $A_s$, the dominant source of additional information when combining all non-Gaussian statistics continues to come from MFs. Focusing on $M_\nu$, we first examine the 68\% upper bound. Relative to the power spectrum alone, which yields an upper limit of $M_\nu^{68}=0.161$ eV, adding MFs tightens this to $0.108$ eV, an improvement of roughly 32\%. None of the other non-Gaussian statistics considered here (PC, MC, MOM, WST) produce a comparable decrease in the upper bound; their constraints remain closer to the power-spectrum-only result, with improvements at the $\sim 10$–20\% level at best. This highlights that MFs are uniquely sensitive to the types of non-Gaussian signatures that correlate with the impact of neutrino mass on the lensing field.

A similar pattern appears when looking at the one-sided standard deviation, which is a more appropriate metric for $M_\nu$ given the hard prior at zero. The power spectrum alone gives $\sigma^{\text{one-sided}}_{M_\nu} = 0.101$~eV, while adding MFs reduces this to 0.032 eV, more than a factor of three improvement. Again, none of the other statistics achieve comparable gains: including PC yields $\sigma^{\text{one-sided}}_{M_\nu} = 0.072$~eV, including MC yields 0.063~eV, and including MOM or WST leave it essentially unchanged. When all non-Gaussian statistics are combined, $\sigma^{\text{one-sided}}_{M_\nu}$ further decreases to 0.030~eV, but this marginal improvement beyond MFs alone indicates that most of the constraining power is already captured by the MFs.

Finally, the FoM for the $(\Omega_m, A_s)$ plane shows that MFs remain the most informative of the non-Gaussian statistics in this parameter subspace. It is important to note that both parameters already have external priors applied, and in the case of $A_s$ the posterior is almost entirely dominated by the prior. Even so, we still observe meaningful improvements when incorporating non-Gaussian statistics, particularly for $\Omega_m$, which has a relatively tight prior of $\sigma_{\rm prior}=0.0086$ but nonetheless benefits from the additional information contained in the MFs. This improvement is evident in the FoM: adding MFs increases the FoM by a factor of 2.4 relative to the power spectrum alone, while the other non-Gaussian statistics yield smaller individual gains. Importantly, combining all non-Gaussian statistics leads to a further enhancement of the FoM beyond that achieved by MFs alone, indicating that additional complementary information is captured by the full set of non-Gaussian probes. This contrasts with the fixed-$M_\nu$ case, where MFs account for a larger fraction (around 75\%) of the total beyond-Gaussian information, and highlights the role of parameter degeneracies in determining the relative impact of different non-Gaussian statistics.

\begin{table*}
\centering
\begin{tabular}{lccccc}
\hline\hline
\textbf{Statistics} & $\sigma(\Omega_m)$ & $\sigma(A_s)$ & $M_\nu$ CDF$_{68/95}$ [eV] & $\sigma(M_\nu)$ [eV] & FoM \\
\hline
$C_\ell$ & $0.0060$ (--) & $0.0242$ (--) & $(0.161,\;0.276)$ & $0.101$ (--) & 6889 (--) \\
$C_\ell+\mathrm{MFs}$ & $0.0027$ (54.7\%) & $0.0225$ (7.1\%) & $(0.108,\;0.147)$ & $0.032$ (68.8\%) & 16660 (141.8\%) \\
$C_\ell+\mathrm{PC}$ & $0.0046$ (23.8\%) & $0.0240$ (0.9\%) & $(0.134,\;0.217)$ & $0.072$ (28.9\%) & 9146 (32.8\%) \\
$C_\ell+\mathrm{MC}$ & $0.0042$ (30.6\%) & $0.0239$ (1.5\%) & $(0.127,\;0.198)$ & $0.063$ (37.7\%) & 10130 (47.0\%) \\
$C_\ell+\mathrm{MOM}$ & $0.0057$ (5.4\%) & $0.0241$ (0.5\%) & $(0.158,\;0.262)$ & $0.095$ (6.6\%) & 7334 (6.5\%) \\
$C_\ell+\mathrm{WST}$ & $0.0055$ (9.1\%) & $0.0239$ (1.2\%) & $(0.147,\;0.263)$ & $0.091$ (10.5\%) & 7672 (11.4\%) \\
$C_\ell+\mathrm{ALL}$ & $0.0026$ (57.4\%) & $0.0219$ (9.5\%) & $(0.109,\;0.146)$ & $0.030$ (70.1\%) & 18150 (163.5\%) \\
\hline\hline
\end{tabular}
\caption{Forecast marginalized $1\sigma$ uncertainties on $\Omega_m$, $A_s$, and the neutrino mass $M_\nu$, with fractional improvements relative to the $C_\ell$-only baseline shown in parentheses. CDF$_{68/95}$ intervals are also provided for $M_\nu$. The final column lists the Figure of Merit (FoM) for the $(\Omega_m, A_s)$ constraints. We impose Gaussian priors $\Omega_m \sim \mathcal{N}(0.3,0.0086^2)$ and $10^9A_s \sim \mathcal{N}(2.1,0.025^2)$.}
\label{tab:3param_clall}
\end{table*}

\subsection{Impact of Each Minkowski Functional on Parameter Constraints}
\begin{figure}
    \centering
    \includegraphics[width=0.95\linewidth]{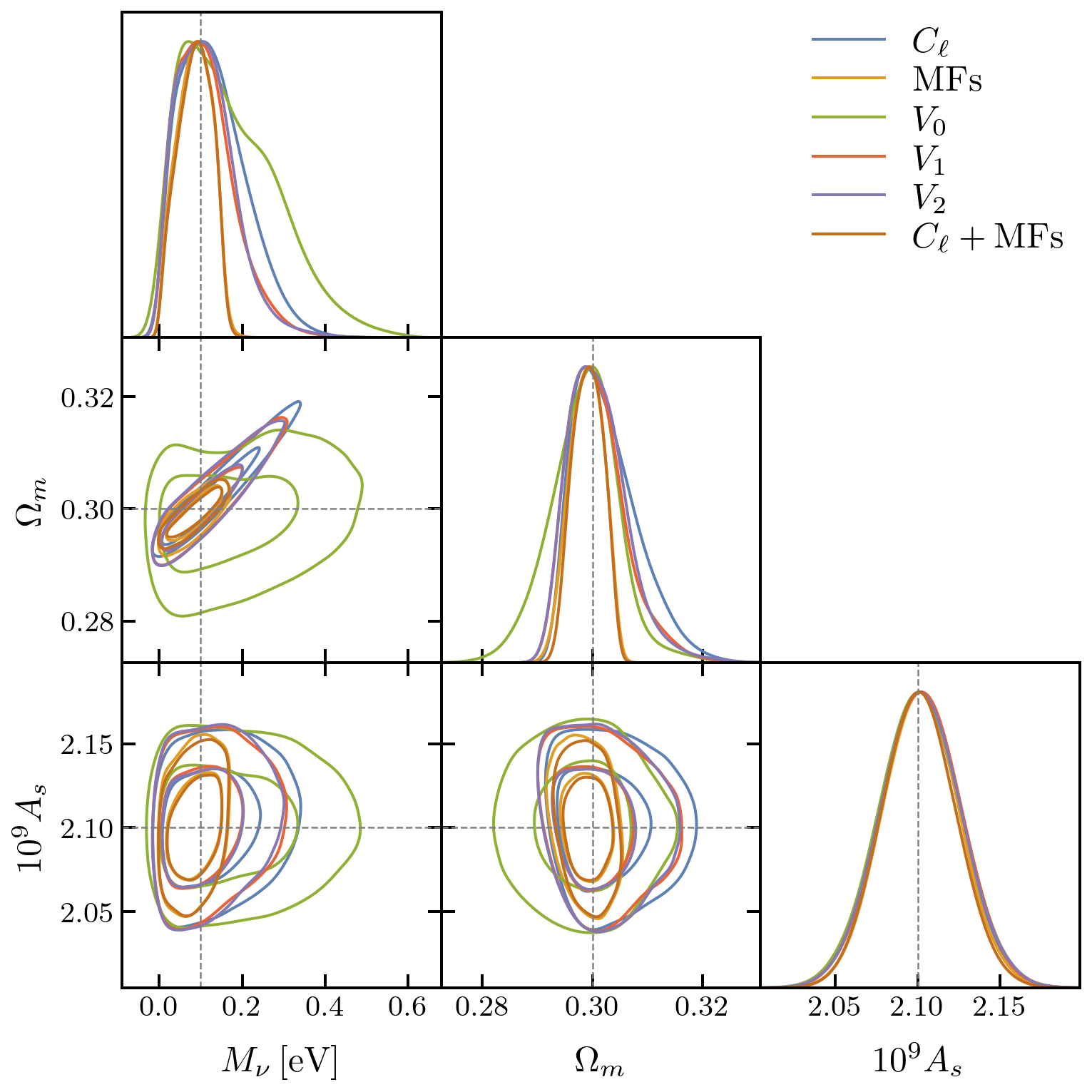}
    \caption{Triangle plot showing the 68\% and 95\% constraints on $M_\nu$, $\Omega_m$, and $A_s$ from the power spectrum ($C_\ell$), the three individual Minkowski functionals ($V_0$, $V_1$, $V_2$), the full MF set, and the combined $C_\ell + \mathrm{MFs}$ analysis. The distinct degeneracy directions of the Minkowski functionals provide complementary information that, when combined with the power spectrum, effectively breaks parameter degeneracies and tightens constraints. The dashed lines mark the fiducial cosmology used to generate the simulations.}
    \label{fig:triangle_3params_clall_mfs}
\end{figure}

As demonstrated earlier, the MFs yield the most stringent constraints among the non-Gaussian statistics considered. To elucidate the origin of this improvement, we examine how the MFs break degeneracies inherent in the power spectrum. Figure~\ref{fig:triangle_3params_clall_mfs} presents the constraints from the individual functionals $V_0$, $V_1$, and $V_2$. Although $V_0$ produces contours that are broader than those from the power spectrum alone, their degeneracy directions differ substantially. On the other hand, although $V_1$ and $V_2$ share similar degeneracy directions with the one from the power spectrum, they can provide tighter constraints. These complementary orientations enable the MFs to break parameter degeneracies when combined with $C_\ell$. This behavior is particularly evident in the $\Omega_m$--$M_\nu$ plane, where the three MFs exhibit similar degeneracy directions that contrast with the strongly positively correlated contour from the power spectrum. Consequently, the joint analysis achieves markedly tighter parameter constraints.

\subsection{Impact of Peak/Minima Counts on Parameter Constraints}
\begin{figure}
    \centering
    \includegraphics[width=0.95\linewidth]{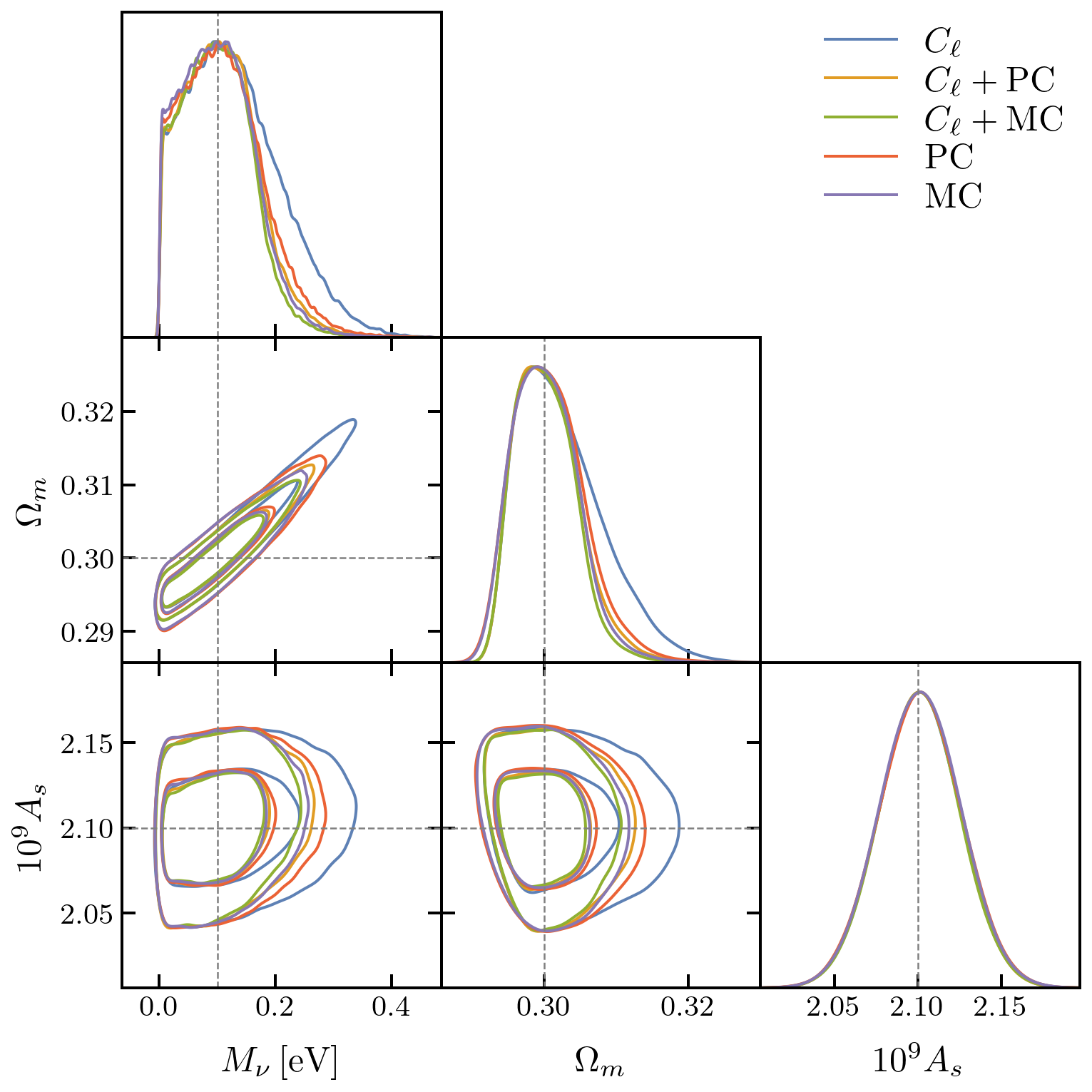}
    \caption{Triangle plot showing the 68\% and 95\% constraints on $M_\nu$, $\Omega_m$, and $A_s$ from the power spectrum ($C_\ell$), the peak and minima counts (PC/MC), and the combined $C_\ell + \mathrm{PC/MC}$ analysis. In contrast to Figure~\ref{fig:triangle_3params_clall_mfs}, the PC/MC contours exhibit degeneracy directions closely aligned with those of $C_\ell$, resulting in more modest improvements when they are combined with the power spectrum. The dashed lines mark the fiducial cosmology used to generate the simulations.}
    \label{fig:triangle_3params_clall_pcmc}
\end{figure}

PC/MC also provide additional information for constraining the parameters, though the improvement is slightly weaker than that achieved by the MFs. As illustrated in Figure~\ref{fig:triangle_3params_clall_pcmc}, this difference arises because, unlike the MFs whose contours differ significantly from those of $C_\ell$, the PC/MC retain a degeneracy direction closely aligned with $C_\ell$. Consequently, they are less effective at breaking parameter degeneracies.

\subsection{Effects of external priors}
In this part, we study how the prior information on $A_s$ or $\Omega_m$ can be used to help the constraint on $M_\nu$. 

\begin{figure}
    \centering
    \includegraphics[width=0.95\linewidth]{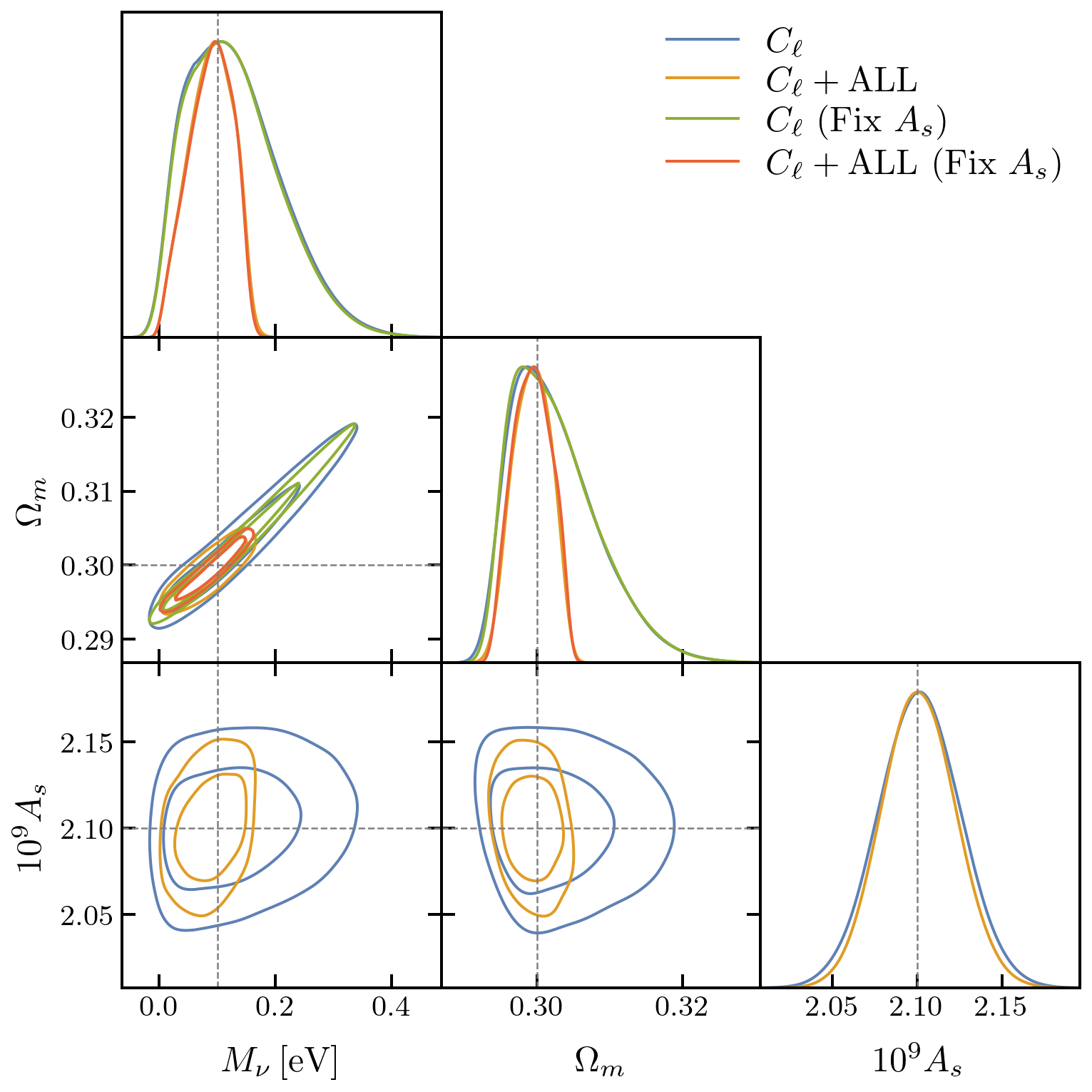}
    \caption{Comparison of the 68\% and 95\% parameter constraints under two different treatments of the amplitude parameter $A_s$. Blue and yellow contours show results obtained using a Planck + ACT DR6–like prior on $A_s$, while green and red contours correspond to fixing $A_s$ to its fiducial value. In both cases, we present constraints from the power spectrum alone ($C_\ell$) as well as from the combination of all non-Gaussian statistics with the power spectrum ($C_\ell + \mathrm{ALL}$). Although fixing $A_s$ substantially reduces the allowed parameter volume in the joint plane, it does not improve the corresponding one-dimensional marginalized constraints. The dashed lines mark the fiducial cosmology used to generate the simulations.}
    \label{fig:triangle_3params_clall_priorAs}
\end{figure}

Figure~\ref{fig:triangle_3params_clall_priorAs} shows the case when we try to narrow the prior on $A_s$ further beyond the information provided by the current combination of Planck and ACT DR6 data~\cite{AtacamaCosmologyTelescope:2025blo}. We consider an extreme case here in which we impose a Dirac delta-function at the fiducial value as a prior for $A_s$. We find that even though we have perfect knowledge of $A_s$ here, the marginalized 1D posterior distributions remain unchanged for both $M_\nu$ and $\Omega_m$. The constraints are mostly improved in the direction orthogonal to the degeneracy between $M_\nu$ and $\Omega_m$. Quantitatively, fixing $A_s$ reduces the 2D $M_\nu$--$\Omega_m$ contour area by $\sim 50\%$ compared to leaving it free when we combine all non-Gaussian statistics, indicating that while perfect knowledge of $A_s$ does tighten the joint constraint, it does not substantially lift the 1D marginal uncertainties on $M_\nu$ or $\Omega_m$.  To check this result, we implement the same forecast using only $C_\ell$ computed analytically with CAMB and we find the same result: a tighter prior on $A_s$ alone does not lead to improved 1D marginalized constraints on both $\Omega_m$ and $M_\nu$.

\begin{figure}
    \centering
    \includegraphics[width=0.95\linewidth]{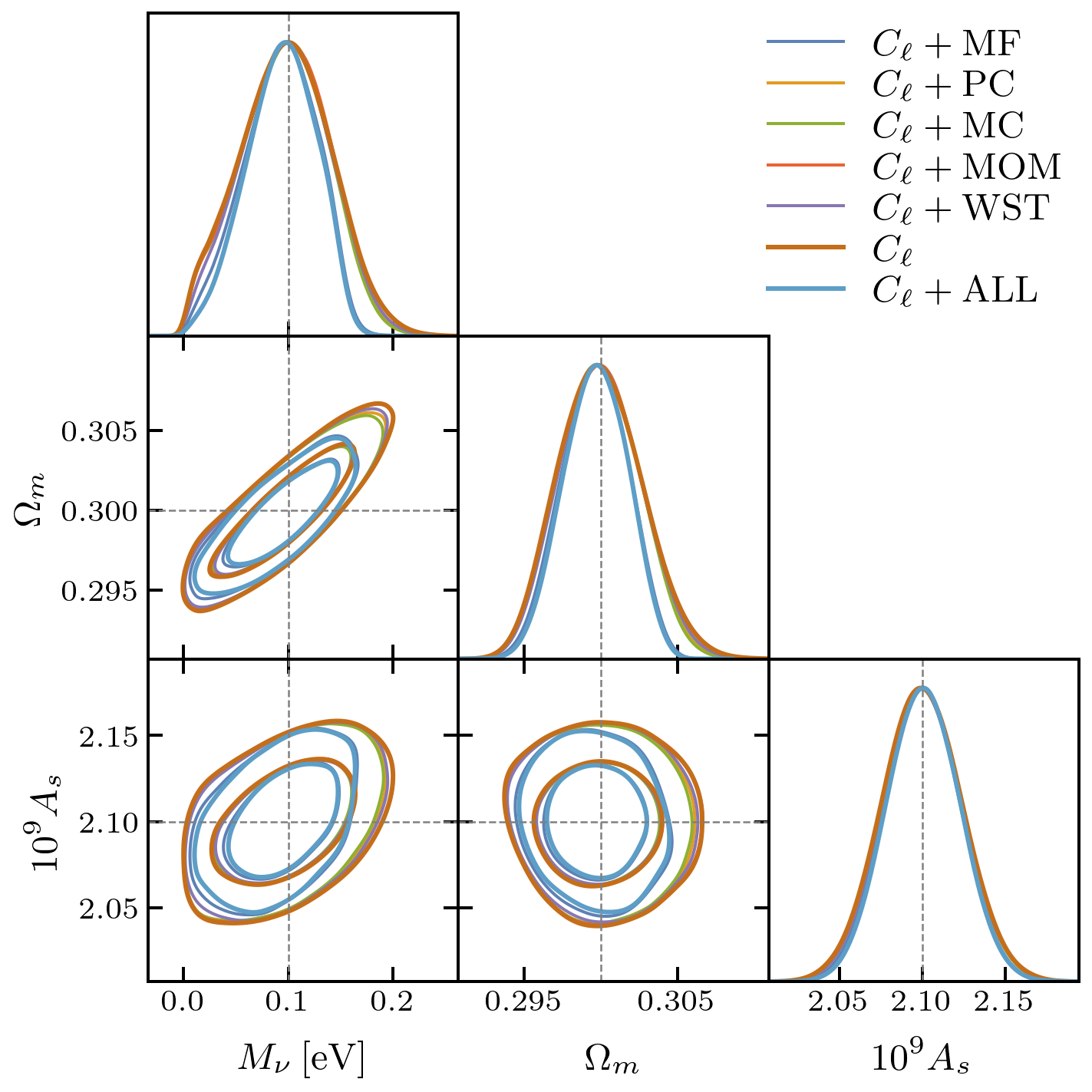}
    \caption{Impact of adding individual non-Gaussian statistics on the 68\% and 95\% cosmological constraints when a tighter prior on $\Omega_m$ is imposed. Blue contours show the baseline constraints from the power spectrum alone ($C_\ell$), while the remaining colors illustrate the effect of augmenting $C_\ell$ with Minkowski functionals (MFs), peak counts (PC), minima counts (MC), moments (MOM), wavelet-scattering coefficients (WST), and the full combination of all statistics ($\mathrm{ALL}$). The tighter $\Omega_m$ prior substantially reduces the allowed parameter volume, making differences among the individual non-Gaussian statistics more subtle. The dashed lines mark the fiducial cosmology used to generate the simulations.}
    \label{fig:triangle_3params_clall_priorOm0}
\end{figure}

This suggests that we should instead focus on the prior on $\Omega_m$. In this case, we run a Fisher forecast with CAMB using power spectra for the primary CMB TT, EE, and TE, CMB lensing, and DESI DR2 BAO, following the configuration in Section 5.2 of Ref.~\cite{SimonsObservatory:2018koc}.\footnote{Note that when adding this prior, we assume no cross-covariance between our statistics and all of these power spectra, including the CMB lensing power spectrum.}  This yields a constraint on $\Omega_m$ of $\sigma(\Omega_m)\approx0.029$, a factor of three improvement over our previous choice, and we adopt this value as the prior in the following analysis. Figure~\ref{fig:triangle_3params_clall_priorOm0} shows the resulting constraints when this prior is imposed, with the corresponding values summarized in Table~\ref{tab:triangle_3params_clall_priorOm0}. We find that tightening the $\Omega_m$ prior substantially reduces the degeneracy between $\Omega_m$ and $M_\nu$, leading to a noticeably improved constraint on the neutrino mass. In particular, the one-sided uncertainty on $M_\nu$ improves by 33.0\% for the full combination $C_\ell+\mathrm{ALL}$, while the individual non-Gaussian statistics yield improvements ranging from 8.4\% (MC) to 30.8\% (MF), with PC providing an intermediate improvement of 5.8\%. As expected, MOM and WST contribute little additional constraining power for $M_\nu$ in this setup.

The prior also enhances the joint $\{\Omega_m, A_s\}$ constraints, as reflected in the increased FoM values. Although the ranking of the statistics remains unchanged, with MF and the combined $C_\ell+\mathrm{ALL}$ statistics producing the largest gains, the prior accentuates the differences, further highlighting the complementarity of non-Gaussian information.

Overall, these results show that part of the remaining uncertainty in $M_\nu$ arises from correlations with $\Omega_m$, and that even a modest external prior can significantly enhance the constraining power of non-Gaussian statistics.

\begin{table*}
\centering
\begin{tabular}{lccccc}
\hline\hline
\textbf{Statistics} & $\sigma(\Omega_m)$ & $\sigma(A_s)$ & $M_\nu$ CDF$_{68/95}$ [eV] & $\sigma(M_\nu)$ & FoM \\
\hline
$C_\ell$ & $0.00270$ (--) & $0.02429$ (--) & $(0.1182,\;0.1680)$ & $0.0421$ (--) & 15250 (--) \\
$C_\ell+\mathrm{MFs}$ & $0.00214$ (20.8\%) & $0.02263$ (6.8\%) & $(0.1104,\;0.1456)$ & $0.0291$ (30.8\%) & 20750 (36.1\%) \\
$C_\ell+\mathrm{PC}$ & $0.00259$ (4.0\%) & $0.02401$ (1.2\%) & $(0.1169,\;0.1637)$ & $0.0397$ (5.8\%) & 16080 (5.4\%) \\
$C_\ell+\mathrm{MC}$ & $0.00255$ (5.6\%) & $0.02400$ (1.2\%) & $(0.1163,\;0.1616)$ & $0.0386$ (8.4\%) & 16340 (7.1\%) \\
$C_\ell+\mathrm{MOM}$ & $0.00268$ (0.7\%) & $0.02422$ (0.3\%) & $(0.1187,\;0.1683)$ & $0.0421$ (0.1\%) & 15390 (0.9\%) \\
$C_\ell+\mathrm{WST}$ & $0.00260$ (3.8\%) & $0.02381$ (2.0\%) & $(0.1178,\;0.1650)$ & $0.0403$ (4.4\%) & 16170 (6.0\%) \\
$C_\ell+\mathrm{ALL}$ & $0.00206$ (23.8\%) & $0.02210$ (9.0\%) & $(0.1105,\;0.1446)$ & $0.0282$ (33.0\%) & 22140 (45.2\%) \\
\hline\hline
\end{tabular}
\caption{Forecast marginalized 1$\sigma$ uncertainties on $\Omega_m$, $A_s$, and the neutrino mass $M_\nu$, with fractional improvements relative to the $C_\ell$ baseline shown in parentheses. CDF$_{68/95}$ intervals are also listed for $M_\nu$. The final column reports the Figure of Merit (FoM) for the $(\Omega_m,A_s)$ constraints. In this forecast, we add priors to both $\Omega_m$ and $A_s$, as discussed in the text.}
\label{tab:triangle_3params_clall_priorOm0}
\end{table*}

\subsection{Impacts from the multipole range of $C_{\ell}^{\kappa\kappa}$}\label{subsec:rangeEllCl}

Since our choice of $\ell_{\rm min}=300$ is set by the sky area of our N-body simulations, we may lose information in the high-S/N regime on larger angular scales (see Figure~\ref{fig:kappa_cl}), where the lensing power spectrum is well-measured and (likely) contains most of the cosmological constraining power.  Properly including $C_\ell$ on large scales could potentially reduce the relative gains that we have found from including non-Gaussian statistics. To assess this, we have run an additional forecast including $C_{\ell<300}^{\kappa\kappa}$ to verify that the improvement remains non-negligible.

As shown in Section~\ref{subsec:validation}, we find that the information content of $C_\ell^{\kappa\kappa}$ is consistent between analytic computations and our N-body simulations when considering the same range of angular scales, and the off-diagonal terms do not contribute significantly to the final constraints. We therefore replace the simulation-based $C_\ell^{\kappa\kappa}$ with one computed using {\tt CAMB}, allowing us to extend the analysis to lower multipoles. For the non-Gaussian statistics, we retain information only from $\ell>300$, which is not a significant limitation since CMB lensing is predominantly linear on large scales. We assume a diagonal covariance matrix and neglect correlations between the power spectrum and non-Gaussian statistics. For the primary CMB and BAO, we follow Ref.~\cite{SimonsObservatory:2018koc}, using the fractional distance errors from Table~3 of Ref.~\cite{DESI:2013agm} for DESI. We vary the six standard $\Lambda$CDM parameters jointly with $M_\nu$, assume that the non-Gaussian statistics carry information only on $\Omega_m$, $A_s$, and $M_\nu$, and impose a prior $\tau \sim \mathcal{N}(0.054, 0.01^2)$.

\begin{figure}
    \centering
    \includegraphics[width=\linewidth]{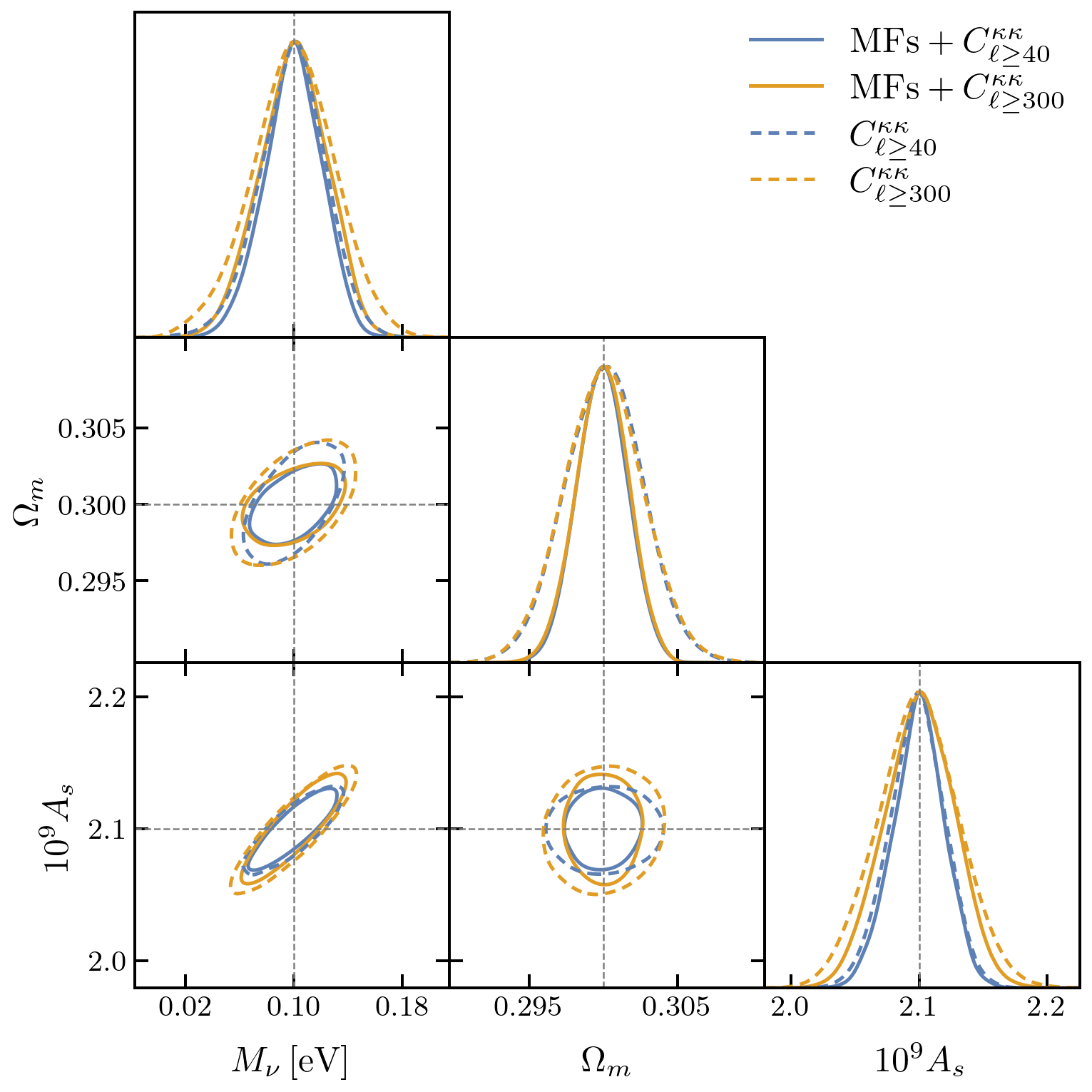}
    \caption{Forecasted 68\% marginalized constraints on $\Omega_m$, $10^9A_s$, and $M_\nu$ from CMB lensing power spectra $C_\ell^{\kappa\kappa}$ combined with Minkowski functionals (MFs). Blue and yellow contours correspond to $\ell_{\rm min} = 40$ and $\ell_{\rm min} = 300$ for $C_\ell^{\kappa\kappa}$, respectively, illustrating the information content of large-scale lensing modes. Solid and dashed contours include and exclude MFs, respectively. All cases also include primary CMB power spectra and DESI-like BAO measurements, jointly vary the six standard $\Lambda$CDM parameters together with $M_\nu$, and adopt a Gaussian prior $\sigma(\tau) = 0.01$ on the optical depth to reionization.}
    \label{fig:triangle_3param_rangeOfEll}
\end{figure}

Figure~\ref{fig:triangle_3param_rangeOfEll} and Table~\ref{tab:triangle_3param_rangeOfEll} show results for four cases, combining two choices of $\ell_{\rm min}$ for $C_\ell^{\kappa\kappa}$ with and without MFs, which are the dominant source of non-Gaussian information. Without MFs, extending the power spectrum to large scales improves $\sigma(M_\nu)$ and $\sigma(A_s)$ by 23\% and 31\% respectively, due to better sensitivity to the matter power spectrum suppression across multiple redshifts, while $\sigma(\Omega_m)$ remains largely unchanged. With MFs included, comparing the $\ell_{\rm min}=40$ cases shows a further 35\% improvement in $\sigma(\Omega_m)$, again reflecting the sensitivity of MFs to the matter density, along with additional gains of 11\% and 7\% for $\sigma(M_\nu)$ and $\sigma(A_s)$. Overall, we find that non-Gaussian statistics in CMB lensing remain an important and non-negligible source of complementary information even when large-scale power spectrum modes are included.

\begin{table*}
\centering
\renewcommand{\arraystretch}{1.2}
\begin{tabular}{lccccc}
\hline\hline
\textbf{Statistics} & $\sigma(M_\nu)$ & $\sigma(\Omega_m)$ & $\sigma(10^9A_s)$ \\
\hline
$C_{\ell\geq300}^{\kappa\kappa}$ + CMB T/E + BAO & $0.0300$ (--) & $0.00270$ (--) & $0.0319$ (--) \\

$C_{\ell\geq40}^{\kappa\kappa}$ + CMB T/E + BAO & $0.0231$ (23.0\%) & $0.00265$ (1.9\%) & $0.0220$ (31.0\%) \\

${\rm MFs}+C_{\ell\geq300}^{\kappa\kappa}$ + CMB T/E + BAO & $0.0239$ (20.3\%) & $0.00175$ (35.2\%) & $0.0274$ (14.1\%) \\

${\rm MFs}+C_{\ell\geq40}^{\kappa\kappa}$ + CMB T/E + BAO & $0.0205$ (31.7\%) & $0.00171$ (36.7\%) & $0.0204$ (36.1\%) \\
\hline\hline
\end{tabular}
\caption{Marginalized $1\sigma$ constraints on $\Omega_m$, $A_s$, and $M_\nu$ for the extended multipole range analysis, with fractional improvements relative to the baseline case shown in parentheses. The power spectrum $C_\ell^{\kappa\kappa}$ is computed analytically using {\tt CAMB}, allowing us to assess the impact of including large-scale modes with $\ell < 300$.}
\label{tab:triangle_3param_rangeOfEll}
\end{table*}

\section{Conclusions}\label{sec:conclusion}
In this work, we have carried out the first comprehensive, fully field-level assessment of the cosmological information encoded in higher-order statistics of the CMB lensing convergence field, focusing on an experiment with noise levels similar to those of the nominal SO survey. By combining ray-traced {\tt MassiveNuS} simulations with realistic SO-like CMB lensing reconstruction, we constructed a forward-modeling pipeline that naturally incorporates nonlinear structure formation, post-Born corrections, and the full hierarchy of reconstruction-noise biases. This framework enables us to evaluate non-Gaussian summary statistics directly on reconstructed maps, ensuring that the resulting forecasts faithfully reflect the information content accessible in real CMB data in the future.

We examined a broad suite of higher-order statistics, including Minkowski functionals (MFs), peak and minima counts (PC/MC), low-order moments (MOM), and wavelet-scattering coefficients (WST).  We modeled their cosmological dependence on $\{\Omega_m, A_s, M_\nu\}$ using Gaussian-process emulators. Across all statistics, we find consistent qualitative behavior: variations in $\Omega_m$ induce the strongest signal, followed by $A_s$, while the signatures of massive neutrinos are intrinsically smaller and more difficult to isolate in the presence of SO-level reconstruction noise.

Our forecasts demonstrate that morphology-based statistics provide substantial gains beyond the lensing power spectrum alone. In particular, the MFs capture the richest non-Gaussian information, improving constraints on $\Omega_m$ by 55\% and increasing the FoM by 141\%. PC/MC also provide meaningful improvements at the $\sim$30-50\% level. By contrast, MOM and WST coefficients contribute little information at SO noise levels, largely because these statistics depend on highly correlated or real-space-averaged quantities that are especially sensitive to reconstruction noise and do not access phase information that encodes large-scale structure morphology.

When all non-Gaussian statistics are combined with the power spectrum, the gains are substantial. We obtain a 163\% enhancement in the FoM and 57\% reductions in the marginalized uncertainty on $\Omega_m$, corresponding to an information increase comparable to roughly half of what is captured by the power spectrum itself. Equivalently, the combined non-Gaussian statistics nearly double the FoM relative to the power spectrum-only case. These improvements persist when $M_\nu$ is included as a free parameter. Under reasonable external priors on $\Omega_m$ and $A_s$, the non-Gaussian information --- dominated by the MFs --- tightens the 68\% upper limit on $M_\nu$ by $\sim$25\% relative to the power spectrum alone and reduces the one-sided uncertainty by more than a factor of two. Adding additional non-Gaussian statistics beyond the MFs yields only incremental gains, underscoring the dominant role of morphology in CMB lensing non-Gaussianity.

Several promising avenues for future work emerge from our results. Including large-scale lensing modes, which are not accessible in our small-area simulations, should increase the utility of MOM and WST and may reveal additional non-Gaussian information carried by long-wavelength structure (e.g., squeezed bispectrum configurations involving one large-scale mode and two small-scale modes~\cite{Goldstein:2023brb}). Expanding the simulation suite, both by increasing the number of independent realizations and by sampling a larger and more finely gridded set of cosmologies, would further improve the emulator accuracy, mitigate boundary effects in parameter space, and enable reliable modeling of higher-order statistics such as the bispectrum. Cross-correlation with external tracers or combination with galaxy surveys may further enhance the sensitivity to neutrino mass and structure growth. We note that baryonic effects and certain observational systematics are not included in the present simulations and should be incorporated in future analyses to fully assess their impact on non-Gaussian statistics, as has been studied for the CMB lensing power spectrum and lensed CMB power spectra~\cite{McCarthy:2020dgq,McCarthy:2021lfp,Mirmelstein:2020pfk}. More broadly, integrating our field-level forward-modeling framework with likelihood-free or simulation-based inference methods may enable a fuller exploitation of non-Gaussian information in upcoming CMB surveys.

Overall, our study demonstrates that non-Gaussian statistics contain a wealth of untapped cosmological information that is well within reach for SO, SPT-3G, and other CMB experiments. Beyond tightening constraints on $\Omega_m$, $A_s$, and $M_\nu$, morphology-based probes such as MFs provide a powerful framework for characterizing nonlinear structure formation and information beyond two-point statistics. The methodology developed here is broadly applicable to other large-scale structure observables, including galaxy weak lensing, galaxy clustering, and line-intensity mapping, where similar non-Gaussian features arise. As a robust complement to the traditional power spectrum, these statistics will play a key role in maximizing the scientific return of forthcoming cosmological surveys.

\section{Acknowledgments}
We thank Will Coulton, Adrian Bayer, and Max Lee for useful discussions.  SFC, JCH, and ZH acknowledge support from NASA grant 80NSSC24K1093 [ATP]. JCH also acknowledges support from NASA grant 80NSSC23K0463 [ADAP].  JCH thanks the Kavli Institute for Theoretical Physics (KITP) for hospitality during the completion of this work; this research was supported in part by grant NSF PHY-2309135 to the KITP. The authors also acknowledge the Texas Advanced Computing Center (TACC) at The University of Texas at Austin for providing computational resources that have contributed to the research results reported within this paper. We acknowledge computing resources from Columbia University’s Shared Research Computing Facility project, which is supported by NIH Research Facility Improvement Grant 1G20RR030893-01, and associated funds from the New York State Empire State Development, Division of Science Technology and Innovation (NYSTAR) Contract C090171, both awarded April 15, 2010. This is not an official Simons Observatory collaboration paper.

\appendix

\section{Convergence of the Covariance Matrix}\label{appendix:convergenceCov}

\begin{figure*}[h!]
    \centering
    \includegraphics[width=0.85\linewidth]{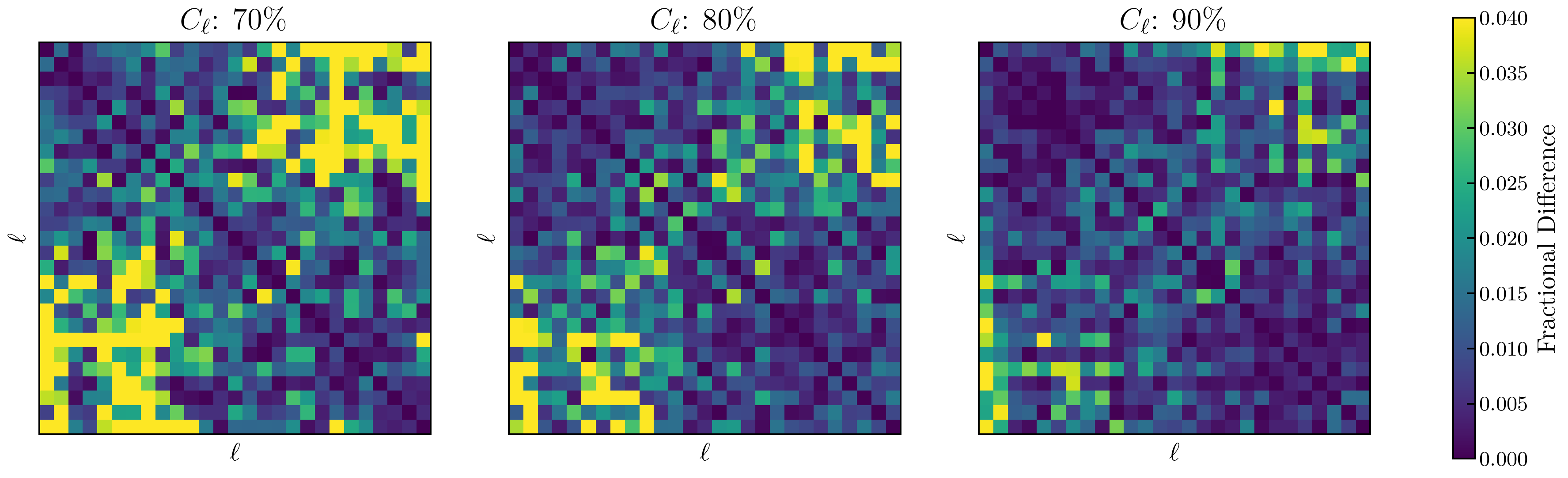}
    \includegraphics[width=0.85\linewidth]{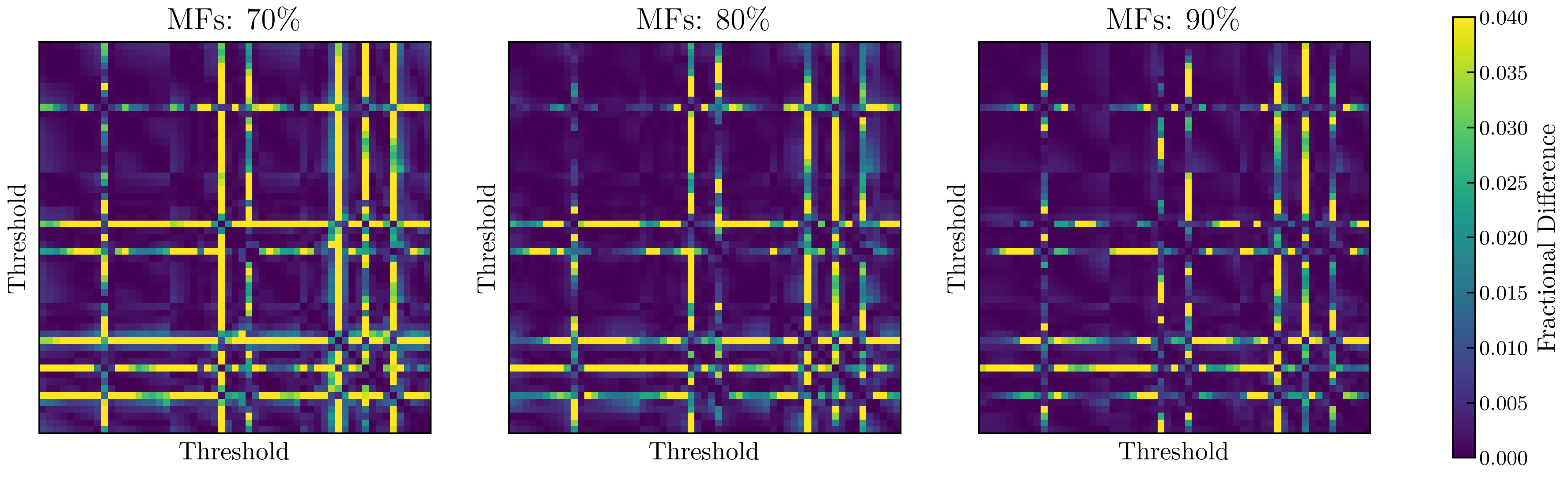}
    \includegraphics[width=0.85\linewidth]{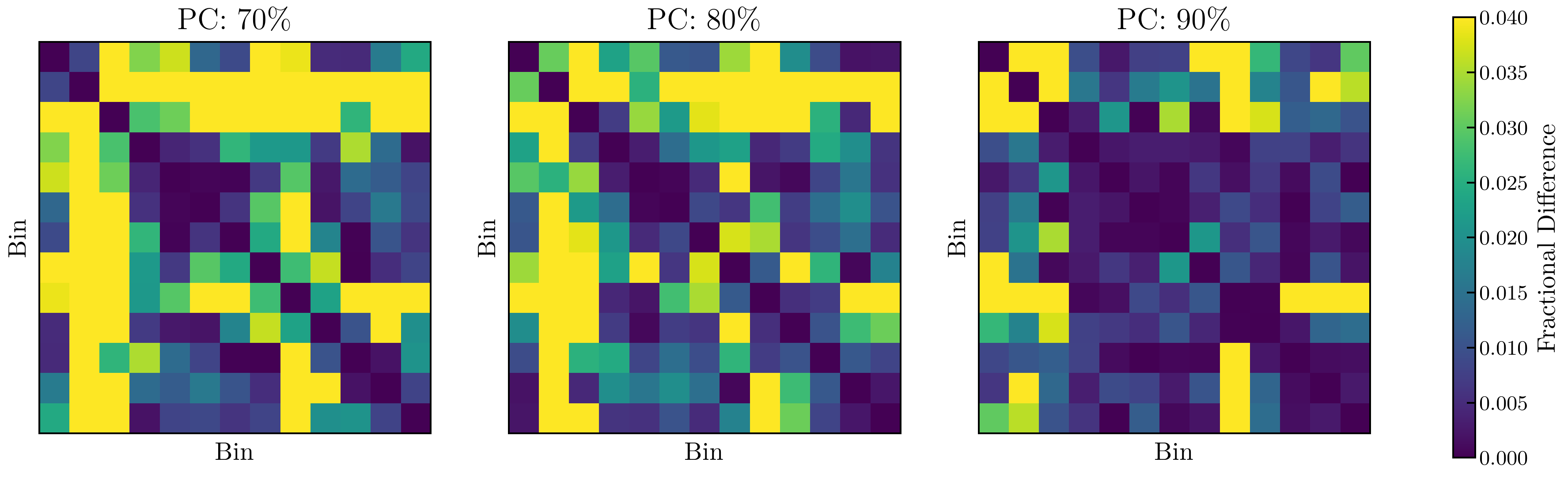}
    \includegraphics[width=0.85\linewidth]{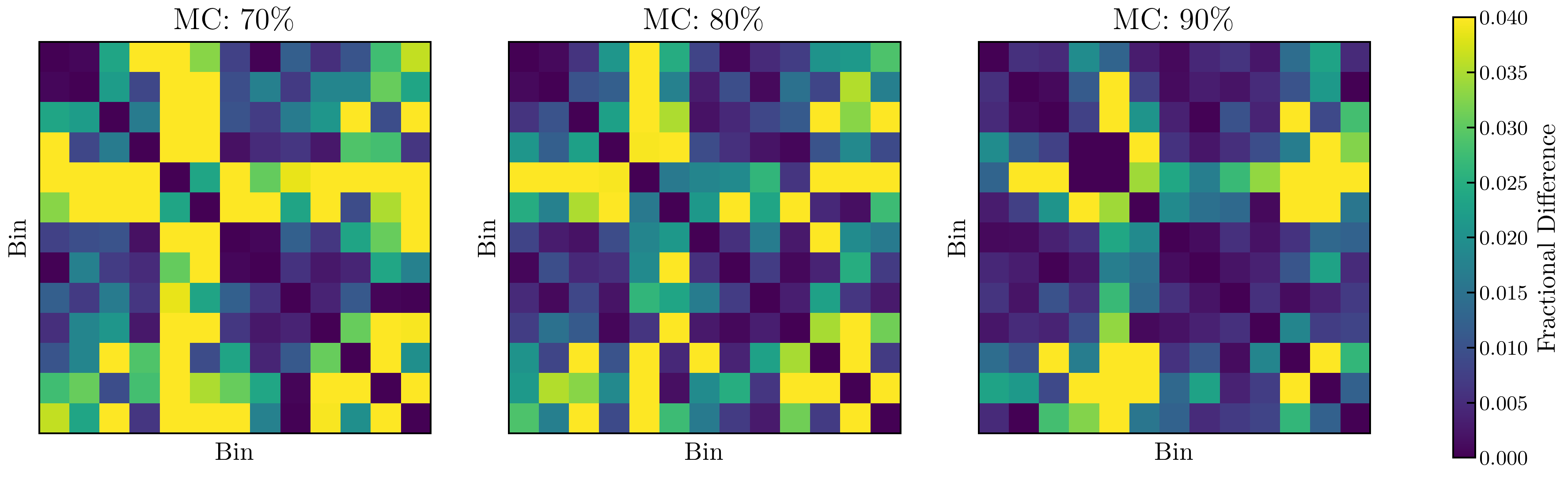}
    \includegraphics[width=0.85\linewidth]{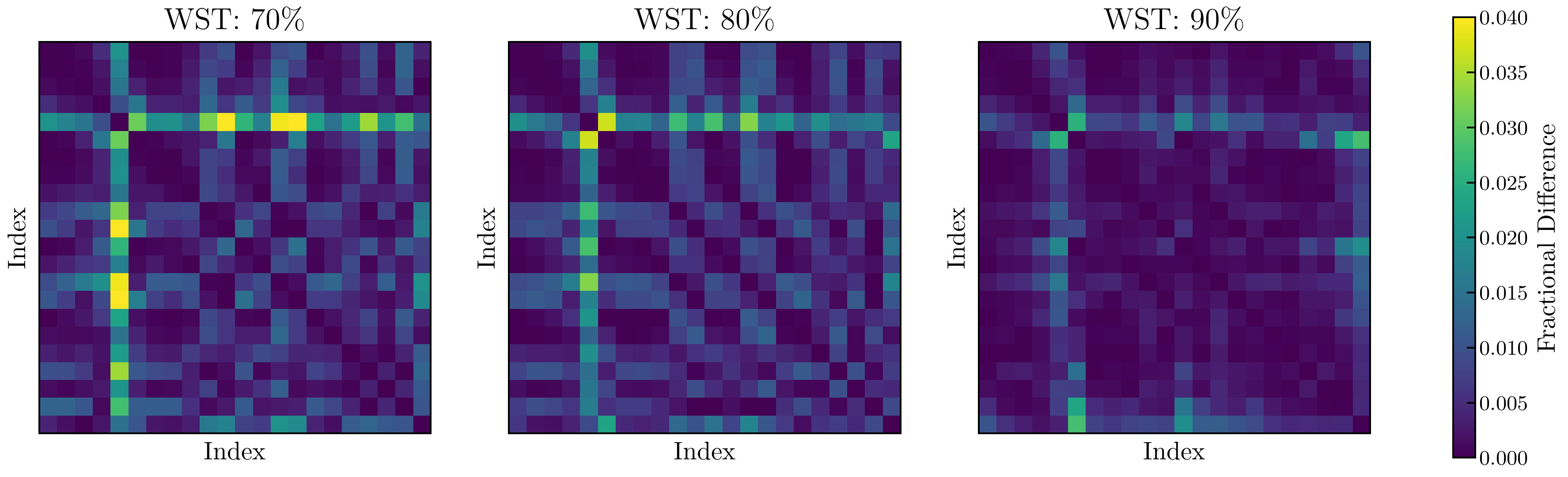}
    \caption{Fractional differences between the covariance matrix estimated from the full set of 10,000 simulations and those estimated from subsets of 90\%, 80\%, and 70\% of the full number of simulations. Each panel shows the convergence behavior for a different statistic, illustrating that the covariance estimates stabilize well before the full sample size is reached.}
    \label{fig:convergence_cov}
\end{figure*}

In this appendix, we present a convergence test for the covariance matrices used in our analysis. Figure~\ref{fig:convergence_cov} shows the fractional differences between the covariance estimated from 10,000 samples and those estimated from 9,000, 8,000, and 7,000 samples. For covariance matrices that are nearly diagonal, such as those for $C_\ell$, PC, and MC, we find that the diagonal elements are well converged. Although the off-diagonal elements exhibit larger fractional differences, we verify that excluding these off-diagonal terms yields consistent parameter constraints. In contrast, the covariance matrices for MFs and WST exhibit strong correlations across modes. In these cases, the off-diagonal elements also converge well, with noticeable deviations only in entries whose true values are close to zero.

\section{Effect of Noise on Non-Gaussian Statistics}\label{appendix:noiseOnStatistics}
In this appendix, we present a comparison of summary statistics, focusing in particular on MFs and PC/MC, computed from three classes of maps: noiseless maps directly drawn from the {\tt MassiveNuS} simulations, and noisy maps after lensing reconstruction, with and without Wiener filtering. In particular, the inclusion of noise introduces additional small-scale fluctuations, which enhances the amplitude of higher-order MFs and increases the abundance of peaks and minima. In contrast, Wiener filtering suppresses noise-dominated high-$\ell$ modes, resulting in a smoother field with reduced small-scale structure and a corresponding suppression of higher-order MFs and PC/MC.

\begin{figure*}[h!]
    \centering
    \includegraphics[width=\linewidth]{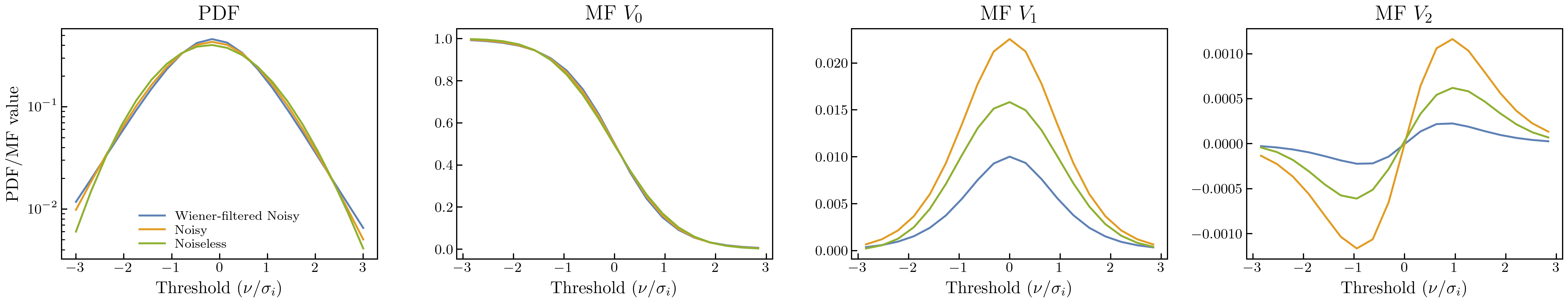}
    \includegraphics[width=0.5\linewidth]{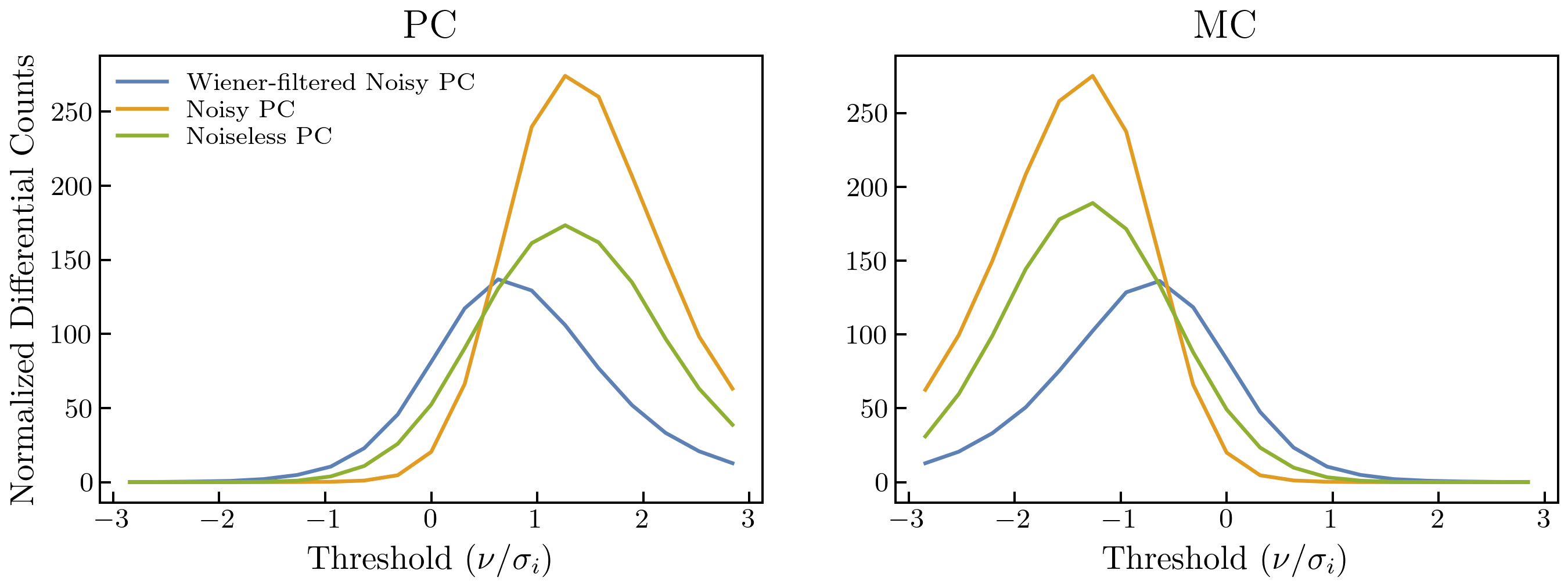}
    \caption{Minkowski functionals (MFs), one-point probability distribution function (PDF), and peak/minima counts (PC/MC) for noiseless, noisy, and Wiener-filtered noisy fields. The MFs/PDF (top row) and PC/MC (bottom row) are shown as functions of the normalized threshold $\nu = \delta / \sigma_i$, where $\sigma_i$ denotes the standard deviation of each field. The presence of noise significantly enhances small-scale structure and increases the abundance of peaks and minima, while Wiener filtering suppresses noise-dominated high-$\ell$ modes, leading to a reduction in topological complexity and extrema counts.}
    \label{fig:compare_stats}
\end{figure*}

\clearpage
\newpage

\bibliography{ref.bib}
\end{document}